\documentclass[numberedappendix]{emulateapj}
\shorttitle{Luminosity Function of Abell Clusters}
\shortauthors{Barkhouse et al.}
\submitted{Accepted for publication in ApJ}
\begin{document}
\title{The Luminosity Function of Low-Redshift Abell Galaxy Clusters}
\author{Wayne A. Barkhouse,\altaffilmark{1,2,4} H.K.C. Yee,
\altaffilmark{2,4} and Omar L\'{o}pez-Cruz\altaffilmark{3,4}}
\altaffiltext{1}{Department of Astronomy, University of Illinois, Urbana, IL 
61801; email: wbark@astro.uiuc.edu}
\altaffiltext{2}{Department of Astronomy and Astrophysics, University 
of Toronto, Toronto, ON, Canada, M5S 3H4; email: hyee@astro.utoronto.ca}
\altaffiltext{3}{Instituto Nacional de Astrof\'{i}sica, Optica y 
Electr\'{o}nica, Tonantzintla, Pue., M\'{e}xico; email: omarlx@inaoep.mx}
\altaffiltext{4}{Visiting Astronomer, Kitt Peak National Observatory. 
KPNO is operated by AURA, Inc.\ under contract to the National Science
Foundation.}

\begin{abstract}

We present the results from a survey of 57 low-redshift Abell galaxy 
clusters to study the radial dependence of the luminosity function (LF). 
The dynamical radius of each cluster, $r_{200}$, was estimated from the 
photometric measurement of cluster richness, $B_{gc}$. The shape of the 
LFs are found to correlate with radius such that the 
faint-end slope, $\alpha$, is generally steeper on the cluster outskirts. 
The sum of two Schechter functions provides a more adequate fit to 
the composite LFs than a single Schechter function. LFs based on the 
selection of red and blue galaxies are bimodal in appearance. The red LFs 
are generally flat for $-22\leq M_{R_{c}}\leq -18$, with a radius-dependent 
steepening of $\alpha$ for $M_{R_{c}}> -18$. The blue LFs contain a larger 
contribution from faint galaxies than the red LFs. The 
blue LFs have a rising faint-end component ($\alpha\sim -1.7$) for 
$M_{R_{c}}> -21$, with a weaker dependence on radius than the red LFs. 
The dispersion of $M^{\ast}$ was determined to be 0.31 mag, which is 
comparable to the median measurement uncertainty of 0.38 mag. This suggests 
that the bright-end of the LF is universal in shape at the 0.3 mag level. 
We find that $M^{\ast}$ is not correlated with cluster richness when using 
a common dynamical radius. Also, we find that $M^{\ast}$ is weakly 
correlated with BM-type such that later BM-type clusters have a brighter 
$M^{\ast}$. A correlation between $M^{\ast}$ and radius was found for 
the red and blue galaxies such that $M^{\ast}$ fades towards the cluster 
center.

\end{abstract}


\section{Introduction}

The study of the formation and evolution of galaxies is a fundamental 
avenue of research in the process of understanding astrophysical and 
cosmological issues. How galaxies form and evolve can be studied using 
a variety of techniques, one of those being the galaxy luminosity 
function (LF). The galaxy LF, assuming that galaxy 
mass-to-light ratios are nearly constant for similar types of galaxies, 
can potentially provide a direct link to the initial mass function and 
hence the distribution of density perturbations that are thought to 
give rise to galaxies \citep{Press74}. Since most galaxies are not 
isolated entities, evolutionary processes, in addition to those 
expected for an aging stellar population, can occur as galaxies 
interact with their environment. 

The galaxy LF --- the number of galaxies per unit volume 
in the luminosity interval $L$ to $L+dL$ --- can be used as a diagnostic 
tool to search for changes in the galaxy population. In particular, the 
LF for cluster galaxies can help ascertain the influence of the cluster 
environment on the galaxy population. For example, a change in the shape 
of the LF with respect to cluster-centric radius provides important 
insight into the dynamical processes at work in the cluster environment. 

A central theme in the early studies of the galaxy cluster LF 
has been to determine whether the LF is universal in shape 
\citep[e.g.,][]{Hubble36,Abell62,Oemler74}. While introducing the modern 
form of the LF, the so-called ``Schechter Function'', \citet{Schechter76} 
suggested that the cluster LF is universal in shape, and can be characterized 
with a turnover of $M_{B}^{\ast}=-20.6+5\,\mbox{log}\,h_{50}$ and a 
faint-end slope of 
$\alpha=-1.25$. Further support for a universal LF has been provided by 
several studies such as \citet{Lugger86}, \citet{Colless89}, 
\citet{Gaidos97}, \citet{Yagi02}, and \citet{DePropris03}. In contrast, 
several studies have shown that the shape of the cluster LF is not universal 
\citep[e.g.,][]{Godwin77,Dressler78,Lopez97b,Piranomonte01,Hansen05,
Popesso06}. One expects that the LF depends on cluster-centric radius since 
the mixture of galaxy morphological types should vary with radius, as implied 
by the morphology--density relation \citep{Dressler80}. Since different 
morphological types are characterized by different LFs \citep{Binggeli88} 
the cluster LF (integrated over all galaxy types) should 
not be universal. Indeed, some studies have provided evidence that the 
cluster LF does vary with cluster-centric radius 
\citep[e.g.,][]{Beijersbergen02a,Goto05,Hansen05,Popesso06}.

The main goal of this paper is to investigate the change in the cluster 
$R_c$-band LF as a function of cluster-centric radius. To avoid the inherent 
bias that has plagued numerous studies, the cluster LF will be compared 
based on scaling relative to the dynamical radius, $r_{200}$. The use of 
a dynamical radius for comparing galaxy populations for a sample of 
clusters, provides one of the most robust, least-biased, photometric 
survey yet published on the LF of Abell clusters. Directly comparing LFs 
that sample only the cluster core region with other cluster LFs that 
extend to the outskirts, will suffer from radial sampling bias given 
that the shape of the LF has been shown to depend on cluster-centric radius 
\citep[e.g.,][]{Christlein03,Hansen05,Popesso06}. A direct comparison of 
the galaxy population with respect to cluster-centric radius based on 
$r_{200}$, will help to accurately measure the change in the properties 
of cluster galaxies as a function of global environment. These data will 
also provide information to help settle the long-standing debate regarding 
the universality of the cluster galaxy LF and the properties of the faint 
dwarf galaxy component. 

This paper is the third in a series resulting from a multi-color imaging 
survey of low-redshift ($0.02\leq z \leq 0.2$) Abell clusters. The paper 
is organized as follows. In \S 2 we present a brief overview of the sample 
selection and data reduction procedure. In \S 3 we describe the methodology 
for generating the galaxy cluster LF. In \S 4 we examine the LF for 
different color-selected galaxy populations of our cluster sample. 
Discussion and conclusions are presented in \S 5. Finally, various selection 
effects and biases are explored in the Appendix. Further details regarding 
sample selection, observations, image preprocessing, catalogs, and 
finding charts can be found in \citet{Lopez97}, \citet{Barkhouse03}, and 
Barkhouse et al. (2007a; in preparation, Paper I of this series). 
A detailed discussion of the color-magnitude relation (CMR) of early-type 
galaxies using this survey can be found in Paper II \citep{Lopez04}. 
Paper IV characteristics the cluster galaxy luminosity and color distribution 
by examining the dwarf-to-giant ratio and the galaxy red-to-blue count ratio 
(Barkhouse et al. 2007b, in preparation). Recent observations suggest 
that the best cosmological model is characterized by $\Omega_{M}\simeq 0.3$, 
$\Omega_{\lambda}\simeq 0.7$, and 
$H_{0}\simeq 70~\mbox{km}~\mbox{s}^{-1}~\mbox{Mpc}^{-1}$ 
\citep[e.g.,][]{Spergel03}. Since the effects of curvature and dark energy 
are negligible at low-redshifts ($z< 0.2$) and to allow direct comparisons 
with previous studies, we have set for convenience, unless otherwise 
indicated, $H_{0}=50~\mbox{km}~\mbox{s}^{-1}~\mbox{Mpc}^{-1}$ and 
$q_{0}=0$ throughout this paper.

\section{Observations and Data Reductions}

The galaxy cluster sample utilized for this paper is identical to that 
described in Paper II of this series \citep{Lopez04}. We summarize the 
observations and data reductions below.

The galaxy cluster sample is composed of Abell clusters selected mainly 
from the X-ray compilation of \citet{Jones99}. The primary cluster sample 
was selected 
based on the following criteria; (1) clusters should be at high galactic 
latitude, $|b| \geq 30\degr$; (2) their redshifts should lie within the 
range $0.04\leq z\leq 0.20$; (3) the Abell richness class (ARC) should, 
preferably, be $> 0$; and (4) the declination $\delta\geq -20\degr$. 
Some $\mbox{ARC}=0$ clusters were included in the final sample due to 
the lack of suitable clusters at certain right ascensions during the 
observations. This sample includes 47 clusters of galaxies observed in 
$B$, and Kron-Cousins $R_c$ and $I$ at KPNO with the 0.9 m telescope using the 
$2048 \times 2048$ pixel T2KA CCD ($0.68\arcsec~\mbox{pixel}^{-1}$) 
\citep[hereafter the LOCOS sample; L\'opez-Cruz et al. 2007, in 
preparation]{Lopez97,Yee99,Lopez01}.

A sub-sample of eight clusters from \citet{Barkhouse03} is included to 
complement our 47-cluster sample by covering the low-redshift interval 
from $0.02\leq z\leq 0.04$. These data were obtained at KPNO with the 0.9 m 
telescope using the 8K MOSAIC camera ($8192\times 8192$ pixels; 
$0.423\arcsec~\mbox{pixel}^{-1}$). The clusters for this sample were 
selected using the previous criteria except that $\mbox{ARC}=0$ clusters 
were not preferentially excluded. In addition, two clusters imaged in $B$ 
and $R_c$ are included from \citet{Brown97} using the same instrumental 
setup as the LOCOS sample and selection criteria as the MOSAIC data. All 
clusters in the sample are detected in X-rays and are found to have a 
prominent CMR \citep{Lopez04}. 

The integration times for our 57-cluster sample varies from 250 to 9900 s, 
depending on the filter and the redshift of the cluster (only the $B$- and 
$R_c$-band data are considered here). Control fields are also an integral 
part of this survey. For this study we use a total of six control fields 
in both the $B$ and $R_c$ filters. The control fields were chosen at random 
positions on the sky at least $5\degr$ away from the clusters in our sample. 
These control fields were observed using the MOSAIC camera to a comparable 
depth and reduced in the same manner as the cluster data. All observations 
included in this study were carried out during 1992-1993 and 1996-1998. 

Processing of the 8k mosaic images were done using the {\tt mscred} package 
within the IRAF environment. The photometric reduction was carried out using 
the program PPP \citep[Picture Processing Package;][]{Yee91}, which includes 
algorithms for performing automatic object finding, star/galaxy 
classification, and total magnitude determination. A series of improvements 
to PPP described in \citet{Yee96} that decreases the detection of false 
objects and allows star/galaxy classification in images with a variable 
point-spread-function (PSF) was utilized.

The object list for each cluster is compiled from the $R_c$ frames. The $R_c$ 
frames are chosen because they are deeper than the images from the other 
filters. Galaxy {\it total magnitudes} are measured with PPP using a 
curve-of-growth analysis. The maximum aperture size ranged from 
$20\arcsec$ for faint galaxies ($R_c> 18.5$) to as large as $120\arcsec$ for 
cD galaxies in $z\sim 0.02$ clusters. An optimal aperture size for each 
object is determined based on the shape of the curve-of-growth using 
criteria described in \citet{Yee91}. The photometry of galaxies near the 
cluster core was carried out after the cD and bright early-type galaxies 
had been removed using profile-modeling techniques developed by 
\citet{Brown97}. 

Galaxy colors were determined using fixed apertures of 
$11.0\,h^{-1}_{50}\,\mbox{kpc}$ on the images of each filter at the redshift 
of the cluster, sampling identical regions of galaxies in different filters, 
while imposing a minimum color aperture of $\sim 3$ times the 
full-width-half-maximum (FWHM) in order to 
avoid seeing effects (average seeing $\sim 1.5\arcsec$). The overall 
internal accuracy in the color determinations is $\sim 0.005$ mag in $B-R_c$ 
for bright objects. The errors for faint objects can be as large as 
0.5 mag in $B-R_c$. We note that the total magnitude of a galaxy 
is determined using the growth curve from the $R_c$ image, while the total 
magnitude in the $B$ image is determined using the color difference with 
respect to the $R_c$ image \citep[for more details, see][]{Yee91}.
 
Star/galaxy classification was performed within PPP using a classifier 
that is based on the comparison of the growth curve of a given object to 
that of a reference PSF. The reference PSF is generated as the average of 
the growth curves of high signal-to-noise ratio (S/N), non-saturated stars 
within the frame. The classifier measures the ``compactness'' of an object 
by effectively comparing the ratio of the fluxes of inner and outer parts 
of an object with respect to the reference PSF.

Instrumental magnitudes are calibrated to the Kron-Cousins system by 
observing standard stars from \citet{Landolt92}. Due to the large field, 
up to 45 standard stars can be accommodated in a single frame. The color 
properties of the standard stars cover a large color range that encompasses 
those of elliptical and spiral galaxies. The standard stars are measured 
using a fixed aperture of 30 pixels for the LOCOS frames and 32 pixels for 
the MOSAIC data. These aperture sizes are selected as being the most 
stable after measuring the magnitudes using a series of diameters. We 
adopt the average extinction coefficients for KPNO and fit for the 
zero points and color terms. The rms in the residuals of individual 
fittings is in the range $0.020-0.040$ mag, which is comparable to the 
night-to-night scatter in the zero points. This can be considered as the 
systematic calibration uncertainty of the data.

The final galaxy catalogs were generated using the information and 
corrections derived previously. For data obtained under non-photometric 
conditions, single cluster images were obtained during photometric nights 
in order to calibrate the photometry (three clusters in total). The 
completeness limit for each field is based on a fiducial $5\sigma$ limit 
determined by calculating the magnitude of a stellar object with a brightness 
equivalent to having a $\mbox{S/N}=5$ in an aperture of $2\arcsec$. This 
is done by scaling a bright unsaturated star in the field to the $5\sigma$ 
level. Since the $5\sigma$ limit is fainter than the peak of the galaxy 
count curve, and hence below the $100\%$ completeness limit for 
galaxies, a conservative $100\%$ completeness limit is in general 
reached at $0.6-1.0$ mag brighter than the $5\sigma$ detection. See 
\citet{Yee91} for a detailed discussion of the completeness limit relative 
to the $5\sigma$ detection limit. 

Galaxy colors and magnitudes were corrected for the extinction produced by 
our Galaxy. The values of the Galactic extinction coefficients were 
calculated from the \citet{Burstein82} maps using the reported $E(B-V)$ 
values, or directly from the $A_{B}$ tabulations for bright galaxies 
\citep{Burstein84} with coordinates in the vicinity of our pointed 
observations using NED. Extinction values used for each cluster are provided 
in Paper I.

\section{The Luminosity Function}
\subsection{Fitting the Schechter Function}

The galaxy cluster LF in modern times has mainly been 
parameterized using the Schechter function \citep{Schechter76}. This 
function has the form
\begin{equation}
\phi(L)\,dL = \phi^{\ast}(L/L^{\ast})^{\alpha}\,\exp(-L/L^{\ast})\,
d(L/L^{\ast}),
\end{equation}
where $\phi(L)\,dL$ is the number of galaxies per unit volume in the 
luminosity interval $L$ to $L + dL$, $\phi^{\ast}$ is the number per 
unit volume, and $L^{\ast}$ is the ``characteristic'' luminosity. 
This function 
is characterized by having an exponential shape at the bright-end and 
a power-law like feature, whose slope is measured by $\alpha$, at the 
faint-end (for example, see Figure 1). By introducing the change of 
variables $M-M^{\ast}=-2.5\log(L/L^{\ast})$, equation 1 can 
be written in terms of absolute magnitude as
\begin{equation}
n(M)\,dM = kN^{\ast}e^{[k(\alpha +1)(M^{\ast}-M)-
\exp\{k(M^{\ast}-M)\}]}dM,
\end{equation} 
where $k=0.4\ln 10$ \citep[cf.,][]{Colless89}.

The Schechter function is fit to the cluster galaxy counts following 
the procedure in \citet{Lopez97}. In summary, the function parameters 
$M^{\ast}$, $N^{\ast}$, and $\alpha$ are estimated by performing a 
$\chi^{2}$ minimization of the form
\begin{equation}
\chi^{2}=\sum_{i}\frac{(N_{i}-N_{i}^{e})^{2}}{\sigma^{2}_{i}},
\end{equation}
where $N_{i}$ is the net galaxy counts in the $i^{th}$ bin of the 
observed LF, $N_{i}^{e}$ is the expected number 
of counts in the $i^{th}$ bin of width $\Delta M$, and $\sigma_{i}$ 
is the uncertainty of the counts in the $i^{th}$ data bin. The 
expected number of counts in the $i^{th}$ bin, corrected for the loss 
of information due to the finite bin size, is given by \citep{Schechter76}
\begin{equation}
N_{i}^{e}=n(M_{i})\,\Delta M +n_{i}^{''}(M_{i})\,\Delta M^{3}/24,
\end{equation}
where the derivative is with respect to absolute magnitude, and 
$\Delta M$ is the bin width. This correction is derived by Taylor-expanding 
$n(M)$ to a third order about the bin's center and integrating 
$N_{i}^{e}=\int_{\scriptscriptstyle M-\frac{1}{2}{\Delta}M}^
{\scriptscriptstyle M+\frac{1}{2}{\Delta}M}n(M)dM$. The uncertainty, 
$\sigma_{i}$, is taken to be \citep[cf.,][]{Lugger86}
\begin{equation}
\sigma_{i}=[(N_{i}^{e}+N_{bi})+1.69\,N_{bi}^{2}]^{1/2},
\end{equation} 
where $N_{bi}$ is the background counts in the $i^{th}$ bin, 
$(N_{i}^{e}+N_{bi})^{1/2}$ is the Poisson uncertainty in the uncorrected LF 
counts, and the second term in the square root expression is the 
measured field-to-field variation per bin in the background field 
counts (see \S 3.2). The uncertainty of the observed net galaxy counts 
due to cosmic variance is thus taken into account when fitting the Schechter 
function. For a single Schechter function fit, $N^{\ast}$ is fixed so that 
the total number of observed galaxies in the data set is equal to the 
number predicted by the Schechter function. The parameters $M^{\ast}$ 
and $\alpha$ are obtained by minimizing equation 3 using the 
Levenberg-Marquardt method \citep{Press92}.

\subsection{Background Galaxy Correction}

The $R_c$-band LF is constructed statistically by subtracting a background 
galaxy population from the cluster galaxy counts. This method, in contrast to 
measuring the redshift of individual galaxies in the cluster field, relies 
on an accurate determination of the background field population. This 
statistical approach has been used in numerous studies to date 
\citep[e.g.,][]{Oemler74,Schechter76,Colless89,Driver98,Yagi02,Andreon04}. 
The modal field-to-field variation per magnitude bin has been measured to be 
$\sim 30\%$ above Poisson statistics among the six background fields. We 
use this value to approximately account for the additional uncertainty in 
the background counts due to field-to-field variations in equation 5. 
Although the use of galaxy redshifts would provide a more robust 
determination of the cluster LF, the relatively modest variation of the 
background counts, with respect 
to the cluster counts, makes the statistically derived cluster LF valid. A 
similar conclusion has also been reached by a number of independent 
studies \citep[e.g.,][]{Driver98}.

Several studies have examined the effect on the derived cluster LF using a 
global background galaxy field correction versus one measured locally for 
each cluster \citep[e.g.,][]{Goto02,Hansen05,Popesso05,Gonzalez06}. For 
example, \citet{Popesso05} has shown for a study of 69 clusters based on 
SDSS DR2 data, that there is no significant difference in the measured 
cluster LF using either a global or local background subtraction technique.

Contamination from the 2-d projection on the sky of a distant cluster in the 
field-of-view of the target cluster can prove to be problematic by skewing 
the LF, especially at the faint-end where galaxies from the background 
cluster directly add to the desired LF. Fortunately, the CMR can help 
minimize this contamination by identifying the early-type 
red sequence of the target cluster \citep{Lopez04}. The effect of 
background clusters on the desired LF can thus be reduced by selecting an 
appropriate color cut, thus eliminating objects redder than the cluster 
red sequence. This method can also help to locate foreground clusters which, 
given the low redshift of our cluster sample, are not a significant concern 
for this study.

To minimize contamination from background galaxies, we cull galaxies 
that are 0.22 mag redward of the CMR (i.e., $\geq 3.0$ times the average 
$B-R_{c}$ dispersion of the cluster red sequence). The dispersion of the 
cluster red sequences are tabulated in Table 1 of \citet{Lopez04} and 
histogram representations of the rectified $B-R_c$ color distributions are 
presented in Paper I \citep[see also Figure 1 from][]{Lopez04}. In several 
cases (e.g., A2152), a color cut $< 3\sigma$ is used if a second red sequence 
from a more distant cluster (and hence at a redder $B-R_c$) is apparent in 
the color-magnitude diagram (CMD) of the target cluster. For faint 
magnitudes, we cull galaxies redder than $2.5\sigma_{B-R_{c}}$ if 2.5 times 
the average uncertainty in the galaxy $B-R_{c}$ is redder than 3.0 times 
the dispersion of the cluster red sequence.

\subsection{K-Correction}

K-corrections are applied using a single parametrization based on early-type 
galaxies, which dominant the cluster galaxy population \citet{Dressler80}. 
At the low redshift of our sample, the difference between early- and 
late-type galaxies is minimal for the $R_c$-band. In general, the maximum 
K-correction applied was $\sim0.2$ mag. The K-correction adopted for each 
cluster is tabulated in Paper I. All LFs presented in this paper 
have been extinction- and K-corrected, and no attempt has been made to correct 
for individual internal galaxy absorption.

\subsection{Cluster Dynamical Radius}

Studies of the properties of galaxy clusters, such as the CMR and LF, have 
been routinely compared on a 
cluster-by-cluster basis. Nearly all of these studies define a ``cluster'' 
based on the total area covered by the telescope detector 
\citep[e.g.,][]{Dressler78,Lopez97} or by using a specific physical 
length \citep[e.g., $1\,h^{-1}_{100}$ Mpc;][]{Yagi02}. The large variation in 
cluster richness \citep{Yee99} inhibits the usefulness of the above 
techniques for robustly comparing cluster properties. Some authors have 
attempted to ``normalize'' clusters by directly comparing only clusters of 
comparable class (e.g., Bautz-Morgan type) or by weighting each cluster 
according to richness \citep{Garilli99,Piranomonte01,DePropris03}. The 
use of a variety of different methods have certainly contributed to 
conflicting results that have emerged from past investigations regarding 
measurements such as the universality of the LF \citep{Lugger86,Driver98,
Popesso06}.   

As a means of computing a ``dynamical'' radius within which cluster 
characteristics can be robustly compared, the $r_{200}$ radius was 
calculated for each cluster. The $r_{200}$ radius marks the size of 
a cluster within which the average density is 200 times the critical 
density, and follows from the definition used in \citet{Carlberg97} and 
\citet{Yee03}. The $r_{200}$ radius is expected to contain the bulk of 
the virialized mass of a cluster \citep[e.g.,][]{Cole96} and is used in 
this study as a scaling factor to compare cluster features. The use of 
this type of ``normalization'' allows 
us to compare galaxy cluster populations in a less biased fashion, 
especially those properties which are a function of cluster richness and 
cluster-centric radius. This approach has recently been implemented 
by \citet{Hansen05} and \citet{Popesso06} to study the properties of a 
sample of clusters/groups from the SDSS.

The procedure for determining $r_{200}$ involves the calculation of the 
velocity dispersion, which requires redshift measurements for a number of 
cluster galaxies. The data available for this study do not include 
redshift information for cluster members, thus an alternative method was 
employed to estimate $r_{200}$ for each cluster.\footnote{Only 
approximately 30\% of the 57 clusters have published robust 
velocity dispersions.} This method relies on the correlation between 
$B_{gc}$ and $r_{200}$ as measured for the CNOC1 sample \citep{Yee03}. The 
$B_{gc}$ parameter is a measure of the cluster-center galaxy correlation 
amplitude, and has been shown to be a robust estimator of cluster 
richness \citep{Yee99,Yee03}. The measured values of $B_{gc}$ for our 
cluster sample are calculated using the method outlined in \citet{Yee99}. 
Figure~\ref{Bgc-r200} depicts the relationship 
between $r_{200}$ and $B_{gc}$ for 15 clusters from the CNOC1 survey 
\citep{Yee96}, adopted from \citet{Yee03}. We note that the $r_{200}$ vs. 
$B_{gc}$ figure in Yee \& Ellingson (see their Figure 5) used the less 
well-determined $r^{\prime}_{200}$ from \citet[see their Table 1]{Carlberg97}. 
This explains the decrease in the scatter of $r_{200}$ vs. $B_{gc}$ for our 
Figure~\ref{Bgc-r200} compared to the corresponding figure in Yee \& 
Ellingson. A fit to these data yields 
\begin{equation}
\mbox{log}~r_{200}=(0.48\pm 0.10)~\mbox{log}~B_{gc}-(1.10\pm 0.31), 
\end{equation}  
where $r_{200}$ has units of Mpc, and $B_{gc}$ units of $\mbox{Mpc}^{1.8}$. 
The rms scatter in the derived values of $r_{200}$ is on the 
order of 15\%. The fit was performed using the bisector bivariate correlated 
errors and intrinsic scatter (BCES) estimator \citep{Akritas96} since this 
algorithm accounts for uncertainties in both variables. The cluster 
MS 1455+22 (open circle in Figure~\ref{Bgc-r200}) was excluded from the 
fit since it is $\sim3\sigma$ from the expected relation (see 
Yee \& Ellingson 2003 for a detailed discussion of MS 1455+22). 

The estimation of $r_{200}$ for each cluster is accomplished by 
calculating the value of $B_{gc}$ directly from the galaxy cluster 
photometric catalog and then applying equation 6 to determine $r_{200}$. 
Table 1 lists the values of $B_{gc}$ and $r_{200}$ for each cluster used 
in this study. The tabulated uncertainty in the value of $r_{200}$ is 
calculated directly from the 15\% rms scatter. 

\citet{Yee99} showed that a subset of our Abell cluster sample has a 
similar relationship between cluster velocity dispersion ($\sigma_{v}$) 
and $B_{gc}$ as the CNOC1 sample, and since $r_{200}$ is estimated from 
$\sigma_{v}$, the $r_{200}$--$B_{gc}$ relation for the Abell clusters 
should be similar to that of the CNOC1 sample.

To test the validity of our results, we compare the values of $r_{200}$ from 
Table 1 for those clusters in common with \citet{Rines03}, \citet{Miller05}, 
\citet{Popesso07}, and \citet{Aguerri07}. For the 3 clusters in 
common with Rines et al., we find a mean difference in $r_{200}$ 
of $-0.17\pm 0.17$ Mpc (all physical length scales have been converted to 
our distance scale), where our values are greater on average (the rms is 
given as the uncertainty). For the 10 clusters in common with Miller et al., 
we find a mean difference of $0.73\pm 0.73$ Mpc. The mean difference for 
the 9 clusters in common with Popesso et al. is $0.25\pm 0.32$ Mpc, while 
the average difference for the 12 clusters in common with Aguerri et al. is 
$-0.26\pm 0.62$ Mpc. The larger discrepancy with the Miller et al. sample is 
due to the uncommonly large $r_{200}$ values ($r_{200}>4$ Mpc) for the three 
richest clusters. In fact, Miller et al. cautions that the radius within 
which the density measurements have been made to determine $r_{200}$ may be 
inaccurate. We believe that the Miller et al. $r_{200}$ values are 
biased-high for the more massive clusters in our comparison sample. If we 
restrict our analysis to the combined Rines et al., Popesso et al. and 
Aguerri et al. samples, we find a mean $r_{200}$ difference of 
$-0.07\pm 0.53$ Mpc. Thus, the values of $r_{200}$ derived via equation 7 
are reasonable. (See the discussion in \S A.5 regarding the effect on the 
derived LF for a 15\% scatter in $r_{200}$.)
 
\section{Results}

\subsection{Individual Cluster Luminosity Functions}

The $R_c$-band LFs for the 57 clusters presented in this 
paper are depicted in Figure 1 for galaxies brighter than the 100\% 
completeness limit. To help facilitate the comparison between clusters, 
the LFs are generated using galaxy counts within a radius of 
$(r/r_{200})\leq 0.4$ from each cluster center. The cluster center is 
normally selected using the brightest cluster galaxy (BCG) or, when 
some doubt exists, the brightest early-type galaxy that is closest to the 
X-ray centroid as given, for example, by \citet{Jones99}. Figure 1 
includes the $R_c$-band LF for A496 and A1142 from data obtained by 
\citet{Brown97}. The small size of the detector (2k$\times$2k) and the 
low redshift, limits the LF coverage to within a cluster-centric radius 
of $(r/r_{200})\leq 0.2$ for Abell 496 and $(r/r_{200})\leq 0.3$ for 
Abell 1142.

The cluster LFs presented in Figure 1 have each been fit with a Schechter 
function in the range $-24\leq M_{R_c}\leq -20$, with the faint-end slope 
fixed 
at $\alpha=-1$. This has been done to help serve as a reference point for 
comparing individual clusters. The fitted value of $M^{\ast}$ for each 
cluster is tabulated in Table 1 and the best-fit Schechter function is 
represented by the solid line in Figure 1. The fitting of the LFs does not 
include the BCG. The presence of these galaxies is easily noticed by their 
affect on the brightest magnitude bin, whose value is usually offset from 
the best-fit Schechter function. \citet{Schechter76} remarked that BCGs do 
not seem to be a natural extension of the cluster LF 
\citep[see also,][]{Sandage76,Dressler78,Loh06}. A fit to his composite LF 
for a sample of 13 clusters was more robust when the BCG was excluded from 
each cluster. This has led to numerous debates on the formation mechanism 
of BCGs \citep[e.g.,][]{Geller76,Bhavsar89,Bernstein01}. Since most studies 
of cluster LFs exclude the BCG from the LF fit, BCGs will not be included 
in subsequent LF analysis unless otherwise noted (see additional discussion 
in \S A.6).
 
Visual inspection of the individual cluster LFs in Figure 1 indicates 
that, in general, the bright-end appears to be well-fit by a Schechter 
function, with a rising faint-end ($M_{R_c}\gtrsim -19$) of various strength. 
The presence of a ``flat'' LF ($\alpha=-1.0$) at the faint-end usually 
occurs for clusters in which the uncertainty in the net galaxy counts at 
the faint-end is large, or the LF has not been sampled to a sufficient 
depth to reveal a rising faint-end component, although we can not rule 
out intrinsically flat LF clusters.

An additional feature, visible for several cluster LFs, is a ``dip'' in 
the galaxy counts at $-20\lesssim M_{R_c}\lesssim -19$. This characteristic is 
most-prominent for A84, A154, A634, A690, A1291, A1569, A1656, A1795, A2384, 
and A2556. A possible cause of this feature is the variation in the ratio of 
galaxy types that comprise the individual cluster galaxy population. This may 
result from the fact that the elliptical and spiral galaxy LF is better 
described using a Gaussian function, while the dwarf galaxy LF is 
``Schechter-like'' in shape \citep{Binggeli88}.

\subsection{Composite Luminosity Function}

\subsubsection{Total Luminosity Function}

The statistical subtraction of a background galaxy population can drastically 
affect the accuracy of determining the shape of the cluster LF when the 
net number of cluster galaxies is small. Generally, this will be an 
important factor for galaxies located in the outskirts of clusters. To 
reduce the uncertainty in the shape of the LF at large cluster-centric 
radii, cluster galaxy counts have been combined to form a composite LF. 
This also averages out any apparent variations in the shape of the 
individual LFs due to cosmic variance in the background counts. To 
provide an adequate coverage of the faint-end of the LF, we have selected a 
sub-sample of 29 clusters for the composite LF that are 100\% complete to 
$M_{R_c}=-16.5$ (absolute $R_c$-band completeness limits, $M_{R_c}^{Com}$, are 
tabulated in Table 1). Following the procedure for combining cluster counts 
in adjacent magnitude bins \citep{Schechter76}, clusters complete to 
$M_{R_c}=-17.0$ (13 additional ones) are also included in our composite LF. 
This has been accomplished by scaling the number of net galaxy counts in the 
faintest magnitude bin ($-17.0\leq M_{R_c}\leq -16.5$) by the ratio 
$(N_{2}/N_{1})$; where $N_{1}$ is the total net galaxy count to 
$M_{R_c}=-17.0$ for the 29 clusters complete to 
at least $M_{R_c}=-16.5$, and $N_{2}$ is the total net galaxy count to 
$M_{R_c}=-17.0$ for the 42 clusters complete to at least $M_{R_c}=-17.0$. 
By following this method, we are able to construct a composite LF which is 
complete to $M_{R_c}=-16.5$ and contains galaxy counts from 42 individual 
clusters.

To search for differences in the LF which may correlate with cluster 
properties, we present the composite $R_c$-band LF in 
Figure~\ref{Comp-All-LF} for four different radial bins, centered on the 
BCG, extending out to $(r/r_{200})=1$. Examination of 
Figure~\ref{Comp-All-LF} clearly shows that a single Schechter function 
is inadequate to fully describe the shape of the composite LF at any 
radius. The sum of two Schechter functions has therefore been used to 
model the shape of the LF, with the resultant fit given by the solid line 
in Figure~\ref{Comp-All-LF}. 

Several recent studies have determined 
that the sum of two Schechter functions provides a more adequate fit to 
the cluster LF than a single Schechter function 
\citep[e.g.,][]{Driver94,Hilker03,Gonzalez06,Popesso06}. Alternative 
LF fitting functions include a Gaussian for the bright-end and a single 
Schechter function for the faint-end 
\citep[e.g.,][]{Thompson93,Biviano95,Parolin03}, a single power-law fit 
to the faint-end \citep[e.g.,][]{Trentham01}, and an Erlang plus a Schechter 
function \citep{Biviano95}.

For Figure~\ref{Comp-All-LF}, a single Schechter function was first fit 
to the bright-end from $-24\le M_{R_c}\le -20$, and a second Schechter 
function from $-24\le M_{R_c}\le -16.5$. Due to the degeneracy involved in 
fitting two Schechter functions, the faint-end slope of the first 
Schechter function (for the bright-end) was fixed at $\alpha_{1}=-1.0$. 
This procedure is justified since a fit to all 57 clusters for 
$-24\le M_{R_c}\le -20$ yields $\alpha=-0.96\pm 0.04$. Limiting the 
analysis to the 42 clusters used to form the composite LF gives 
$\alpha=-0.95\pm 0.05$. In addition, we imposed a second constraint 
that $N^{*}_{2}=2N^{*}_{1}$, where $N^{*}_{1}$ and $N^{*}_{2}$ are the scale 
factors of the Schechter function fit to the bright- and faint-end of the 
LFs, respectively. When $N^{*}_{1}$ and $N^{*}_{2}$ are free to vary, we 
find that the geometric mean of the ratio ($N^{*}_{2}/N^{*}_{1})=2.12$ when 
measured over the four radial bins from $(r/r_{200})=0.0$ to 1.0. Due to the 
relatively bright absolute magnitude limits ($-16.5$) and the strong 
coupling of the Schechter function parameters, we have chosen to derive the 
faint-end slope $\alpha$ as the primary parameter of interest and thus 
set ($N^{*}_{2}/N^{*}_{1})=2$.

The resultant Schechter function fit parameters for the composite LFs are 
tabulated in Table 2, along with the $1\sigma$ uncertainties, where 
$M_{1}^{\ast}$ and $M_{2}^{\ast}$ are the turnover in the bright and faint 
Schechter functions, respectively. Since the faint-end slope of the first 
function has been fixed at $\alpha_{1}=-1.0$, only $\alpha_{2}$ is presented. 
The individual clusters that comprise the sample of 42 clusters used to 
construct the composite LF cover various fractions of $(r/r_{200})$. This 
results in a different number of clusters contributing to the LF counts in 
each of the four separate radial bins (see column 7 from Table 2). 

Examination of Figure~\ref{Comp-All-LF} and Table 2 clearly indicates that 
the faint-end slope tends to become steeper as cluster-centric distance 
increases. The faint-end slope is significantly steeper, as measured out 
to $(r/r_{200})=1$, than the traditional value of $\alpha=-1.25$ 
\citep{Schechter76}. To facilitate the comparison between the different LFs, 
the LF for each of the four radial bins has been plotted 
together in Figure~\ref{Comp-All-LF-Scale}. The outer three LFs, 
$0.2\leq (r/r_{200})\leq 1.0$, have been scaled to match the counts 
of the inner-most LF in the $-22\leq M_{R_c}\leq -21$ magnitude range. This 
figure clearly demonstrates the trend of a steepening of the faint-end 
slope with increasing cluster-centric radius.

To determine what effect imposing the constraint 
 ($N^{*}_{2}/N^{*}_{1})=2$ has on the robustness of the measured 
radial-dependent change in the faint-end slope, 
we refit our LFs by allowing $N^{*}_{1}$ and $N^{*}_{2}$ to vary. Due 
to the degenerate nature of fitting for $N^{*}$, $M^{\ast}$ and 
$\alpha$ simultaneously, we measure the ratio $(N^{*}_{2}/N^{*}_{1})$ for 
the total cluster sample from four annuli by; 
i) fixing $\alpha_{2}=-2.01$, $M_{1}^{\ast}=-22.31$ 
(mean values averaged over radial bins) and $\alpha_{1}=-1.0$, and then 
solving for $N^{*}_{1}$, $N^{*}_{2}$ and 
$M_{2}^{\ast}$, and ii) fixing $M_{2}^{\ast}=-18.04$, $M_{1}^{\ast}=-22.31$ 
(mean values averaged over radial bins) and $\alpha_{1}=-1.0$, and 
then solving for $N^{*}_{1}$, $N^{*}_{2}$ and 
$\alpha_{2}$. The increase in the steepness of the faint-end slope with 
increasing cluster-centric radius will be manifest by an increase in the 
$(N^{*}_{2}/N^{*}_{1})$ ratio with increasing radius.

For case (i) we find that the ratio 
$(N^{*}_{2}/N^{*}_{1})=(0.49,0.50,5.10,13.00)$ for 
$(r/r_{200})=(0.0-0.2,0.2-0.4,0.4-0.6,0.6-1.0)$. For case (ii) we 
measure $(N^{*}_{2}/N^{*}_{1})=(0.85,2.76,2.60,4.19)$ for 
$(r/r_{200})=(0.0-0.2,0.2-0.4,0.4-0.6,0.6-1.0)$. These results are 
consistent with the increase in the steepness of the faint-end slope 
with increasing radius. Due to the degeneracy in fitting the LF, the 
relative fraction of bright and faint cluster galaxies can be best 
studied by integrating over the respective LFs. This will be discussed in 
detail in the context of the dwarf-to-giant ratio of red- and blue-selected 
galaxies in Paper IV of this series.

We compare our composite LF with two recent studies by converting to 
$R_{c}$ using \citet{Fukugita95} and adopting our distance scale. 
\citet{Gonzalez06} used a double Schechter function to characterize the 
composite LF of 728 groups/clusters selected from SDSS DR3. They find 
$M^{\ast}=-23.12\pm 0.12$ with $\alpha=-1.89\pm 0.04$ for galaxies 
selected within 1.0 Mpc of the group center. Dividing their sample into 
two radial bins (0.0--0.6 and 0.6--1.2 Mpc), Gonz\'alez et al. determined 
that $M^{\ast}$ gets brighter with increasing radius ($-22.9\pm 0.2$ to 
$-23.5\pm 0.2$), while the faint-end slope becomes steeper 
($\alpha=-1.80\pm 0.03$ to $\alpha=-1.99\pm 0.03$). The measured range in 
$M^{\ast}$ for this study is significantly brighter ($\sim0.6-1.1$ mag) 
than our results tabulated in Table 2. This may be related to the fact that 
the Gonz\'alez et al. sample consists predominately of group systems. 
For the faint-end, the measured slope values are consistent with our 
results, including the trend for a steepening of $\alpha$ with increasing 
radius. 
 
\citet{Popesso06} studied the composite LF of 69 clusters from the 
$ROSAT$ All Sky Survey/SDSS galaxy cluster catalog (RASS-SDSS) using the 
sum of two Schechter functions. The faint-end slope was found to increase 
from $\alpha=-2.02\pm 0.06$ to $-2.19\pm 0.09$ when the sampling radius 
increased from $r_{500}$ to $r_{200}$. This compares well with 
our measured range in $\alpha$ even though we use annuli to 
determine $\alpha$ rather than a circular region with a changing radius. 
For the bright-end LF, Popesso et al. finds $\alpha$ varies from 
$-1.05\pm 0.07$ for $r_{500}$ to $-1.09\pm 0.09$ for $r_{200}$. This 
result is consistent with our use of a fixed $\alpha_{1}=-1.0$. Also, 
$M^{\ast}$ of the bright-end was found to brighten slightly from 
$M_{R_{c}}=-22.54\pm 0.13$ at $r_{500}$ to $M_{R_{c}}=-22.64\pm 0.16$ 
for $r_{200}$. These results are consistent within $1\sigma$ of our measured 
values of $-22.26\pm 0.06$ and $-22.38\pm 0.15$ for the inner- and outer-most  
radial bin, respectively. In addition, we also detect a slight brightening 
of $M^{\ast}$ with increasing cluster-centric radius as reported by 
Popesso et al. for their sample.

\subsubsection{Red Sequence Luminosity Function}

To explore the dependency of the composite LF on the physical properties 
of cluster galaxies such as color, the cluster galaxy catalogs have been 
divided into various sub-samples that populate different regions of the 
CMD.

The composite LF of galaxies located on the red sequence of the CMR was 
constructed by selecting galaxies using the following criteria: 1) galaxies 
are selected if they are within $\pm 3.0$ times the average dispersion 
of the Gaussian fit to the CMR \citep[i.e., $\pm 0.22$ mag;][]{Lopez04} 
or, depending on which is greater, 2) within 2.5 times the average 
uncertainty in the galaxy $B-R_c$ color redward of the red sequence fit 
(see Figure~\ref{redseq-region}). In practice, galaxies at the bright-end 
of the red sequence will be selected according to the first criterion; 
as fainter galaxies are chosen, the uncertainty in the $B-R_c$ color will 
increase to a value where galaxies will be selected via criterion 2. 

The first criterion defines the loci of the red sequence in the 
CMD. The second criterion helps insure that faint red sequence galaxies near 
the completeness limit are selected, even though they will be scattered 
further from the CMR than the limit imposed by criterion 1 due to the 
increased uncertainty in the measured color. As stated in the second 
criterion, only galaxies redward of the CMR were chosen. This was 
implemented to minimize the inclusion of galaxies blueward of 
the red sequence (what we refer to as the ``blue'' galaxy population), 
which are scattered into the red sequence region due to the color 
uncertainty at faint magnitudes. To compensate for the ``missing'' members 
of the red galaxy population that are located blueward of the 
red sequence, the net galaxy counts for those galaxies fainter than the 
magnitude at which objects are culled based on criterion 2, have been 
increased by a factor of two, since in the absence of a blue galaxy 
population one might expect the red sequence members to be symmetrically 
distributed on either side of the CMR.

Figure~\ref{redseq-region} illustrates the region of the CMD where 
red sequence galaxies have been selected for Abell 260. For 
galaxies brighter than $R_c=18.8$ (indicated by the vertical dashed line), 
the $2.5\sigma_{B-R_c}$ is less than 3.0 times the average dispersion of the 
red sequence (0.22 mag), thus galaxies are chosen according to criterion 1. 
For galaxies fainter than $R_c=18.8$, the uncertainty in the color 
measurement invokes selection by criterion 2. As depicted in 
Figure~\ref{redseq-region}, only galaxies redder than the red sequence are 
selected for magnitudes fainter than $R_c=18.8$ mag.

The net red sequence galaxy counts were calculated by subtracting the 
background field counts from the ``raw'' red galaxy counts. The background 
field galaxies were selected using identical color cuts as imposed on 
the red sequence galaxy sample.
  
The composite red sequence galaxy LF for four different 
radial bins (same annuli as that used for the total composite LF) is 
depicted in Figure~\ref{Comp-Red-LF}. The LFs have been constructed by 
scaling and combining clusters, complete to $M_{R_c}=-16.5$, using the same 
method as that described for the total composite LF (\S 4.2.1). Visual 
inspection of the red sequence LFs shows that a single Schechter function, 
in general, would provide a reasonable fit to the overall shape of the LF 
except for the faintest magnitude bins. To compare the red sequence LF 
to the total LF, we fit the sum of two Schechter functions to the red 
sequence LFs. By allowing $N^{*}_{1}$ and $N^{*}_{2}$ to vary, we find 
that the geometric mean of the ratio ($N^{*}_{2}/N^{*}_{1}$)=0.4, averaged 
over the four radial bins. We thus impose that ($N^{*}_{2}/N^{*}_{1}$)=0.4 
and fit for $M_{1}^{\ast}$, $M_{2}^{\ast}$, and $\alpha_{2}$. These fits 
are depicted in Figure~\ref{Comp-Red-LF} and the best-fit parameters 
tabulated in Table 3. We note that the LF fit to the faint component for 
the inner-most region is poorly constrained due to the lack of data points 
from the rising faint-end slope. We also find that when holding 
$M_{2}^{\ast}$ fixed and allowing ($N^{*}_{2}/N^{*}_{1}$) and $\alpha_{2}$ 
to vary, ($N^{*}_{2}/N^{*}_{1}$) increases with increasing cluster-centric 
radius.

In Figure~\ref{Scale-Comp-Red-LF} the four red sequence LFs have been 
superimposed by scaling the outer three LFs to match the inner-most LF 
in the magnitude range $-22\leq M_{R_c}\leq -21$. All four LFs appear to be 
equivalent to each other for magnitudes brighter than $M_{R_c}\simeq -18$. 
For magnitudes fainter than $M_{R_c}\simeq -18$, the inner-most LF appears to 
contain the smallest contribution from the faint galaxy component. In 
addition, a dip in the LF at $M_{R_c}\sim -18$ is apparent for all 
four radial bins (see also Figure~\ref{Comp-Red-LF}). 

The red sequence LFs depicted in 
Figure~\ref{Scale-Comp-Red-LF} are remarkably similar in shape to the 
composite LF for galaxies selected as having an $r^{1/4}$-like surface 
brightness profile given in \citet{Yagi02} (see their Figure 7). This is 
not unexpected since the red sequence is dominated by early-type galaxies 
whose surface brightness distribution can be approximated by an 
$r^{1/4}$-like profile. The Yagi et al. composite LF is sampled to 
$M_{R_c}\sim -16.5$, for our adopted cosmology, and thus compares directly 
with the red sequence LFs presented in this paper. An upturn in the red 
sequence LF at the faint-end is present in our sample 
(see Figure~\ref{Comp-Red-LF} and \ref{Scale-Comp-Red-LF}) and 
in the composite $r^{1/4}$ LF from Yagi et al. This upturn may be the 
result of scattering from the blue galaxy population; however, if this 
is the case, the scattered galaxies must also contaminate the sample 
of $r^{1/4}$-like surface brightness profiles measured by Yagi et al., 
which seems unlikely.

In addition to \citet{Yagi02}, we also compare our red sequence LFs with the 
2dF Galaxy Redshift Survey cluster sample from \citet{DePropris03}. This study 
presents composite LFs for a sample of 60 clusters based on the selection 
of galaxies relative to a classification parameter derived from the principal 
component analysis of spectra from \citet{Madgwick02}. LFs are constructed 
for early-, mid-, and late-type classes, with spectral classification based 
on star formation rates rather than morphological type. Assuming that the 
early-type spectral class is associated with red sequence cluster 
galaxies, we compare our results from Table 3 with De Propris et al. 
A single Schechter function fit to the early-type LF yields 
$M^{\ast}_{b_{J}}=-20.04\pm 0.09$ and $\alpha=-1.05\pm 0.04$ for 
$M_{b_{J}}<-15$. Converting to $R_{c}$ and using our distance scale, we find 
$M^{\ast}_{R_{c}}=-22.92\pm 0.09$ for $M_{R_{c}}\lesssim -18$. Since 
De Propris et al. samples galaxies out to 3.0 Mpc, we compare our 
value for $M^{\ast}$ measured in the outer-most radial bin. From Table 3 
we have $M^{\ast}_{R_{c}}=-22.61\pm 0.19$, which is within $1.5\sigma$ of the 
De Propris et al. result. The De Propris et al. faint-end slope, 
$\alpha=-1.05\pm 0.04$, is consistent with the value of $\alpha=-1.0$ that 
we assume for the bright-end Schechter function given that De Propris et al. 
only samples to $M_{R_{c}}\sim -18$. Examination of our 
Figure~\ref{Scale-Comp-Red-LF} indicates that $\alpha$ becomes steeper for 
magnitudes fainter than $-18$. Thus, De Propris et al. does not probe faint 
enough to sample the increasing faint galaxy component.

In \S 4.2.1 we compared our total composite LF with \citet{Popesso06}. That 
study also presents the composite LF for early-type cluster galaxies based 
on selecting galaxies with $2.22\leq u-r\leq 3$. Converting 
to $R_{c}$ and our distance scale, Popesso et al. finds 
$M^{\ast}_{R_{c}}=-22.27\pm 0.14$, with a faint-end slope of 
$\alpha=-2.01\pm 0.11$ for galaxies measured within $r_{200}$. Our 
$M^{\ast}$ values range from $-22.28\pm 0.07$ to $-22.61\pm 0.19$, thus 
we are in good agreement with Popesso et al. given that we measure LFs in 
a series of concentric annuli. The faint-end slope of Popesso et al. is 
flatter than our values tabulated in Table 3 ($-5.26\leq \alpha\leq -2.83$). 
This discrepancy is most-likely due to the poor fit at the faint-end of 
our LFs as indicated by the large reduced-$\chi^{2}$ values in Table 3. 
In fact, Popesso et al. samples $\gtrsim 1$ mag deeper than our data set, 
and thus is able to place a better constraint on the faint-end slope. 
We also note that the rise in the faint-end slope of the early-type LF 
from Popesso et al. occurs at approximately the same magnitude as our study 
($M_{R_{c}}\sim-18.5$).

\subsubsection{Blue Galaxy Luminosity Function}

We construct the LF of blue cluster galaxies by selecting galaxies in the 
CMD blueward of the red galaxy sample. The selection of the galaxies is 
illustrated in Figure~\ref{Blue-region}, using Abell 260 as an example. 
Here, galaxies brighter than $R_c=18.8$ are designated as blue galaxies 
if their $B-R_c$ color is bluer than the blueward limit used to select the 
red sequence population. The color criterion for blue galaxies at $M^{\ast}$ 
is $B-R_c=1.4$ for Abell 260.  This corresponds to an Sab-type galaxy using 
$B-R_c$ galaxy colors tabulated in \citet{Fukugita95}. Galaxies fainter 
than $R_c=18.8$ mag are selected if they are located in the region of the 
CMD that is bluer than the area used to select the red sequence galaxies and 
brighter than the completeness magnitude. An additional correction to the 
galaxy counts is made by subtracting the net red sequence galaxy counts 
(measured from the region of the CMD fainter than $R_c=18.8$) from the net 
blue cluster galaxy counts.\footnote{The 
net blue cluster galaxy counts were calculated by subtracting background 
field galaxies, selected using the same color criteria as the blue 
cluster galaxy population, from the raw blue cluster galaxy counts.}
This correction is necessary to account for the red sequence galaxies 
that inhabit the region in the CMD that is bluer than the area used to 
select the red galaxy population at the faint-end. 

The blue galaxy LF for four different annuli (same radial 
bins as that used for the total and red composite LFs) is depicted in 
Figure~\ref{Comp-Blue-LF}. The LFs have been constructed by scaling and 
combining clusters, complete to $M_{R_c}=-16.5$, as described previously 
for the total and red sequence LFs. As determined for the total and 
red sequence LFs, a single Schechter function is unable to adequately 
describe the shape of the blue galaxy LF. A sum of two 
Schechter functions (solid line in Figure~\ref{Comp-Blue-LF}) is fit to 
the blue LFs, with the best-fit Schechter parameters given in Table 4. 
For the LF fit procedure we set ($N^{*}_{2}/N^{*}_{1}$)=3, which is equal 
to the geometric mean of the ratio averaged over all four radial bins when 
$N^{*}_{1}$ and $N^{*}_{2}$ are allowed to vary. 

In Figure~\ref{Scale-Comp-Blue-LF} the blue LFs have been superimposed 
by scaling the outer three LFs to match the inner-most LF in the magnitude 
range $-22\leq M_{R_c}\leq -21$. In general, all four LFs appear to be very 
similar in shape, with a slight tendency for a steeper faint-end slope 
at a larger cluster-centric distance. This trend can be ascertained 
by examining the value of $\alpha_{2}$ measured for the four LFs 
(see Table 4). We note that if we fix $M_{2}^{\ast}$ and solve for 
($N^{*}_{2}/N^{*}_{1}$) and $\alpha_{2}$, ($N^{*}_{2}/N^{*}_{1}$) decreases 
slightly with increasing cluster-centric radius. We also note that the 
blue LFs are similar in shape to the composite LF composed of galaxies 
having an exponential-like surface brightness profile from 
\citet{Yagi02} (see their Figure 9). 

Analogous to the comparison of our red sequence LF with \citet{DePropris03} 
and \citet{Popesso06} in \S 4.2.2, we also compare our blue LF with the 
late-type LFs presented in these studies. Correcting 
for filter and cosmology difference, the single Schechter function fit 
to the late-type LF (Type 3+4) from De Propris et al. yields 
$M^{\ast}_{R_{c}}=-22.02\pm 0.19$ with $\alpha=-1.30\pm 0.10$. This value 
for $M^{\ast}$ agrees well with our measured range 
($-21.96\leq M^{\ast}\leq -21.81$), while our faint-end slope is much steeper 
($-1.62\leq \alpha\leq -1.82$). As described in \S 4.2.2, the De Propris 
et al. study does not probe the faint-end of the LF to the same depth 
as our study ($\Delta M\sim1.5$ mag). The fit of the 
faint-end slope is further complicated by the large completeness corrections 
required for the faintest two magnitude bins. Thus, it is not surprising that 
the faint-end slope of the late-type LF from De Propris et al. is flatter 
than our value tabulated in Table 4. The trend of a steeper $\alpha$ for 
late-type galaxies compared to early-type systems, is consistent with our 
results for blue versus red galaxies.

\citet{Popesso06} presents the late-type composite LF constructed by selecting 
galaxies with $u-r\leq 2.22$ for a sampling radius of $r_{200}$. Unlike our 
results for the blue LF, a single Schechter function is found to be an 
adequate fit to the late-type LF. Converting the SDSS $r$-band to $R_{c}$ 
and employing our distance scale, the late-type LF from Popesso et al. 
has $M^{\ast}_{R_{c}}=-23.41\pm 0.52$ and $\alpha=-1.87\pm 0.04$. The value 
of $M^{\ast}$ from Popesso et al. is significantly brighter ($\sim 1.5$ mag; 
$2.7\sigma$ level) than the average value measured for our blue LF. 
The Popesso et al. result is also $\sim 1.4$ mag brighter ($2.5\sigma$ 
level) then De Propris et al. for their late-type LF, and $\sim0.8$ mag 
brighter than the exponential composite LF from \citet{Yagi02}. We also 
note, based on De Propris et al., Yagi et al., and this study, that 
$M^{\ast}$ is brighter for the red (early-type, $r^{1/4}$-like) LF than 
for the blue (late-type, exp-like) LF. This is not the case for Popesso 
et al., where the early-type LF has a fainter $M^{\ast}$ ($\sim 1.1$ mag) 
than the late-type LF. We do not understand this discrepancy, but 
speculate that it may be due to the selection of galaxy types by Popesso 
et al. based on $u-r$ color. The faint-end slope of the Popesso et al. 
late-type LF ($\alpha=-1.87\pm 0.04$) is consistent with the range in 
slope values measured for our blue LF ($-1.62\pm 0.05$ to $-1.82\pm 0.10$).

In Figure~\ref{Scale-Compare-LF} the total, red sequence, and blue 
LFs for the four radial bins are directly compared. 
To aid this comparison, the red sequence and blue LFs have been matched to 
the total LF in each radial bin by scaling the red and blue galaxy counts 
such that the number of galaxies between $-24\leq M_{R_c}\leq -16.5$ is equal 
to the total LF in the same magnitude range. Inspection of 
Figure~\ref{Scale-Compare-LF} shows that the total and red sequence LFs 
are composed of a bright galaxy component that is much more significant 
than the contribution from the blue LFs. For the faint-end, the blue 
galaxy sample appears to contribute the greatest fraction of the faint 
dwarf galaxy population, with the largest difference apparent for the 
inner-most radial bin where $(r/r_{200})\leq 0.2$. 

It is also interesting to observe that $M^{\ast}$ for the faint LF component 
gets brighter with increasing cluster-centric radius for the red galaxy 
population and fainter for the blue galaxies. We will fully discuss this 
feature in the context of color-selected giant and dwarf galaxies in Paper IV 
of this series.

\subsection{Bright-End of the Luminosity Function}

\subsubsection{The Distribution of $M^{\ast}$}

The variation in the value of $M^{\ast}$ of the cluster LF can be used to 
gauge whether the LF is universal in shape at the bright-end, or whether 
luminosity segregation takes place (i.e., bright galaxies occupy 
preferentially the central regions of clusters). The measurement of 
$M^{\ast}$ has been important, historically, for potential use as a 
``standard candle'' \citep[e.g.,][]{Schechter76,Dressler78,Colless89} and 
quantifying any variation in its value is important. We adopt the premise 
that the bright-end of the LF is universal if the median uncertainty of 
$M^{\ast}$ is comparable to the measured dispersion of the distribution.

To measure the variation in the bright-end of the LF, the value of 
$M^{\ast}$ was determined for our sample by fitting each 
cluster with a single Schechter function as described in \S 4.1; 
i.e., fitting the net galaxy counts 
in the magnitude range $-24\leq M_{R_c}\leq -20$, with the faint-end slope 
fixed at $\alpha=-1$. Only galaxies measured within a cluster-centric 
radius of $(r/r_{200})\leq 0.4$ were selected. This allows us to maximize the 
number of clusters in our sample while normalizing each one to the same 
dynamical radius without being adversely affected by small number 
statistics. Values for $M^{\ast}$ for 55 clusters satisfying the magnitude 
and radius criteria are tabulated in Table 1.

In Figure~\ref{Mstar-Gauss} we plot the distribution of $M_{R_c}^{\ast}$, 
which appears approximately Gaussian with a mean 
$\langle M_{R_c}^{\ast}\rangle=-22.24\pm 0.06$ and a dispersion of 
$\sigma=0.31$ mag. The median uncertainty in the measurement of 
$M_{R_c}^{\ast}$ for all 55 clusters is 0.38 mag, thus supporting the 
universality of $M^{\ast}$ within the measured range of 
$\Delta M^{\ast}\sim 0.3$ magnitude. 
\citet{Popesso05} presents the histogram distribution of $M^{\ast}$ for 
LFs measured within a radius of $2h^{-1}_{50}$ Mpc. Their distribution is 
Gaussian-like in shape with an estimated mean of $\sim -22.6$ and a 
dispersion of $\sim 0.5$ mag (transformed to our filter and distance scale). 
Given that the Popesso et al. LFs are measured with a fixed aperture, and 
at a larger radius than ours, it is not unexpected that they obtain a 
somewhat larger dispersion and a brighter mean $M_{R_c}^{\ast}$ than our 
results, which may arise from a mild dependence of $M^{\ast}$ on 
cluster-centric radius (see \S 4.3.2).

Given the size of the dispersion, our results are not too dissimilar 
even though Popesso et al. used a fixed physical length scale for the 
counting aperture size rather than scaling relative to a 
dynamical radius like $r_{200}$.

Since the uncertainty in $M_{R_c}^{\ast}$ is expected to be a function 
of cluster richness (assuming Poisson statistics), we measure the median 
uncertainty and dispersion of $M^{\ast}$ for three bins based on cluster 
richness; i) $B_{gc}>1500$, ii) $1000<B_{gc}<1500$, and iii) $B_{gc}<1000$. 
The dividing $B_{gc}$ values are equivalent to line-of-sight velocity 
dispersions of $\sigma_{v}\sim 960~\mbox{km}~\mbox{s}^{-1}$ for 
$B_{gc}=1500$ and $\sigma_{v}\sim 750~\mbox{km}~\mbox{s}^{-1}$ for 
$B_{gc}=1000$ \citep{Yee03}.

For clusters with $B_{gc}>1500$, we find a median uncertainty in 
$M_{R_c}^{\ast}$ of 0.22 mag with a corresponding dispersion of 0.25 mag. 
For the cluster sample with $1000<B_{gc}<1500$, we obtain a median 
uncertainty of 0.35 mag and a dispersion of 0.35 mag. Finally, for the 
poorest group we measure a median uncertainty of 0.51 mag and a dispersion 
of 0.32 mag. A comparison of the median uncertainty and dispersion of 
$M_{R_c}^{\ast}$ for our three cluster sub-samples supports the 
hypothesis that the bright-end of the cluster LF is universal at the 
$\sim 0.3$ mag level.

A direct comparison of $\langle M_{R_c}^{\ast}\rangle$ with other published 
values is 
problematic given that $M_{R_c}^{\ast}$ is not independent of $\alpha$ 
\citep{Schechter76}. \citet{Piranomonte01} finds 
$M_{r}^{\ast}=-22.02\pm 0.16$ for the combined 
LF from a sample of 80 clusters using a Schechter function fit with 
$\alpha=-1.01$. The composite LF for this study was constructed by weighting 
clusters according to their richness. Transforming $M_{r}^{\ast}$ to the 
$R_c$ passband \citep{Fukugita95} and our adopted cosmology, we find 
$M_{R_c}^{\ast}= -22.39\pm 0.16$, which is in very good agreement with our 
result. From a sample of 69 clusters extracted from the RASS-SDSS, 
\citet{Popesso06} found 
$M_{r}^{\ast}=-20.94\pm 0.16$ and $\alpha=-1.09$. The cluster galaxies 
were measured within a cluster-centric radius of $r_{200}$. Transforming 
to our cosmology and passband yields $M_{R_c}^{\ast}=-22.64\pm 0.16$, which 
is consistent with our measurement. 

\citet{Hansen05} presents $M_{r}^{\ast}$ for a sample of clusters/groups 
selected from the SDSS Early Data Release. The Hansen et al. study tabulates 
$M_{r}^{\ast}$ for several cluster samples of various richness with a 
faint-end slope fixed at $\alpha=-1.0$ (see their Table 2). The value of 
$M^{\ast}$ for their richest sub-sample, which compares more directly with our 
Abell clusters, was determined to be $M_{r}^{\ast}=-20.86\pm 0.05$. 
Transforming to our adopted cosmology and $R_c$-band filter, we find 
$M_{R_c}^{\ast}=-22.53\pm 0.05$. This result is based on cluster galaxies 
measured within $r_{200}$ and is in excellent agreement with our results. 
Hansen et al. also suggests that $M^{\ast}$ is correlated with 
cluster richness in the sense that richer clusters have a brighter 
$M_{r}^{\ast}$. For their three richest bins, $M_{r}^{\ast}$ differs by 
$\sim 4\sigma$ between the poorest and richest sub-samples. To search for a 
similar correlation in our Abell sample, we have divided our clusters into 
three bins based on $B_{gc}$; i) $B_{gc} < 1000$, ii) $1000\leq B_{gc}<1500$, 
and iii) $B_{gc}>1500$. For these three sub-samples we find  
$M_{R_c}^{\ast}=(-22.36\pm 0.08, -22.25\pm 0.07, -22.28\pm 0.05)$ for 
$B_{gc}=(< 1000, 1000-1500,>1500)$. No correlation between $M_{R_c}^{\ast}$ 
and cluster richness (as measured by $B_{gc}$) is evident.

To search for a possible correlation with richness for individual 
clusters, we plot in Figure~\ref{Mstar-Bgc} the distribution of 
$M_{R_c}^{\ast}$ as a function of $B_{gc}$. A Spearman rank-order correlation 
coefficient \citep{Press92} indicates that a correlation between 
$M_{R_c}^{\ast}$ and $B_{gc}$ is not significant ($r_{s}=-0.08$), with a 
45\% probability that the variables are correlated. It is not too surprising 
that cluster richness, as measured by $B_{gc}$, does not have a strong 
influence on $M_{R_c}^{\ast}$ since we have normalized our cluster sample in 
terms of $r_{200}$. In Figure~\ref{Mstar-Bgc} we do see a trend where the 
spread in the measured value of $M_{R_c}^{\ast}$ decreases with increasing 
$B_{gc}$. This is most-likely due to the increased uncertainty in the 
measurement of $M_{R_c}^{\ast}$ for poorer clusters. For the ten clusters with 
the smallest value of $B_{gc}$, the average uncertainty in $M_{R_c}^{\ast}$ is 
$0.69$ mag, while for the ten clusters with the largest value of $B_{gc}$, 
the mean uncertainty is $0.22$ mag. 

\citet{Hilton05} finds a correlation between $M^{\ast}_{b_{j}}$ and X-ray 
luminosity such that low-$L_{X}$ clusters have a brighter $M^{\ast}$ than 
high-$L_{X}$ systems ($\Delta M\sim 0.51$ mag). To compare our results, we 
use $L_{X}$ measurements from \citet{Ebeling96,Ebeling00} for a sub-sample of 
41 clusters in common and construct composite LFs for a low-$L_{X}$ 
($L_{X}<3\times 10^{44}~\mbox{ergs}~\mbox{s}^{-1}$; 22 clusters) and a 
high-$L_{X}$ sample ($L_{X}\geq3\times 10^{44}~\mbox{ergs}~\mbox{s}^{-1}$; 
19 clusters). Fitting a single Schechter function to the LFs yields 
$M^{\ast}_{R_{c}}=-22.25\pm 0.06$ for the low-$L_{X}$ sample and 
$M^{\ast}_{R_{c}}=-22.22\pm 0.05$ for the high-$L_{X}$ group. Thus, there 
is no significant difference in $M^{\ast}$ when dividing our sample in terms 
of $L_{X}$. The discrepancy between our result and Hilton et al. may be 
due to the higher fraction of low-mass systems in their sample. For 
example, using the same dividing $L_{X}$ as Hilton et al. 
($0.36\times 10^{44}~\mbox{erg}~\mbox{s}^{-1}$), our low-$L_{X}$ sample 
contains only 3 clusters (7\% of the total), while the corresponding 
Hilton et al. sample contains 49 clusters (72\% of the total).

To search for a correlation of LF parameters with cluster mass, 
\citet{DePropris03} divided their sample into two groups based on 
velocity dispersion ($\sigma \gtrless 800~\mbox{km}~\mbox{s}^{-1}$). A 
Schechter function fit to the low- and high-$\sigma$ group yields the 
same $M^{\ast}$ at the $1\sigma$ level. This is consistent with our 
findings that $M^{\ast}$ is not correlated with cluster mass.

The characterization of clusters based on the relative contrast of the 
brightest cluster galaxy defines the Bautz-Morgan type (BM-type) 
classification scheme \citep{BM70}. The evolutionary state of a cluster, as 
characterized by its BM-type, is expected to be correlated with 
$M^{\ast}$ if the BCGs are created via the merger of giant galaxies through 
a cannibalism-like process \citep[e.g.,][]{Dressler78}. Galactic cannibalism 
would reduce the number of bright galaxies and therefore shift 
$M_{R_c}^{\ast}$ to a fainter value. To test this scenario, we plot in 
Figure~\ref{Mstar-Bgc-BM} 
$M^{\ast}$ versus BM-type for 54 clusters (A1569 has no published BM-type). 
The solid line in Figure~\ref{Mstar-Bgc-BM} depicts a 
least-squares fit to the data and suggests that a weak correlation exits in 
the sense that early BM-type clusters have a fainter $M^{\ast}$ than later 
BM-types. A Spearman rank-order test gives a correlation coefficient 
of $r_{s}=-0.28$, with a 96\% probability that these two variables are 
correlated. This result lends support to the theoretical study of 
\citet{merritt84} in which the formation of the BCG occurs while the 
cluster is collapsing. It is expected that relatively little evolution 
of the BCG happens after the cluster is virialized. If we associate the 
early BM-type clusters with fully relaxed virialized systems and late 
BM-types with unrelaxed systems in the process of collapsing, the trend 
of the correlation between $M^{\ast}$ and BM-type from 
Figure~\ref{Mstar-Bgc-BM} can be explained.

Since $M^{\ast}$ is correlated with BM-type, we correct the values 
of $M^{\ast}$ for this correlation and find that the dispersion in 
$M_{R_c}^{\ast}$ decreases from 0.31 mag to 0.24 mag. This result 
is consistent with the universality of $M^{\ast}$ within the 0.3 mag 
range.

To investigate the possible correlation between $M^{\ast}$ and $B_{gc}$ for 
selected BM-types, we divide our 54 clusters into early BM-types 
(I and I-II) and late BM-types (II, II-III, and III). A Spearman rank-order 
correlation analysis for the early BM-type sample yields $r_{s}=-0.27$, 
with a 72\% probability that $M^{\ast}$ and $B_{gc}$ are correlated. For 
the late BM-type clusters we find $r_{s}=-0.08$ with only a 39\% probability 
that $M^{\ast}$ and $B_{gc}$ are correlated. These results indicate that 
no significant correlation exists between $M^{\ast}$ and cluster richness 
among similar BM-type clusters when measuring galaxies within an equivalent 
dynamical radius and thus further supports the notion that $M^{\ast}$ is 
universal within the measured uncertainty.

\citet{DePropris03} compares their composite LF fit parameters for two 
sub-samples divided according to BM-type. The difference in $M^{\ast}$ 
between the early BM-type group (I, II-II, II) and the late BM-type sample 
(II-III, III) is $0.06\pm 0.22$ mag, where the uncertainties in $M^{\ast}$ 
are added in quadrature. The difference in $M^{\ast}$ between our BM-type I 
and III systems is $\Delta M=0.3$ mag, which is significant at the $3\sigma$ 
level. Using identical BM-type bins as De Propris et al., we find 
$\Delta M=0.2$ mag at the $3\sigma$ level. The discrepancy between our 
result and De Propris et al. may be related to differences in the 
technique used to construct the composite LFs. For example, De Propris 
et al. includes the BCG in their LFs. In addition, instead of using a 
dynamically scaled radius, De Propis et al. used all galaxies within a 
fixed aperture of $1.5h^{-1}_{100}$ Mpc, which may add to the dispersion 
of $M^{\ast}$. Also, galaxy magnitudes from 
De Propris et al. are based on $m_{b_{j}}$ magnitudes, which are more 
susceptible to recent star formation and dust attenuation than $R_{c}$ 
magnitudes. In addition, De Propris et al. fits for the value of $\alpha$ 
while we impose the constraint that $\alpha=-1$. Since $M^{\ast}$ and 
$\alpha$ are correlated, a steeper $\alpha$ will in general yield a fainter 
$M^{\ast}$. The faint-end slope of the early BM-type LF from 
De Propris et al. is flatter than the slope of the late BM-type LF. 
Forcing $\alpha$ to be the same will increase $\Delta M$ between these 
two samples. Although the difference in $M^{\ast}$ between the BM-type 
samples in De Propris et al. is not significant, it is interesting to note 
that the late BM-type sample has a slightly brighter $M^{\ast}$, 
equivalent to the trend we measure.

\subsubsection{The Cluster-Centric Radial Dependence of $M^{\ast}$}

The variation of $M^{\ast}$ as a function of cluster-centric radius was 
examined by measuring the bright-end of the composite LF constructed from 
our entire 57 cluster sample. These clusters are 100\% photometrically 
complete to $M_{R_c}=-20$, and thus maximize the number of clusters used to 
determine $M^{\ast}$ in order to minimize its measured uncertainty. 
(Recall that the $M^{\ast}$ values tabulated in Tables 2--4 are 
measured for a sample of 42 clusters that are complete to $M_{R_c}=-16.5$.) 

The dependence of $M^{\ast}$ on cluster-centric radius was determined by 
fitting a single Schechter function to the net galaxy counts from 
$-24\leq M_{R_c}\leq -20$ for the four radial bins used previously (for 
example, see Figure~\ref{Comp-All-LF}). The faint-end slope of the 
Schechter function was fixed at $\alpha=-1.0$ as was done when measuring 
$M^{\ast}$ for the individual cluster LFs. Since different clusters 
cover various fractions of $r_{200}$, the total number of clusters 
contributing to the composite sample will vary with cluster-centric radius. 

In Figure~\ref{Mstar-Radius} we plot $M^{\ast}$ as a function of 
$(r/r_{200})$ for the total composite cluster sample, as well as 
samples compiled by selecting red and blue galaxies from our 
57-cluster sample (see \S 4.2 for color selection criteria). Inspection 
of Figure~\ref{Mstar-Radius} reveals that for the red and blue composite 
samples, $M^{\ast}$ gets brighter with increasing cluster-centric radius. 
For the total composite sample, we find a similar correlation, although the 
trend is much weaker. This weaker trend is due to the interplay between the 
relative dominance of the red and blue galaxies as a function of 
cluster-centric radius. $M^{\ast}$ near the center is dominated by red 
galaxies; whereas at larger radii, blue galaxies with a fainter $M_{R_c}$ 
become an increasingly more important component, flattening the 
$M_{R_c}^{\ast}$ dependence on radius. 
The difference in $M^{\ast}$ between the inner- and outer-most radial bin 
is $\Delta M=0.48$ mag for the red sequence LF ($3.3\sigma$ level) and 
$\Delta M=0.27$ mag for the blue LF ($1.2\sigma$ level).

To quantify the correlation for each of our three composite cluster 
samples, we calculate the linear correlation coefficient $r$ 
statistic \citep{Press92}. For the total composite cluster sample, we find 
that the linear correlation coefficient is $r=-0.91$, with a 92\% 
probability that $M^{\ast}$ and radius are correlated. The red sequence 
composite cluster sample yields $r=-0.98$, with a 98\% probability of a 
correlation. For the blue galaxy cluster sample, the linear correlation 
coefficient is $r=-0.98$ and a 98\% probability of a correlation. 


\section{Discussion and Conclusions}
\subsection{Universality of the Luminosity Function}

We have examined the distribution of $M^{\ast}$ for the bright-end of 
the LF in the core of Abell clusters. We found that the dispersion in 
$M^{\ast}$ is comparable to the average measured uncertainty, even when 
dividing the cluster sample into different richness groups based on 
$B_{gc}$. We use this result to indicate that $M^{\ast}$ is universal 
at the 0.3 mag level for a restricted magnitude range and when measured 
within a specific dynamical radius. In addition, we find a weak trend 
in which early BM-type clusters have a fainter $M^{\ast}$ than late 
BM-types.

One of the primary goals of this paper is to explore the change in the cluster 
$R_c$-band LF as a function of cluster-centric radius. From a sample of 
57 low-$z$ Abell clusters, we have measured LFs covering various fractions 
of $r_{200}$ for both the red sequence and blue cluster galaxy populations. 
Our results indicate that the overall shape of the LF is dependent 
upon distance from the cluster center. In general, the LFs exhibit an 
increase in the steepness of the faint-end slope with increasing 
radius. The radial dependence of the rate-of-change in $\alpha$ is greatest 
for the total cluster sample (i.e., red plus blue galaxies), while the blue 
LF is less dependent. The red sequence LF is mostly flat ($\alpha\sim -1$) 
over a span of $\sim 5$ magnitudes ($-23<M_{R_c}<-18$), with a rising 
faint-end for $M_{R_{c}}> -18$. In contrast, the blue 
galaxy LF is much steeper ($\alpha\sim -1.8$) over this same luminosity 
range, with minimal change in shape out to $r_{200}$. The very rapid 
increase in $\alpha$ for the total LF is likely due to a combination of 
steepening slope for both red and blue LFs and the increasing 
dominance of the blue population at large cluster-centric radii. In addition, 
the red sequence LF has a much brighter characteristic magnitude 
($\sim 0.6$ mag) over all radii than the blue galaxy luminosity distribution. 
These results lend support to several recent studies that have observed 
similar characteristics for red and blue galaxy LFs drawn from low- to 
high-density environments \citep[e.g.,][]{Beijersbergen02a,Goto02,Baldry04}.

The general trend for the LF to become flatter with decreasing 
cluster-centric radius supports the hypothesis that dwarf galaxies are 
tidally disrupted near the cluster center. This idea has been 
used by \citet{Lopez97b} to help explain the formation of BCG halos and 
the origin of a large fraction of the gas content in the intracluster medium.

The dependence of the shape of the LF on cluster-centric radius provides 
strong evidence that the relative mixture of giant and dwarf galaxies 
depends on the fraction of the virial radius that is measured. This argues 
against the global universality of the cluster LF for the magnitude interval 
$-26\leq M_{R_c}\leq -16.5$ and suggests an environmental influence. The 
non-parametric galaxy dwarf-to-giant ratio will be explored in Paper IV 
of this series.

In Figure~\ref{CompareLF} we plot composite cluster LFs from several published 
sources \citep{Colless89,Piranomonte01,Goto02,DePropris03,Hansen05,
Popesso06}. These LFs have been transformed to $M_{R_c}$ using 
\citet{Fukugita95} and our adopted cosmology. Also depicted in 
Figure~\ref{CompareLF} are the total composite LFs from this study for the 
inner- and outer-most radial bin. In addition, the SDSS field galaxy LF 
from \citet{Blanton03} is included for comparison purposes. 

Figure~\ref{CompareLF} illustrates that the slope of the LFs are very similar 
at bright magnitudes where the giant galaxies have the greatest 
influence on the shape of the LF. The main difference in the slope arises 
at the faint-end where the influence of the dwarf galaxies tends to increase 
the steepness of $\alpha$. The sampling depth and effective cluster-centric 
radius thus has a major influence on the measured shape of the cluster LF 
since the inclusion of different amounts of the dwarf galaxy population will 
directly impact the faint-end slope. Thus, the evidence supports the notion 
that the cluster LF is not universal in shape in a global sense. In 
addition, we measure a dispersion of 0.3 mag in the value of $M^{\ast}$ for 
the depicted cluster LFs; the faintest value is $M_{R_c}^{\ast}=-22.26$ 
for the inner-most annuli from this study and the brightest is 
$M_{R_c}^{\ast}=-23.14$ from \citet{Goto02}. 

\subsection{Cluster Galaxy Population Gradients}

As discussed in \S4.3.2, Figure~\ref{Mstar-Radius} shows that the 
bright-end $M^{\ast}$ becomes brighter with increasing cluster-centric 
radius for both the red, and with a lesser significance, the blue cluster 
galaxies. The observed dimming of 
$M^{\ast}$ toward the 
cluster center can be explained as a simple fading of the galaxy population. 
For the blue cluster galaxies, the truncation of star formation as field 
galaxies fall into the cluster environment 
\citep[e.g.,][]{Abraham96,Ellingson03} would lead to a fading of 
$M^{\ast}$ with decreasing cluster-centric radius. Since clusters are 
believed to have formed via the infall of galaxies along filamentary-like 
structures \citep[e.g.,][]{Dubinski98}, it is expected that a population of 
infalling field galaxies can be detected in addition to the older, 
mainly early-type red galaxies. If star formation for infalling spiral 
galaxies, which 
dominate the field galaxy population \citep[e.g.,][]{Binggeli88}, is 
truncated via some type of dynamical mechanism \citep[e.g., ram pressure 
stripping, galaxy harassment, etc.;][]{Moore98,Goto05a,Roediger06}, then 
a roughly continuous infall (but with an allowed variable rate) would be 
observed as a fading of the population toward the cluster center. Given 
enough time, infalling spiral galaxies may acquire characteristics that 
are similar to S0 galaxies. In fact, this type of mechanism for S0 formation 
has been proposed by numerous authors 
\citep[see][and references therein]{Dressler99}, although the formation of 
field S0s has remained problematic for these models.

The dimming of $M^{\ast}_{2}$, or the decrease of 
($N_{2}^{\ast}/N_{1}^{\ast}$) for the faint-end of the red sequence LF for 
decreasing cluster-centric radii, places constraints on the evolutionary path 
of the faint blue galaxies. If these galaxies simply fade and turn red, we 
would expect them to contribute to the red sequence LF by increasing the 
number of faint red galaxies in the central cluster region. We thus suggest 
that the faint blue cluster galaxies are destroyed in the central cluster 
region. This is not a far-fetched hypothesis since the faint blue galaxies are 
very similar to the low-mass dwarf spheroidal galaxies, which are expected 
to undergo tidal disruption in the central cluster environment 
\citep{Thompson93,Gallagher94,Hilker03}. These low luminosity blue galaxies 
could also be the source of the dwarf galaxies that get tidally disrupted 
and form the halo of BCGs as proposed in a model by \citet{Lopez97b}. 
We will explore this possibility further in Paper IV of this series.

The fading of the bright-end $M^{\ast}$ for the red sequence galaxy 
population may be the result of two separate processes; the continuous 
fading of infalling red 
galaxies (which may have turned red relatively recently due to the 
truncation of star formation) and galactic cannibalism in the high-density 
central cluster region. 
For the infall scenario, red galaxies are expected to originate from a 
population that had arrived in the cluster environment early in its history. 
The observed dispersion of the CMR for the cluster red sequence places 
constraints both on the formation timescale of the early-type galaxy 
population \citep[i.e., $z>2$; e.g.,][]
{Stanford98,dePropris99,Gladders99,Holden04,Lopez04} and any episodes 
of recent star formation \citep[e.g.,][]{Bower98}. Red galaxies which 
have entered the cluster environment during the earliest part of the 
cluster lifetime, would be expected to be the faintest since the time 
between the last phase of major star formation and today would be the 
greatest for these galaxies (after the truncation of star formation, 
galaxies would evolve passively with an associated reddening and dimming 
of their starlight). Under this scenario, we would expect that the blue 
galaxy population would show a greater rate-of-change in $M^{\ast}$ with 
radius due to the more recent decline in the star formation rate. The red 
galaxies would be expected to exhibit a more gradual change in $M^{\ast}$, 
as compared to the blue galaxy population, since the red galaxies are just 
a passively evolving old galaxy population. 

As depicted in Figure~\ref{Mstar-Radius} the bright-end $M^{\ast}$ for the 
red sequence galaxies fades more rapidly toward the cluster center than for 
the blue galaxy population. Since a simple infall scenario as described 
predicts that the blue galaxy population should exhibit the more rapid 
decline in $M^{\ast}$, we hypothesize that an additional mechanism may 
affect the rate-of-change of $M^{\ast}$ with radius for the red galaxy 
population. A clue to this additional effect is gleamed from 
Figure~\ref{Mstar-Bgc-BM} where the correlation between $M^{\ast}$ 
and BM-type is depicted for the total cluster sample. As described in 
\S 4.3.1, the trend for a brightening of $M^{\ast}$ with later BM-type 
can be explained by galactic cannibalism. As first theorized by 
\citet{Ostriker77} and \citet{Hausman78}, as bright galaxies fall into 
the cluster center they will be swallowed by the giant central galaxy, thus 
resulting in the fading of $M^{\ast}$ as bright galaxies are depleted from 
the LF. Since early BM-type clusters have, by definition, a bright central 
dominant galaxy, the correlation depicted in Figure~\ref{Mstar-Radius} 
is expected. Galactic cannibalism will also result in the fading of 
$M^{\ast}$ with decreasing cluster-centric radius, as illustrated in 
Figure~\ref{Mstar-Radius}. We performed a simple test of this hypothesis 
by calculating $M^{\ast}$ for our red cluster galaxy sample for early BM-type 
clusters (I and I-II) and late BM-type clusters (II, II-III, and III). We 
assume that the effect of galactic cannibalism on the radial dependence of 
$M^{\ast}$ will be greatest for the early BM-type sample. Comparing 
$M^{\ast}$ between the inner- and outer-most radial bin shows that for 
the early BM-type sample, $\Delta M=0.71$ mag with the two measurements 
different at the $2\sigma$ level. For the late BM-type sample, we find 
$\Delta M=0.36$ mag at the $2\sigma$ level. Although the significance of the 
difference in $M^{\ast}$ for a given BM-type group is not high (partly due 
to the fact that only four clusters comprise the outer-most radial bin for 
the early BM-type clusters), the larger difference in $M^{\ast}$ for the 
early BM-type sample suggests that the infall and galactic cannibalism 
scenarios may help to explain the observed rate-of-change of $M^{\ast}$ 
with radius for the red sequence galaxy population.

In Paper IV of this series we will further elucidate the nature of the 
cluster galaxy population by examining the radial-dependence of the 
dwarf-to-giant ratio and the blue-to-red galaxy count ratio. These results 
will complement the observations presented in this paper and provide 
additional constraints on the composition of the galaxy population in 
low-redshift Abell clusters.

\acknowledgments

We thank the anonymous referee for useful comments and suggestions. 
Research by W. A. B. at the University of Toronto was supported by the 
Carl Reinhardt Fund, the Walter C. Sumner Fellowship, and NSERC through 
the Discovery grant of H. K. C. Y. W. A. B. also acknowledges support from 
NASA LTSA award NAG5-11415, NASA Chandra X-ray Center archival research 
grant AR7-8015B, and a University of Illinois seed funding award to the 
Dark Energy Survey. Research by H. K. C. Y. is supported by an NSERC 
Discovery grant. O. L.-C research is supported by INAOE and a CONACyT grant 
for Ciencia B{\'a}sica P45952-F. O. L.-C. also acknowledges support from 
a research grant from the Academia Mexicana de Ciencias-Royal Society 
during 2006-2007, taken to the Univesity of Bristol. We thank Huan Lin for 
providing photometric catalogs for five control fields, and James Brown 
for the use of his galaxy profile fitting software and photometric data 
for A496 and A1142.

The Image Reduction and Analysis Facility (IRAF) is distributed by the 
National Optical Astronomy Observatory, which is operated by AURA, Inc., 
under contract to the National Science Foundation. This research has made 
use of the NASA/IPAC Extragalactic Database (NED) which is operated by the 
Jet Propulsion Laboratory, California Institute of Technology, under 
contract with the National Aeronautics and Space Administration.

\appendix

\section{Quantifying Selection Effects and Biases}

A major challenge when conducting any observational study is 
the ability to quantify the impact of selection effects and biases 
on the robustness of the results. In this appendix we explore several 
of these potential sources of systematic errors and quantify their affect 
on the derived galaxy cluster LF.  

\subsection{Chance Projections}

A main concern in interpreting the results of a study based on the 
LF constructed from the statistical subtraction 
of background galaxies is the possible influence of projection effects. 
The chance projection of field galaxies, or unrelated groups, can 
appear as clusters when redshift information is not available 
\citep{val01}. Selection of cluster members using 
color information (see \S 3.2) helps to alleviate some of this concern. 
To determine 
what impact chance projections may have on our LFs, 
we have divided our cluster sample into two groups according to the number of 
published redshifts \citep[e.g.,][]{struble99}. The first group contains 
only those clusters which have at least 25 spectroscopically-confirmed 
cluster members. The second group contains clusters in which the number 
of redshift-confirmed members is $\le 10$. In this comparison it is assumed 
that clusters with a small number of confirmed members may still suffer from 
unknown projection effects.

In Figure~\ref{Redshift10-25} we present LFs for the total composite cluster 
sample for four different radial bins. The cluster sample has been divided 
into two groups according to the number of confirmed cluster members based 
on the number of redshifts (i.e., 20 clusters with $\leq 10$ galaxy redshifts 
and 32 clusters with $\geq 25$ galaxy redshifts). The LFs for each radial 
bin have been scaled to match the LF for that particular cluster-centric 
radius from the total composite LFs depicted in Figure~\ref{Comp-All-LF}. 
The overall shape of the LFs generated for the two redshift-selected cluster 
groups are very similar in shape for all four annuli. Hence, this test 
indicates that the chance projection of background clusters/galaxy groups 
does not have a significant effect on our results.

\subsection{Deprojecting the Luminosity Function}

The LFs presented in this paper are derived from the two-dimensional 
projected distribution of cluster galaxies. Since the full three-dimensional 
galaxy spatial positions cannot be resolved, the presence of contaminating 
galaxies from {\it within} the cluster can adversely affect the accuracy in 
determining the shape of the central cluster LF. Note that in this context, 
the projected galaxies are those which lie in the outskirts of the 
cluster and are projected onto the central region. This is in addition to 
the presence of fore/background galaxies that are unrelated to the 
target cluster, and are corrected using background corrections. The 
projection of galaxies located in the outskirts of clusters onto the central 
region will have its greatest impact on the faint-end of the central LF. 
This is due to the increase in the relative fraction of dwarf galaxies with 
increasing cluster-centric radius, as implied by the steepening of the 
faint-end slope with radius (see Figure~\ref{Comp-All-LF}).

Simulations by \citet{val01} indicate that projection effects can severely 
affect the shape of the LF by artificially producing a steeper 
faint-end slope. Some of this steepening is due to the fact that cluster 
galaxies were not selected in terms of color 
(i.e., using the CMR), thus resulting in contamination 
by projected field galaxies. A certain portion can also be attributed to the 
inclusion of galaxies in cluster outskirts that are projected onto the 
central region. \citet{Beijersbergen02a} published a study of the LF of the 
Coma cluster in which they corrected the LF for 
projection effects by subtracting the contribution of the outer Coma LF that 
is projected onto the central region. The resulting ``deprojected'' LF was 
measured to be marginally flatter than the projected 2-d LF.

To determine the extent to which our LFs are affected by projection 
effects from the galaxy population in the cluster outskirts, we have 
measured the deprojected composite LF for the total, red sequence, and 
blue galaxy samples. We assume that the cluster galaxies are distributed 
symmetrically in a sphere about the cluster center with the method of 
deprojecting the LF depicted in Figure~\ref{Deproject-Draw}. As shown, any 
sight-line to the central 
area of the cluster (region A) must pass through regions B1 and B2. Thus, 
the observed central LF will include galaxies that are located in regions 
B1 and B2. To correct for this effect, the LF determined from region C is 
subtracted from the LF calculated for the regions B1+B2+A. Before subtracting 
the two LFs, the LF measured for region C must be 
normalized to the same volume as that included in regions B1+B2. This 
``correction'' factor is given by $(V_{1}-V_{2})/(V_{3}-V_{1})$, 
where $V_{1}$ is the combined volume of regions B1+B2+A, $V_{2}$ is the 
volume of the inner region A ($V_{2}=(4\pi/3)R_{1}^{3}$), and 
$V_{3}$ is the volume of the outer-most sphere ($V_{3}=(4\pi/3)R_{2}^{3}$). 
The volume $V_{1}$ can be calculated from 
$V_{1}=(4\pi/3)\left[R_{2}^{3}-(R_{2}^{2}-R_{1}^{2})^{3/2}\right]$, where 
$R_{1}$ is the radius of the inner sphere, and $R_{2}$ is the radius of 
the outer sphere.

Using this procedure, the deprojected composite LF for the 
total, red sequence, and blue galaxy samples were constructed for two radial 
bins, $(r/r_{200})\leq 0.2$ and $0.2\leq (r/r_{200})\leq 0.4$. The large 
uncertainty in the net galaxy counts from the deprojection process limits 
our analysis of the LF to these two inner-most radial bins. In 
Figures~\ref{All-Deproj}$-$\ref{Blue-Deproj} comparisons between the 
projected and deprojected LFs for the three cluster galaxy 
samples are presented. For each figure the LFs have been scaled to match in 
the $-22\leq M_{R_c}\leq -21$ magnitude range to facilitate the 
comparison. The top panel for each of the three figures depicts the 
deprojected LF for the two inner-most radial bins. For 
all three galaxy samples, the LF for the $(r/r_{200})\leq 0.2$ radial bin 
(open triangles) has a flatter faint-end slope than for the 
$0.2\leq (r/r_{200})\leq 0.4$ radial bin (solid triangles). Thus, the 
trend of a steepening faint-end slope with increasing 
cluster-centric radius is valid for the deprojected LFs for the total, 
red sequence, and blue galaxy samples.

In the middle panel of Figures~\ref{All-Deproj}$-$\ref{Blue-Deproj}, the 
deprojected and projected LFs are compared for the $(r/r_{200})\leq 0.2$ 
radial bin. For all three galaxy samples, the deprojected LF has a flatter 
faint-end slope than the projected LF. This 
demonstrates that galaxies from the cluster outskirts that are projected 
onto the central cluster region will result in an artificial steepening of 
the faint-end of the central LF.  

The bottom panel of Figures~\ref{All-Deproj}$-$\ref{Blue-Deproj} presents the 
deprojected and projected LFs for the $0.2\leq (r/r_{200})\leq 0.4$ radial 
bin. In general, the shape of the LFs are very similar given the size of 
the uncertainties in the net galaxy counts. No statistically significant 
difference in the shape of the deprojected and projected LFs can be 
discerned for either the total, red sequence, or blue galaxy cluster samples.

The deprojection of the cluster LF indicates that the presence of galaxies 
from the outer cluster region can affect the slope of the faint-end, 
especially for the central region. The basic trend of a steepening faint-end 
slope with increasing cluster-centric radius is still evident from the 
deprojected LF data. 

\subsection{Faint Source Correction}

The shape of the faint-end of the cluster LF can be 
affected by bias as a result of counting galaxies in the faintest 
magnitude bin. The faint galaxy correction \citep[also known as the Eddington 
bias or correction;][]{edd40} is due to the fact that each observed galaxy 
has an associated magnitude uncertainty, causing galaxies to be scattered 
below and above our observed magnitude limit. If the number distribution of 
galaxies is identical over all magnitudes, the number of galaxies scattered 
above and below our magnitude threshold will be statistically equal. A cluster 
LF having a faint-end slope steeper than $\alpha=-1$, 
will have a net number of galaxies scattered brighter than the observed 
magnitude limit. This will artificially enhance the number of detected 
galaxies at the faint-end of the LF. Since our galaxy counts are binned to 
produce the LF, the magnitude bin size relative to the uncertainty of the 
measured galaxy magnitudes will have a direct impact on the importance of 
this bias.

To investigate the impact of the faint source correction on our measured 
LFs, simulated LFs using the shape parameters tabulated in Table 2 for the 
total composite LF were constructed. For each galaxy magnitude we include 
an offset calculated from the observed distribution of magnitude uncertainty 
$\sigma_{R_c}$ as a function of $M_{R_c}$, assuming Gaussian statistics. 
The magnitude offset applied to each simulated galaxy was randomly chosen 
from a Gaussian distribution with a dispersion of $1\sigma_{R_c}$. Simulated 
LFs were assembled using the same selection criteria as that used for the 
observed LFs. The fractional change in the number of galaxies detected in 
the faintest magnitude bin ($-17.0\leq M_{R_c}\leq -16.5$) depends directly 
on the value of $\alpha_{2}$. Comparing simulated LFs to those tabulated  
in Table 2 yields an increase from 0.0\% to +0.4\% in the number of 
detected galaxies in the faintest magnitude bin as the value of $\alpha_{2}$ 
changed from $-1.18$ to $-2.43$. This demonstrates that the increase in the 
number of detected galaxies due to the scattering of faint galaxies into 
the faintest magnitude bin, is insignificant given the measured range in 
$\alpha_{2}$. This result is reasonable given the relatively large width of 
our magnitude bins (0.5 mag) and the average magnitude uncertainty 
($\sim 0.1$ mag) at our imposed faint-end magnitude limit.

\subsection{Color Selection Bias}

The study of the LF of galaxies selected according to their position in the 
color-magnitude plane could be affected by a color bias. This bias is a 
result of scattering in $B-R_c$ and $R_c$ of galaxies due to photometric 
uncertainties. To understand the extent of this effect, an artificial 
galaxy catalog for Abell 260 was constructed by taking the original galaxy 
catalog and adding a small, random magnitude offset based on the measured 
$R_c$ magnitude and its uncertainty for each galaxy, assuming Gaussian 
statistics. By tracking the relative offset in position on the CMD between the 
original cluster galaxies and their simulated counterparts, the impact of the 
color bias can be measured. 

In Figure~\ref{ColorBias} a vector-type CMD for Abell 260 is presented for 
galaxies brighter than the cluster completeness limit ($R_c=20.7$). The 
vectors trace the scattering path of a galaxy from its initial position in 
the observed color-magnitude plane to its position in the artificial catalog. 
For bright galaxies, the change in position is minimal compared to the size 
of the region from where the galaxies are selected 
(see Figures~\ref{redseq-region} and \ref{Blue-region} for selection of 
red sequence and blue galaxies). The largest displacement in the 
color-magnitude plane occurs for the faint red galaxies where the relatively 
large uncertainties in $B-R_c$ produce a larger change in color. The magnitude 
of this displacement is not significant compared to the size of the regions 
used to select red and blue galaxies. This exercise justifies the method used 
to select faint red sequence galaxies (see \S 4.2.2, criterion 2). By 
restricting the selection of faint red sequence galaxies from a region 
redward of the CMR (see Figure~\ref{redseq-region}), 
the contamination from blue galaxies scattered into the red sequence 
region is minimized.

\subsection{Dynamical Radius Variation}

As stated in Section 3.4, the value of $r_{200}$ determined from $B_{gc}$ 
(via equation 7) has an associated rms scatter of $\sim 15\%$. To determine 
whether a 15\% scatter in $r_{200}$ will have a significant influence on our 
conclusions, a simulated composite LF was derived by randomly changing 
$r_{200}$ by $\pm 15\%$. A comparison of the observed LF with the simulated 
LF is presented in Figure~\ref{Allr200} for the four cluster-centric radial 
bins used previously. Inspection of Figure~\ref{Allr200} shows that the 
two LFs are equivalent for each radial bin depicted. Thus, 
the 15\% scatter in the value of $r_{200}$ is expected to have minimal 
impact on our results. This is at least in part due to the expected 
gradual change in the properties of the LF as a function of $r_{200}$.

\subsection{Exclusion of the Brightest Cluster Galaxy}

In \S 4.1 it was noted that the BCG was not included in the construction 
of the cluster LF since they do not appear, statistically, to be a natural 
extension of the LF. In Figure~\ref{CompAll-BCG} the composite LF 
for the inner-most radial bin, $(r/r_{200})\leq 0.2$, is depicted for 
clusters (complete to $M_{R_c}=-20$) with and without the inclusion of the 
BCG. As this figure demonstrates, the BCGs are not a simple extension of 
the Schechter function, which may indicate that BCGs are formed by a 
different process \citep[e.g., mergers, cannibalism, 
etc.;][]{Dressler78} than the fainter cluster galaxy population. 


\clearpage

\begin{deluxetable}{lcccccccc}
\tablecaption{CLUSTER PROPERTIES}
\tablewidth{0pt}
\tablehead{\colhead{Cluster} &\colhead{Redshift} &\colhead{$M_{R_c}^{\ast}$} 
&\colhead{$\Delta M_{R_c}^{\ast}$} &\colhead{$B_{gc}$} 
&\colhead{$\Delta B_{gc}$} &\colhead{$r_{200}$}&\colhead{$\Delta r_{200}$} 
&\colhead{$M_{R_c}^{Com}$}\\ & & & &($\mbox{Mpc}^{1.8}$) 
& & (Mpc) & &}
\startdata
A21 & 0.0946 & $-22.28$ & 0.29 & 1480 & 229 & 2.641 & 0.396 & $-17.5$\tablenotemark{a} \\
A84 & 0.1030 & $-22.46$ & 0.38 & 917 & 184 & 2.098 & 0.315 & $-17.5$ \\
A85 & 0.0518 & $-22.34$ & 0.42 & 780 & 168 & 1.942 & 0.291 & $-16.0$ \\
A98 & 0.1043 & $-22.37$ & 0.24 & 1657 & 243 & 2.788 & 0.418 & $-18.0$ \\
A154 & 0.0638 & $-22.11$ & 0.32 & 1462 & 227 & 2.626 & 0.394 & $-16.5$ \\
A168 & 0.0452 & $-22.56$ & 0.41 & 992 & 187 & 2.179 & 0.327 & $-16.0$ \\
A260 & 0.0363 & $-22.04$ & 0.44 & 855 & 176 & 2.029 & 0.304 & $-16.0$ \\
A399 & 0.0715 & $-21.87$ & 0.26 & 1427 & 224 & 2.595 & 0.389 & $-17.5$ \\
A401 & 0.0748 & $-22.25$ & 0.19 & 2242 & 279 & 3.224 & 0.484 & $-17.0$ \\
A407 & 0.0472 & $-22.39$ & 0.37 & 1327 & 216 & 2.506 & 0.376 & $-16.5$ \\
A415 & 0.0788 & $-21.84$ & 0.59 & 500 & 141 & 1.568 & 0.235 & $-17.0$ \\
A496 & 0.0329 & $-21.45$ & 0.49 & 1114 & 228 & 2.304 & 0.346 & $-15.0$ \\
A514 & 0.0731 & $-22.20$ & 0.40 & 920 & 183 & 2.102 & 0.315 & $-17.0$ \\
A629 & 0.1380 & $-22.80$ & 0.44 & 1154 & 207 & 2.344 & 0.352 & $-18.0$ \\
A634 & 0.0265 & $-22.28$ & 0.80 & 360 & 117 & 1.340 & 0.201 & $-16.5$ \\
A646 & 0.1303 & $-22.62$ & 0.51 & 859 & 182 & 2.034 & 0.305 & $-18.0$ \\
A665 & 0.1816 & $-22.64$ & 0.20 & 2068 & 272 & 3.101 & 0.465 & $-19.0$ \\
A671 & 0.0491 & $-22.32$ & 0.35 & 1253 & 210 & 2.438 & 0.366 & $-16.0$ \\
A690 & 0.0788 & $-21.72$ & 0.58 & 566 & 149 & 1.664 & 0.250 & $-17.0$ \\
A779 & 0.0229 & $-22.82$ & 0.80 & 468 & 131 & 1.519 & 0.228 & $-16.0$ \\
A957 & 0.0437 & $-21.58$ & 0.37 & 1037 & 191 & 2.226 & 0.334 & $-15.5$ \\
A999 & 0.0323 & $-22.22$ & 1.01 & 357 & 117 & 1.334 & 0.200 & $-16.0$ \\
A1142 & 0.0349 & $-22.66$ & 1.24 & 469 & 148 & 1.521 & 0.228 & $-15.0$ \\
A1213 & 0.0469 & $-23.06$ & 0.59 & 966 & 184 & 2.151 & 0.323 & $-16.5$ \\
A1291 & 0.0530 & $-21.35$ & 0.37 & 1146 & 202 & 2.336 & 0.350 & $-16.5$ \\
A1413 & 0.1427 & $-22.36$ & 0.21 & 1737 & 249 & 2.852 & 0.428 & $-18.5$ \\
A1569 & 0.0784 & $-22.76$ & 0.56 & 803 & 173 & 1.969 & 0.295 & $-17.0$ \\
A1650 & 0.0845 & $-21.88$ & 0.26 & 1912 & 257 & 2.986 & 0.448 & $-17.0$ \\
A1656 & 0.0232 & $-22.04$ & 0.31 & 2167 & 292 & 3.171 & 0.476 & $-14.50$ \\
A1775 & 0.0700 & $-21.59$ & 0.38 & 1025 & 192 & 2.214 & 0.332 & $-16.5$ \\
A1795 & 0.0621 & $-21.50$ & 0.28 & 1531 & 232 & 2.684 & 0.403 & $-16.5$ \\
A1913 & 0.0530 & $-22.54$ & 0.40 & 980 & 187 & 2.166 & 0.325 & $-16.5$ \\
A1983 & 0.0430 & $-22.10$ & 0.40 & 903 & 178 & 2.084 & 0.312 & $-16.0$ \\
A2022 & 0.0578 & $-22.91$ & 0.56 & 1061 & 196 & 2.251 & 0.338 & $-17.0$ \\
A2029 & 0.0768 & $-22.07$ & 0.20 & 1777 & 249 & 2.883 & 0.432 & $-17.0$ \\
A2152 & 0.0410 & $-21.92$ & 0.40 & 801 & 169 & 1.967 & 0.295 & $-16.0$ \\
A2244 & 0.0997 & $-22.14$ & 0.22 & 1674 & 243 & 2.802 & 0.420 & $-17.5$ \\
A2247 & 0.0385 & $-23.07$ & 0.78 & 639 & 151 & 1.765 & 0.265 & $-16.5$ \\
A2255 & 0.0800 & $-22.60$ & 0.21 & 2278 & 280 & 3.248 & 0.487 & $-17.0$ \\
A2256 & 0.0601 & $-22.51$ & 0.21 & 2187 & 274 & 3.185 & 0.478 & $-16.5$ \\
A2271 & 0.0568 & $-21.46$ & 0.51 & 669 & 157 & 1.804 & 0.270 & $-16.5$ \\
A2328 & 0.1470 & $-22.01$ & 0.20 & 1935 & 263 & 3.004 & 0.450 & $-18.5$ \\
A2356 & 0.1161 & $-22.25$ & 0.36 & 964 & 189 & 2.150 & 0.322 & $-18.0$ \\
A2384 & 0.0943 & $-22.38$ & 0.29 & 1514 & 232 & 2.670 & 0.400 & $-17.5$ \\
A2399 & 0.0587 & $-22.16$ & 0.54 & 676 & 157 & 1.813 & 0.272 & $-16.5$ \\
A2410 & 0.0806 & $-22.24$ & 0.71 & 546 & 145 & 1.636 & 0.245 & $-17.0$ \\
A2415 & 0.0597 & $-21.74$ & 0.44 & 940 & 184 & 2.123 & 0.318 & $-16.5$ \\
A2420 & 0.0838 & $-21.94$ & 0.27 & 1239 & 210 & 2.425 & 0.364 & $-17.5$ \\
A2440 & 0.0904 & $-22.34$ & 0.35 & 1050 & 196 & 2.240 & 0.336 & $-17.5$ \\
A2554 & 0.1108 & $-22.61$ & 0.35 & 1221 & 211 & 2.408 & 0.361 & $-18.0$ \\
A2556 & 0.0865 & $-22.47$ & 0.49 & 796 & 172 & 1.961 & 0.294 & $-17.0$ \\
A2593 & 0.0421 & $-22.25$ & 0.40 & 1133 & 200 & 2.323 & 0.348 & $-16.0$ \\
A2597 & 0.0825 & $-21.56$ & 0.58 & 696 & 163 & 1.839 & 0.276 & $-17.0$ \\
A2626 & 0.0573 & $-22.65$ & 0.50 & 911 & 181 & 2.092 & 0.314 & $-16.5$ \\
A2634 & 0.0310 & $-22.30$ & 0.32 & 1109 & 197 & 2.299 & 0.345 & $-16.5$ \\
A2657 & 0.0414 & $-22.44$ & 0.51 & 723 & 162 & 1.872 & 0.281 & $-16.0$ \\
A2670 & 0.0761 & $-22.52$ & 0.26 & 1783 & 249 & 2.888 & 0.433 & $-17.0$ \\
\enddata
\tablenotetext{a}{The absolute $R_c$ magnitude represents our adopted 100\% 
completeness limit.}
\end{deluxetable}

\clearpage

\begin{deluxetable}{cccccccr}
\tablecaption{COMPOSITE LUMINOSITY FUNCTION}
\tabletypesize{\footnotesize}
\tablewidth{0pt}
\tablehead{
\colhead{Radius} &\colhead{$M^{\ast}_{1}$} &\colhead{$\chi^{2}_{\nu}$} &\colhead{$M^{\ast}_{2}$} &\colhead{$\alpha_{2}$} &\colhead{$\chi^{2}_{\nu}$} &\colhead{No. of}&\colhead{$<B_{gc}>$}\\
($r/r_{200}$)&($R_c$ mag)& &($R_c$ mag)& & & Clusters & ($\mbox{Mpc}^{1.8}$)}
\startdata
0.0--0.2 & $-22.26\pm 0.06$\tablenotemark{a} & 0.64 & $-17.43\pm 0.07$\tablenotemark{b} & $-1.45\pm 0.10$ & 0.94 & 42 & $1066\pm 397$\tablenotemark{c} \\
0.2--0.4 & $-22.26\pm 0.07$ & 1.04 & $-18.22\pm 0.05$ & $-1.81\pm 0.04$ & 4.50 & 39 & $1052\pm 378$ \\
0.4--0.6 & $-22.36\pm 0.10$ & 0.58 & $-18.14\pm 0.05$ & $-2.32\pm 0.05$ & 3.24 & 28 & $961\pm 347$ \\
0.6--1.0 & $-22.38\pm 0.15$ & 0.67 & $-18.39\pm 0.06$ & $-2.46\pm 0.05$ & 2.99 & 11 & $683\pm 205$ \\
\enddata
\tablenotetext{a}{$M_{1}^{\ast}$ is derived from a Schechter function fit 
to the bright-end of the LF with $\alpha_{1}=-1.0$.}
\tablenotetext{b}{$M_{2}^{\ast}$ and $\alpha_{2}$ are obtained from a 
Schechter function fit to the faint-end of the composite LF.}
\tablenotetext{c}{The uncertainty is calculated from the dispersion of the mean.}
\end{deluxetable}

\clearpage

\begin{deluxetable}{cccccccr}
\tablecaption{COMPOSITE RED SEQUENCE LUMINOSITY FUNCTION}
\tabletypesize{\footnotesize}
\tablewidth{0pt}
\tablehead{
\colhead{Radius} &\colhead{$M^{\ast}_{1}$} &\colhead{$\chi^{2}_{\nu}$} &\colhead{$M^{\ast}_{2}$} &\colhead{$\alpha_{2}$} &\colhead{$\chi^{2}_{\nu}$} &\colhead{No. of}&\colhead{$<B_{gc}>$}\\
($r/r_{200}$)&($R_c$ mag)& &($R_c$ mag)& & & Clusters & ($\mbox{Mpc}^{1.8}$)}
\startdata
0.0--0.2 & $-22.28\pm 0.07$\tablenotemark{a} & 0.69 & $-16.95\pm 0.58$\tablenotemark{b} & $-5.26\pm 15.51$ & 3.96 & 42 & $1066\pm 397$\tablenotemark{c} \\
0.2--0.4 & $-22.36\pm 0.08$ & 1.29 & $-17.81\pm 0.22$ & $-3.30\pm 0.64$ & 1.45 & 39 & $1052\pm 378$ \\
0.4--0.6 & $-22.50\pm 0.12$ & 0.72 & $-18.18\pm 0.24$ & $-3.16\pm 0.51$ & 1.16 & 28 & $961\pm 347$ \\
0.6--1.0 & $-22.61\pm 0.19$ & 1.14 & $-18.60\pm 0.36$ & $-2.83\pm 0.53$ & 1.06 & 11 & $683\pm 205$ \\
\enddata
\tablenotetext{a}{$M_{1}^{\ast}$ is derived from a Schechter function fit 
to the bright-end of the LF with $\alpha_{1}=-1.0$.}
\tablenotetext{b}{$M_{2}^{\ast}$ and $\alpha_{2}$ are obtained from a 
Schechter function fit to the faint-end of the red sequence LF.}
\tablenotetext{c}{The uncertainty is calculated from the dispersion of the mean.}
\end{deluxetable}

\clearpage

\begin{deluxetable}{cccccccr}
\tablecaption{COMPOSITE BLUE LUMINOSITY FUNCTION}
\tabletypesize{\footnotesize}
\tablewidth{0pt}
\tablehead{
\colhead{Radius} &\colhead{$M^{\ast}_{1}$} &\colhead{$\chi^{2}_{\nu}$} &\colhead{$M^{\ast}_{2}$} &\colhead{$\alpha_{2}$} &\colhead{$\chi^{2}_{\nu}$} &\colhead{No. of}&\colhead{$<B_{gc}>$}\\
($r/r_{200}$)&($R_c$ mag)& &($R_c$ mag)& & & Clusters & ($\mbox{Mpc}^{1.8}$)}
\startdata
0.0--0.2 & $-21.96\pm 0.26$\tablenotemark{a} & 0.28 & $-19.30\pm 0.10$\tablenotemark{b} & $-1.62\pm 0.05$ & 0.61 & 42 & $1066\pm 397$\tablenotemark{c} \\
0.2--0.4 & $-21.84\pm 0.17$ & 0.36 & $-19.28\pm 0.09$ & $-1.64\pm 0.05$ & 0.93 & 39 & $1052\pm 378$ \\
0.4--0.6 & $-21.81\pm 0.21$ & 1.21 & $-19.01\pm 0.12$ & $-1.69\pm 0.09$ & 1.05 & 28 & $961\pm 347$ \\
0.6--1.0 & $-21.87\pm 0.27$ & 0.74 & $-18.79\pm 0.12$ & $-1.82\pm 0.10$ & 0.61 & 11 & $683\pm 205$ \\
\enddata
\tablenotetext{a}{$M_{1}^{\ast}$ is derived from a Schechter function fit 
to the bright-end of the LF with $\alpha_{1}=-1.0$.}
\tablenotetext{b}{$M_{2}^{\ast}$ and $\alpha_{2}$ are obtained from a 
Schechter function fit to the faint-end of the blue galaxy LF.}
\tablenotetext{c}{The uncertainty is calculated from the dispersion of the mean.}
\end{deluxetable}

\clearpage

\begin{figure}
\figurenum{1a}
\epsscale{0.8}
\plotone{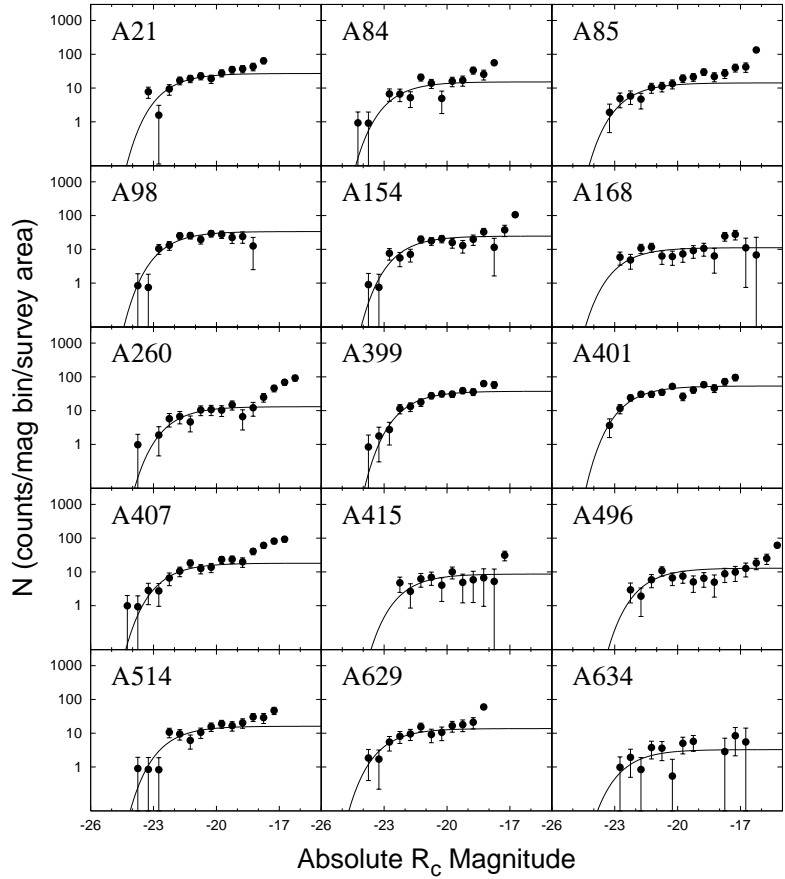}
\caption{Solid points depict the $R_c$-band luminosity function for 57 Abell 
clusters. The solid line represents the best-fit Schechter function with 
a fixed faint-end slope of $\alpha=-1$. The galaxy counts are measured from 
within a cluster-centric radius of $(r/r_{200})= 0.4$, except for A496 and 
A1142 due to the lack of adequate radial coverage. The brightest cluster 
galaxy has been omitted from each LF.}
\label{single-LFa}
\end{figure}

\begin{figure}
\figurenum{1b}
\epsscale{0.9}
\plotone{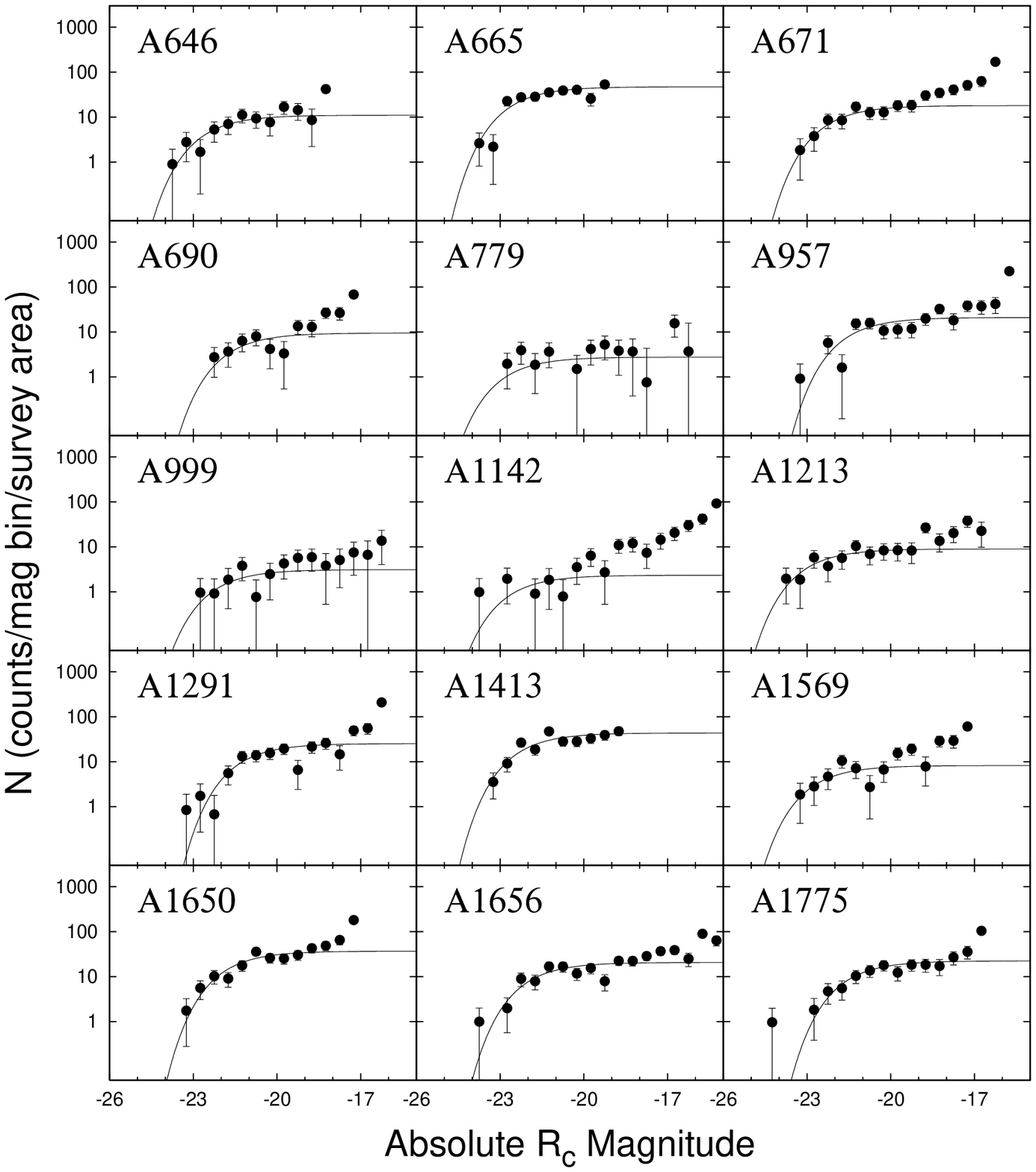}
\caption{Continued}
\label{single-LFb}
\end{figure}

\begin{figure}
\figurenum{1c}
\epsscale{0.9}
\plotone{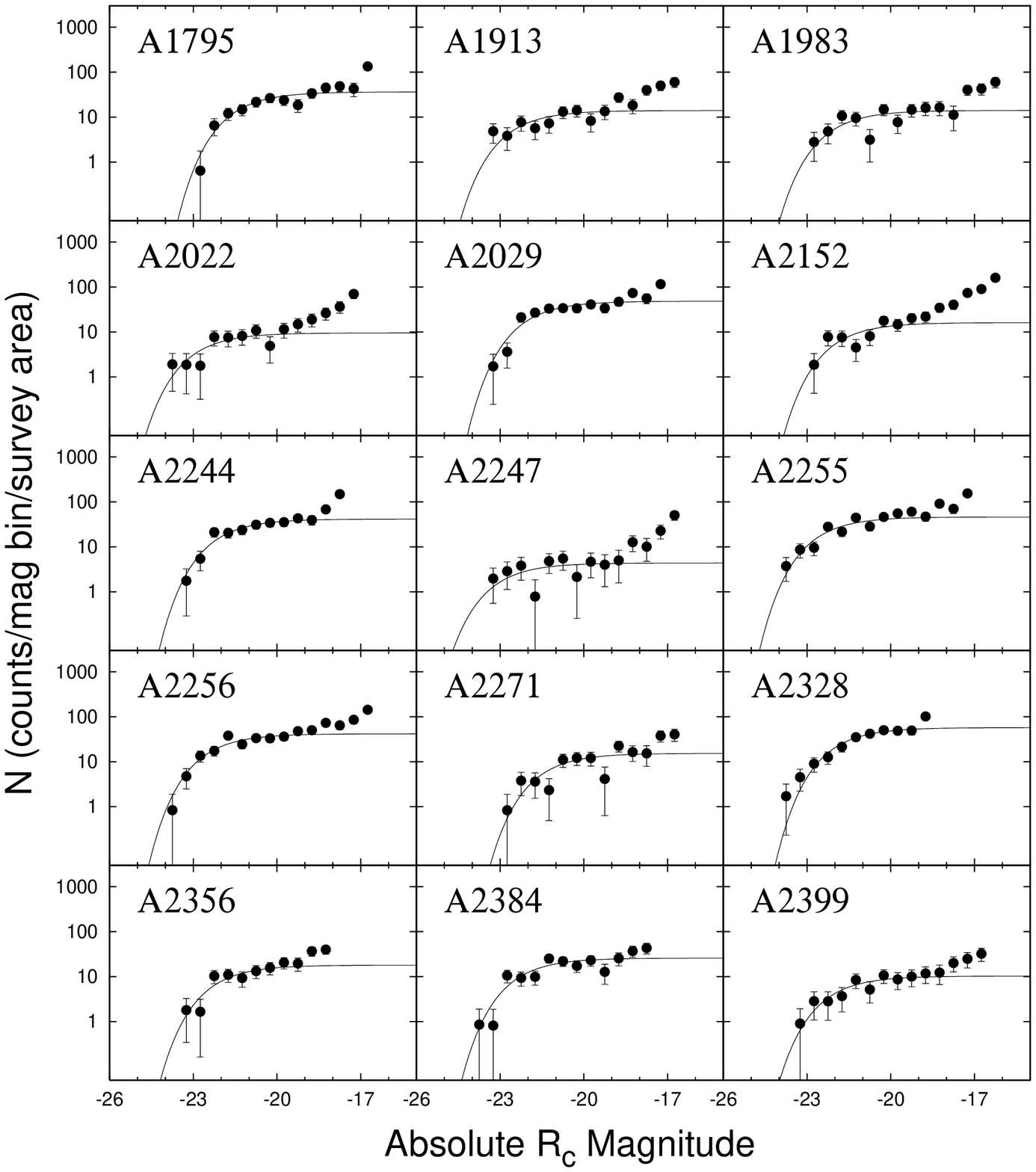}
\caption{Continued}
\label{single-LFc}
\end{figure}

\begin{figure}
\figurenum{1d}
\epsscale{0.9}
\plotone{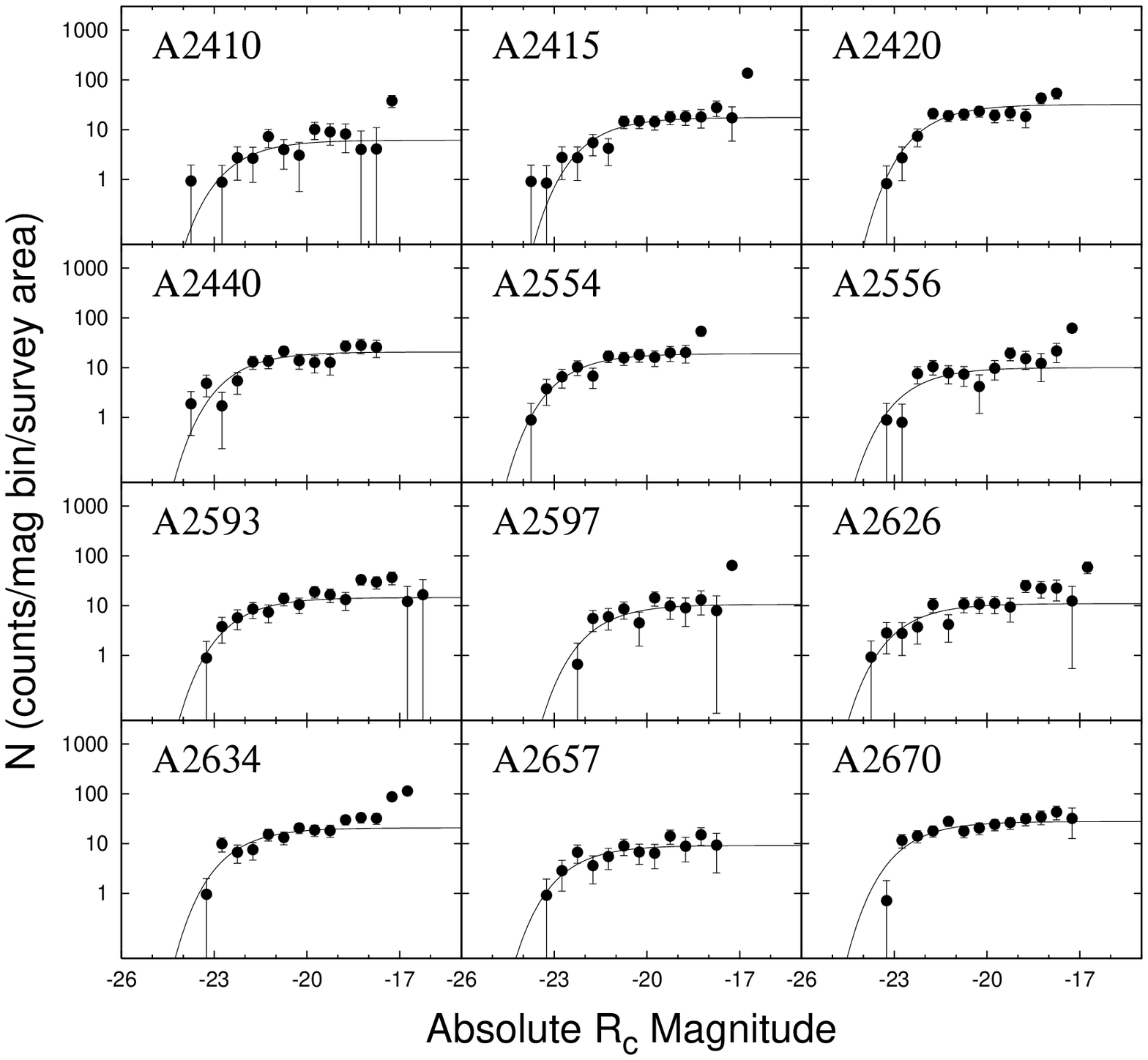}
\caption{Continued}
\label{single-LFd}
\end{figure}

\begin{figure}
\figurenum{2}
\epsscale{1.0}
\plotone{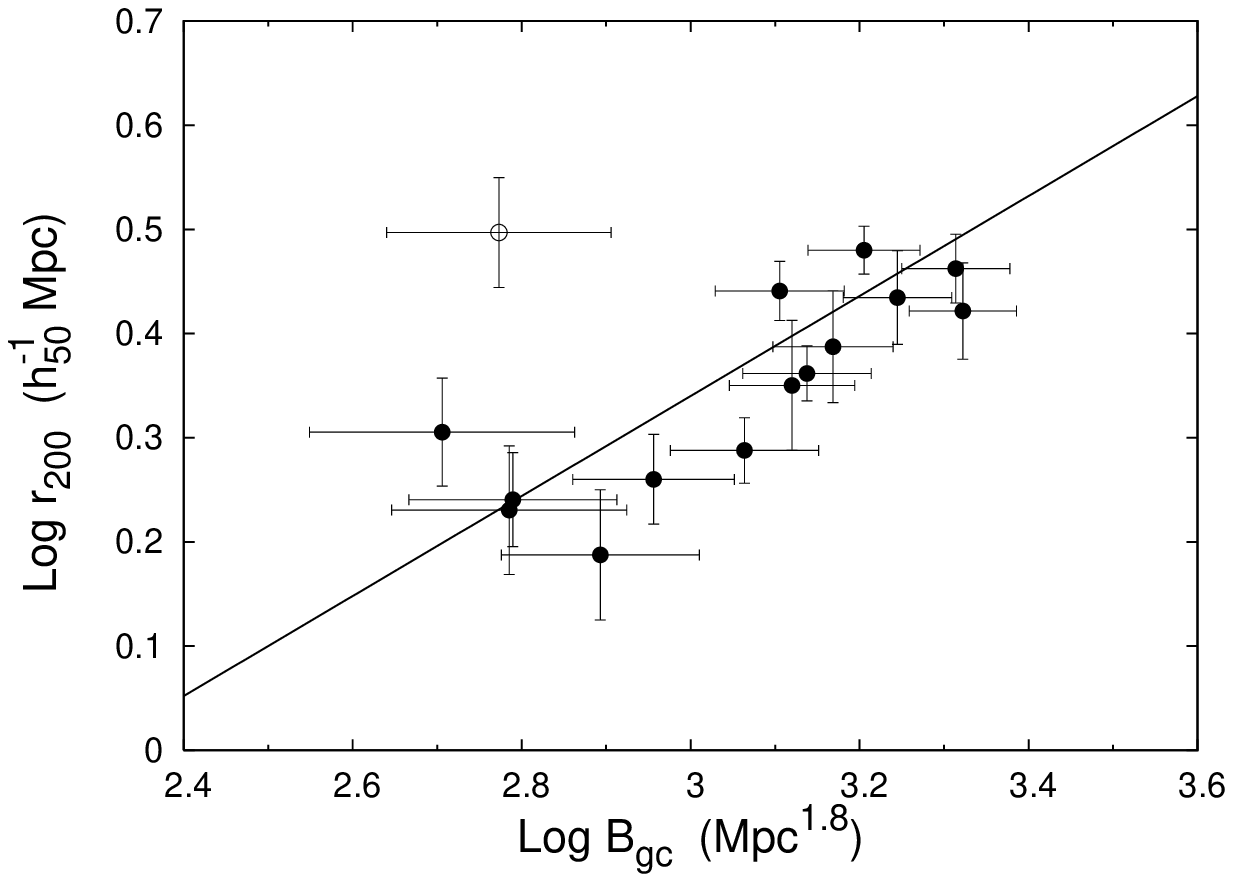}
\caption{Logarithmic correlation between $r_{200}$ and $B_{gc}$ for 
15 clusters from the CNOC1 cluster sample. The solid line is a fit using 
the BCES estimator to the 14 clusters depicted by the solid circles. The 
open circle represents the outlier cluster MS 1455+22, which was not used 
in the fitting process. The rms scatter in the derived values of $r_{200}$ 
is on the order of 15\%.} 
\label{Bgc-r200}
\end{figure}

\clearpage

\begin{figure}
\figurenum{3}
\epsscale{1.0}
\plotone{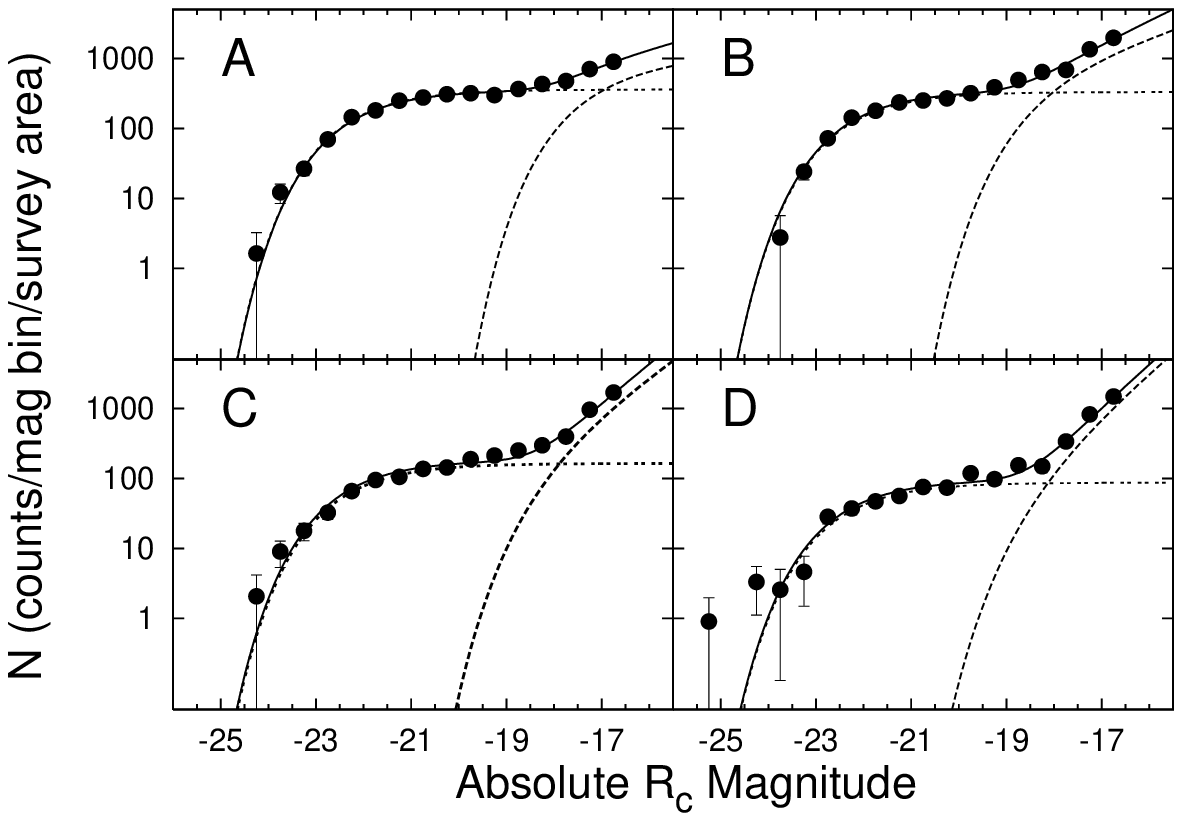}
\caption{Composite total $R_c$-band luminosity function for four cluster-centric 
annuli: A) $(r/r_{200})\leq 0.2$, B) $0.2\leq (r/r_{200})\leq 0.4$, C) 
$0.4\leq (r/r_{200})\leq 0.6$, and D) $0.6\leq (r/r_{200})\leq 1.0$. The 
short dashed line represents a Schechter function fit to bright-end with 
a fixed faint-end slope, $\alpha=-1$. The solid circles depict the combined 
net galaxy counts for all contributing clusters. The long dashed line is a 
Schechter function fit to the faint-end, while the solid line is the sum of 
the two Schechter functions.}
\label{Comp-All-LF}
\end{figure}

\clearpage

\begin{figure}
\figurenum{4}
\epsscale{1.0}
\plotone{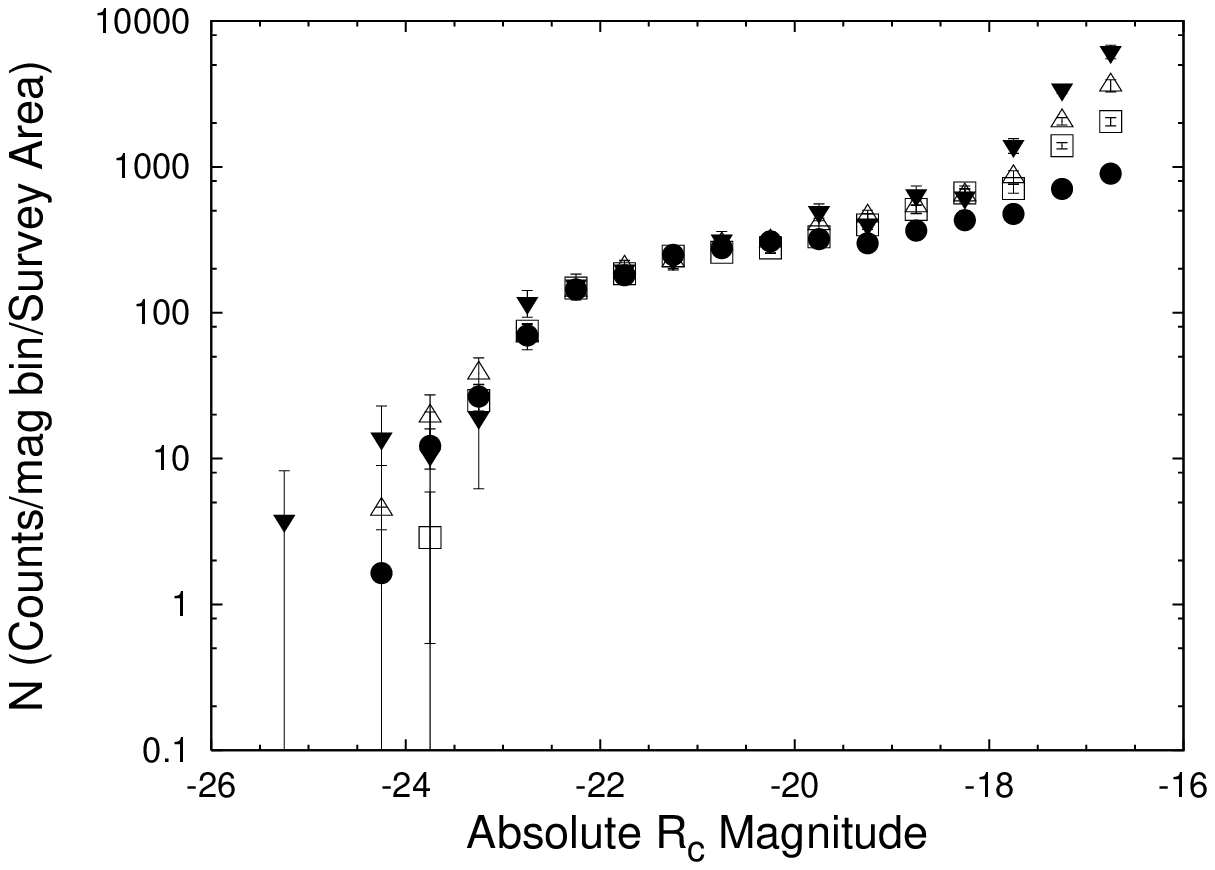}
\caption{Superposition of the total composite LF measured for four 
cluster-centric radial bins. The outer three LFs have been scaled to match 
the inner-most LF in the $-22\leq M_{R_c}\leq -21$ magnitude range. The plot 
symbols depict the following: solid circles --- the net galaxy counts in 
the $(r/r_{200})\leq 0.2$ annulus; open squares --- the counts in the 
$0.2\leq (r/r_{200})\leq 0.4$ annulus; open triangles --- the galaxy 
counts in the $0.4\leq (r/r_{200})\leq 0.6$ radial bin; and the solid 
triangles --- the $0.6\leq (r/r_{200})\leq 1.0$ annulus.}
\label{Comp-All-LF-Scale}
\end{figure}

\clearpage

\begin{figure}
\figurenum{5}
\epsscale{1.0}
\plotone{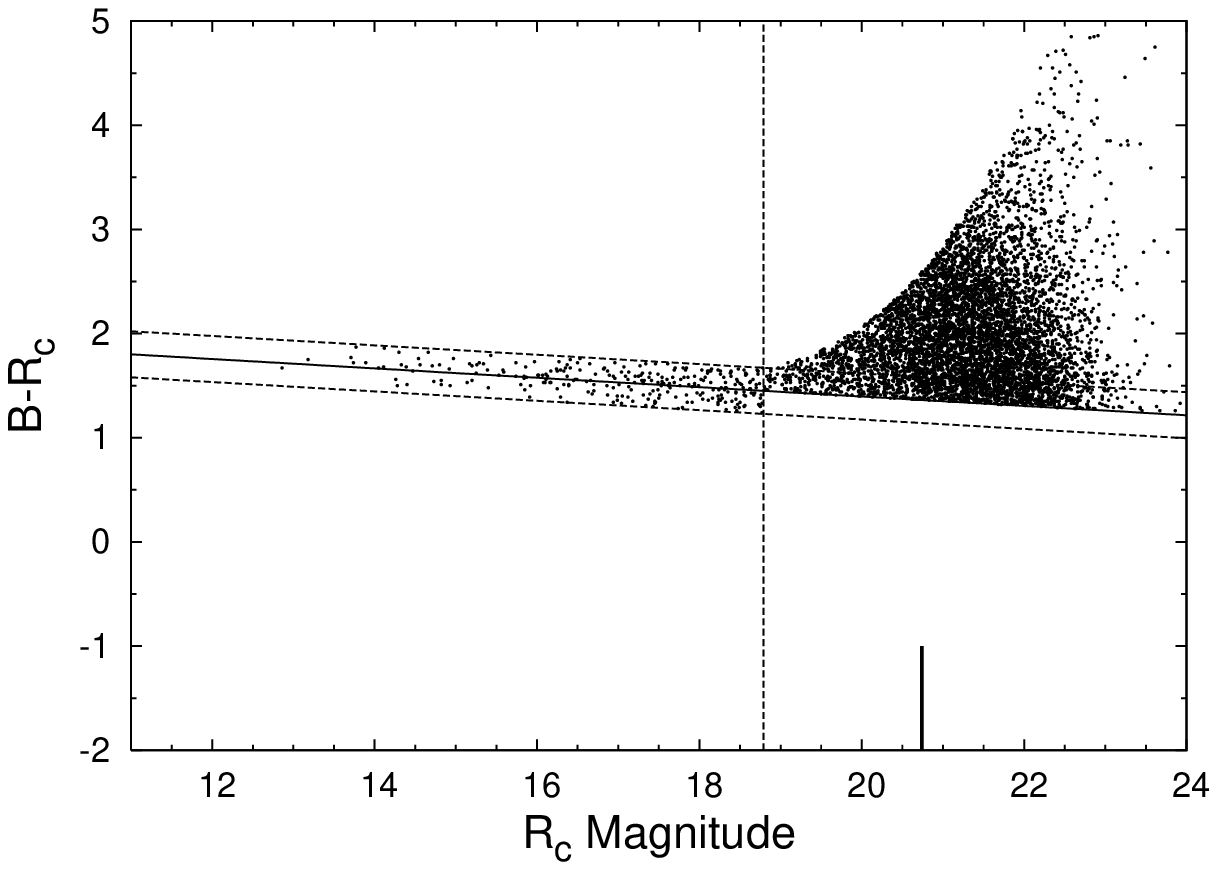}
\caption{Color-magnitude diagram for A260, depicting the region where 
red sequence galaxies were selected. For galaxies brighter than $R_c=18.8$ 
(dashed vertical line), red sequence galaxies are designated as those 
within $\pm 0.22$ mag ($\pm 3\sigma$) of the cluster red sequence relation 
(nearly horizontal solid line). Galaxies fainter than $R_c=18.8$ are selected 
if they lie within the region $2.5\sigma_{B-R_c}$ redward of the 
red sequence and brighter than the magnitude completeness limit (solid 
short vertical line).}
\label{redseq-region}
\end{figure}

\clearpage

\begin{figure}
\figurenum{6}
\epsscale{1.0}
\plotone{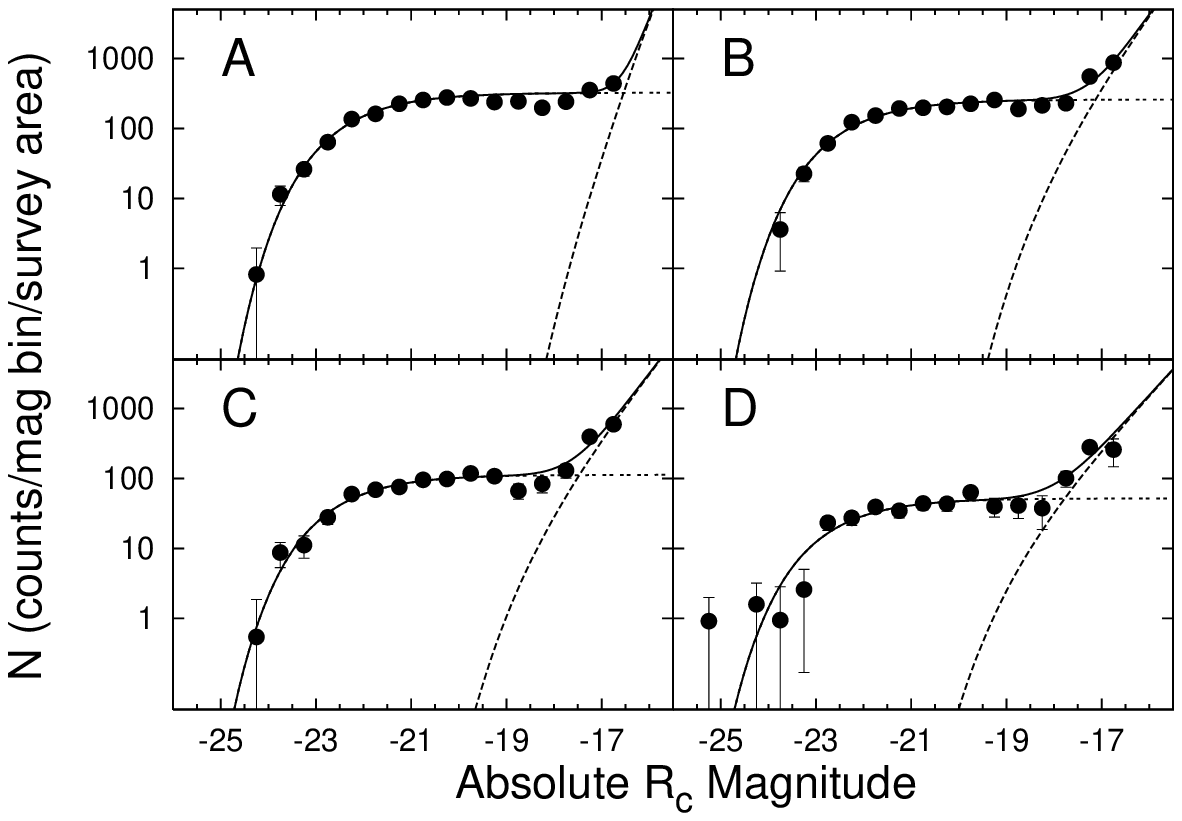}
\caption{Composite red sequence $R_c$-band LF for four cluster-centric annuli: 
A) $(r/r_{200})\leq 0.2$, B) $0.2\leq (r/r_{200})\leq 0.4$, C) $0.4\leq 
(r/r_{200})\leq 0.6$, and D) $0.6\leq (r/r_{200})\leq 1.0$. The short dashed 
line represents a Schechter function fit to bright-end with a fixed 
faint-end slope, $\alpha=-1$. The solid circles depict the combined net 
galaxy counts for all contributing clusters in each annulus. The long dashed 
line is a Schechter function fit to the faint-end, while the solid line is 
the sum of the two Schechter functions.}
\label{Comp-Red-LF}
\end{figure}

\clearpage

\begin{figure}
\figurenum{7}
\epsscale{1.0}
\plotone{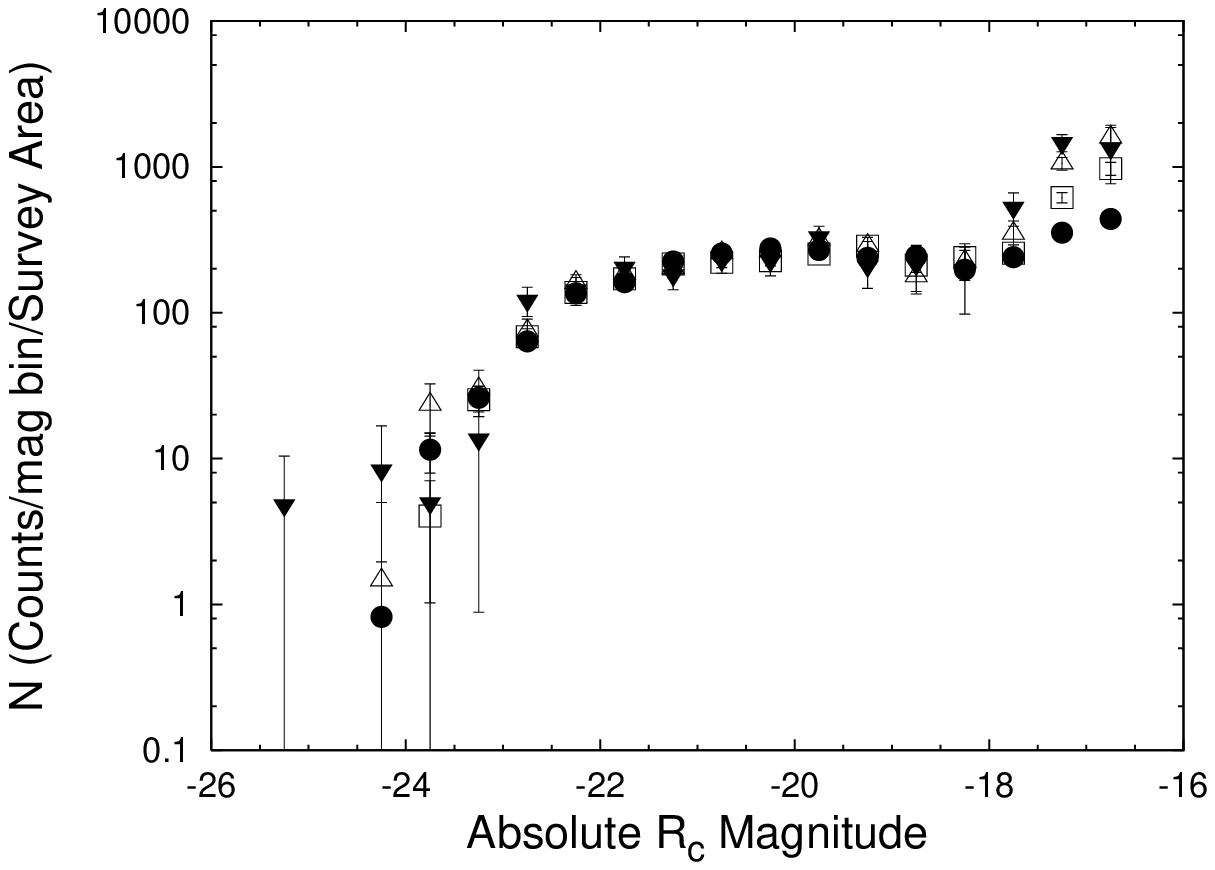}
\caption{Superposition of the red sequence composite LF measured for four 
cluster-centric radial bins. The outer three LFs have been scaled to match 
the inner-most LF in the $-22\leq M_{R_c}\leq -21$ magnitude range. Plot 
symbols are equivalent to those defined in Figure~\ref{Comp-All-LF-Scale}.}
\label{Scale-Comp-Red-LF}
\end{figure}

\clearpage

\begin{figure}
\figurenum{8}
\epsscale{1.0}
\plotone{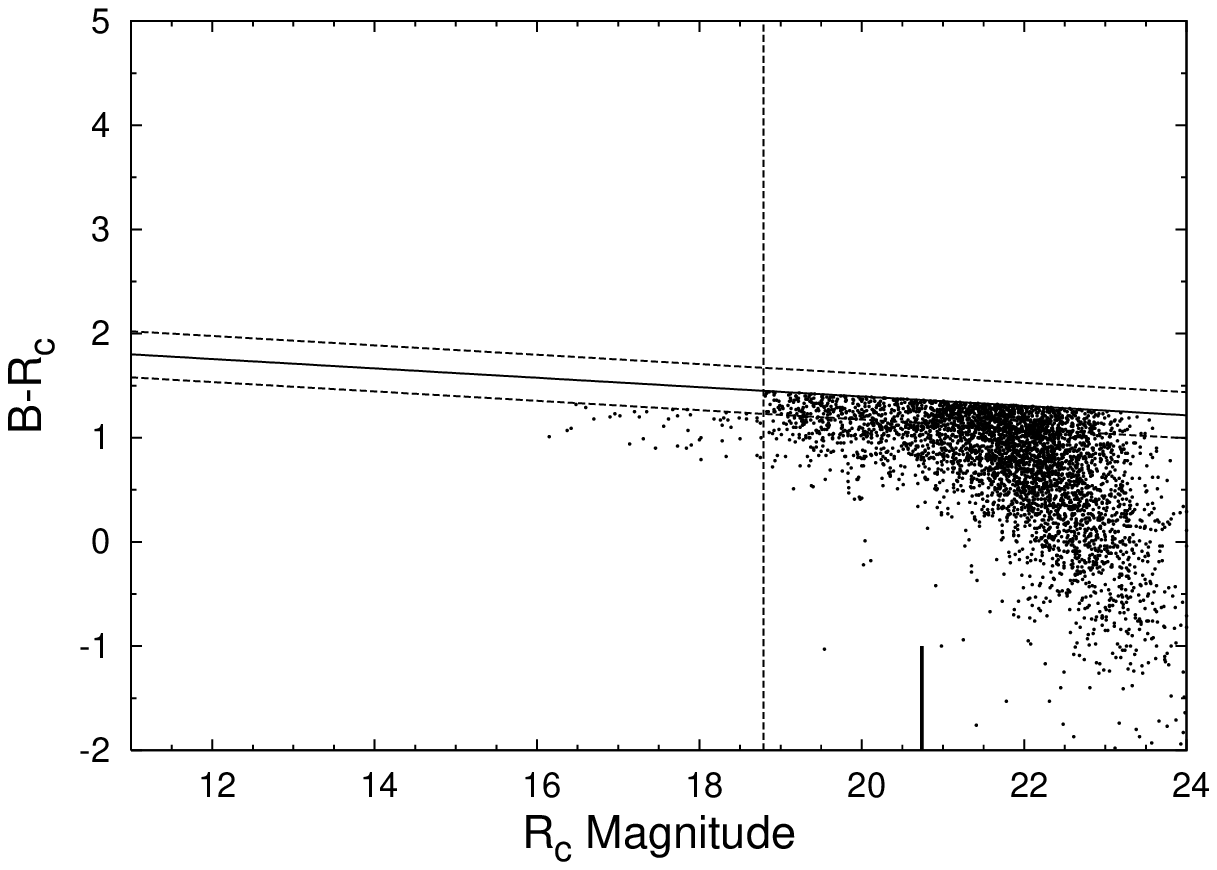}
\caption{Color-magnitude diagram for A260, depicting the region used to 
select the blue cluster galaxy population. For galaxies brighter than 
$R_c=18.8$ (dashed vertical line), blue galaxies are designed as those with 
$B-R_c$ color bluer than the lower envelope used to select the red sequence 
galaxies. Galaxies fainter than $R_c=18.8$ and brighter than the completeness 
limit ($R_c=20.7$, short vertical line), are selected if they are located in 
the region bluer than the lower envelope (blueward of the CMR) defining 
the boundary of the region used to select red sequence galaxies.}
\label{Blue-region}
\end{figure}

\clearpage

\begin{figure}
\figurenum{9}
\epsscale{1.0}
\plotone{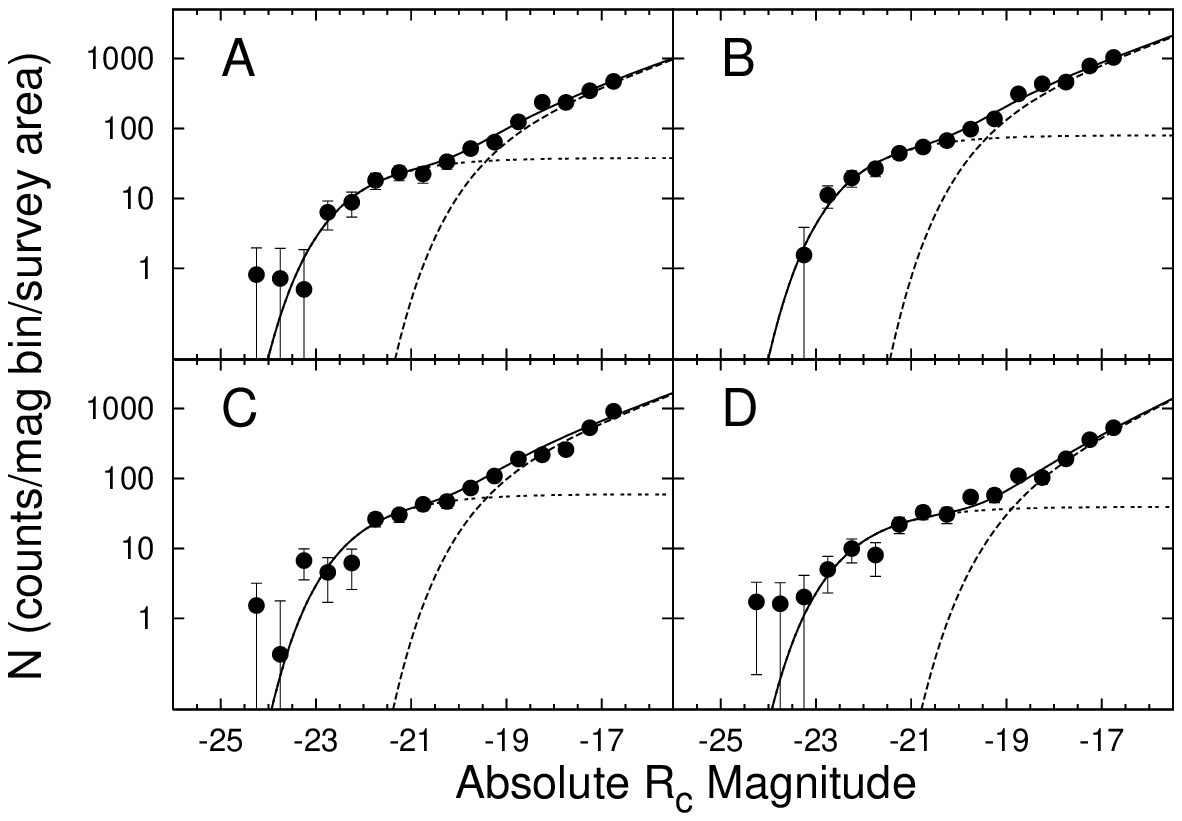}
\caption{Composite blue galaxy $R_c$-band LF for four 
cluster-centric annuli: A) $(r/r_{200})\leq 0.2$, 
B) $0.2\leq (r/r_{200})\leq 0.4$, C) $0.4\leq (r/r_{200})\leq 0.6$, and 
D) $0.6\leq (r/r_{200})\leq 1.0$. The short dashed line represents a 
Schechter function fit to bright-end with a fixed faint-end slope, 
$\alpha=-1$. The solid circles depict the combined net galaxy counts for all 
contributing clusters in each annulus. The long dashed line is a Schechter 
function fit to the faint-end, while the solid line is the sum of the two 
Schechter functions.}
\label{Comp-Blue-LF}
\end{figure}

\clearpage

\begin{figure}
\figurenum{10}
\epsscale{1.0}
\plotone{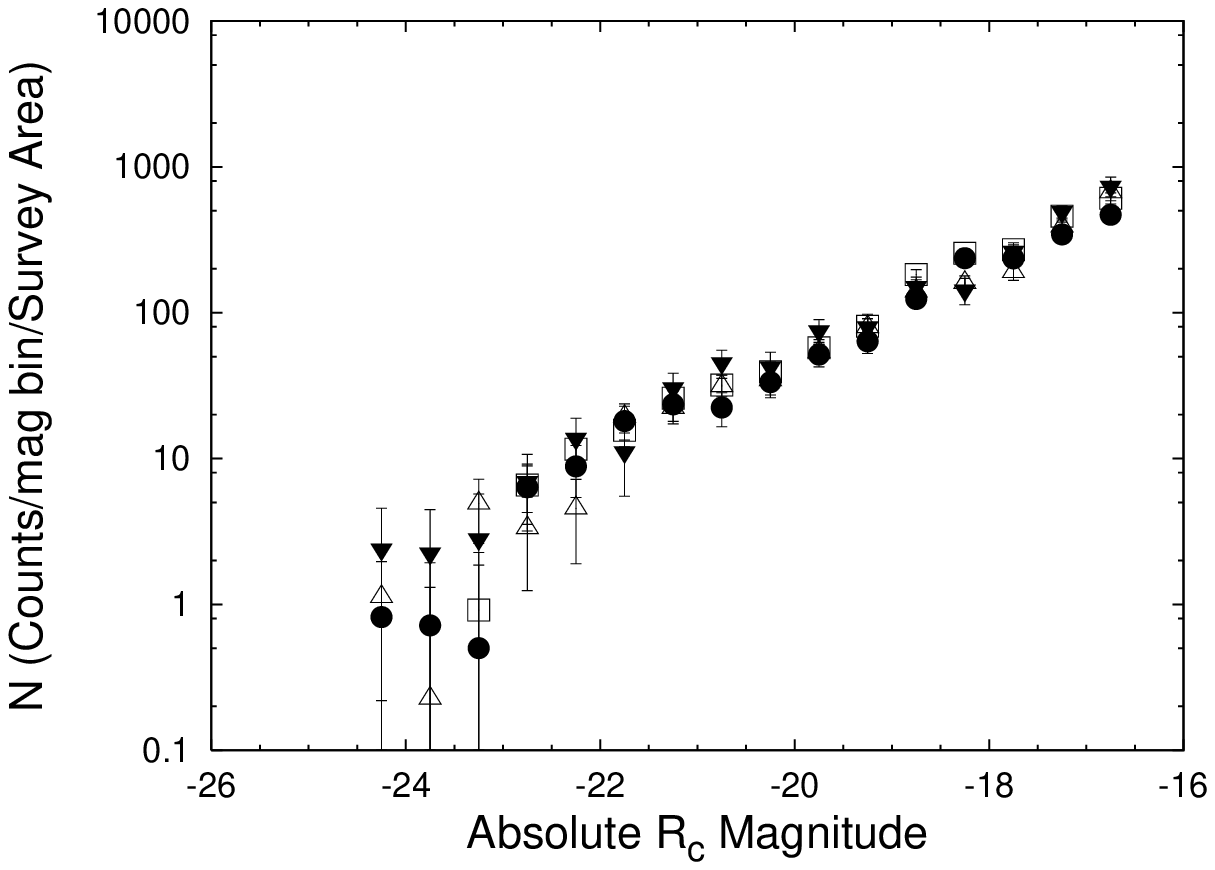}
\caption{Superimposed composite blue galaxy LFs measured for four 
cluster-centric radial bins. The outer three LFs have been scaled to match 
the inner-most LF in the $-22\leq M_{R_c}\leq -21$ magnitude range. Plot 
symbols are equivalent to those defined in Figure~\ref{Comp-All-LF-Scale}.}
\label{Scale-Comp-Blue-LF}
\end{figure}

\clearpage

\begin{figure}
\figurenum{11}
\epsscale{1.0}
\plotone{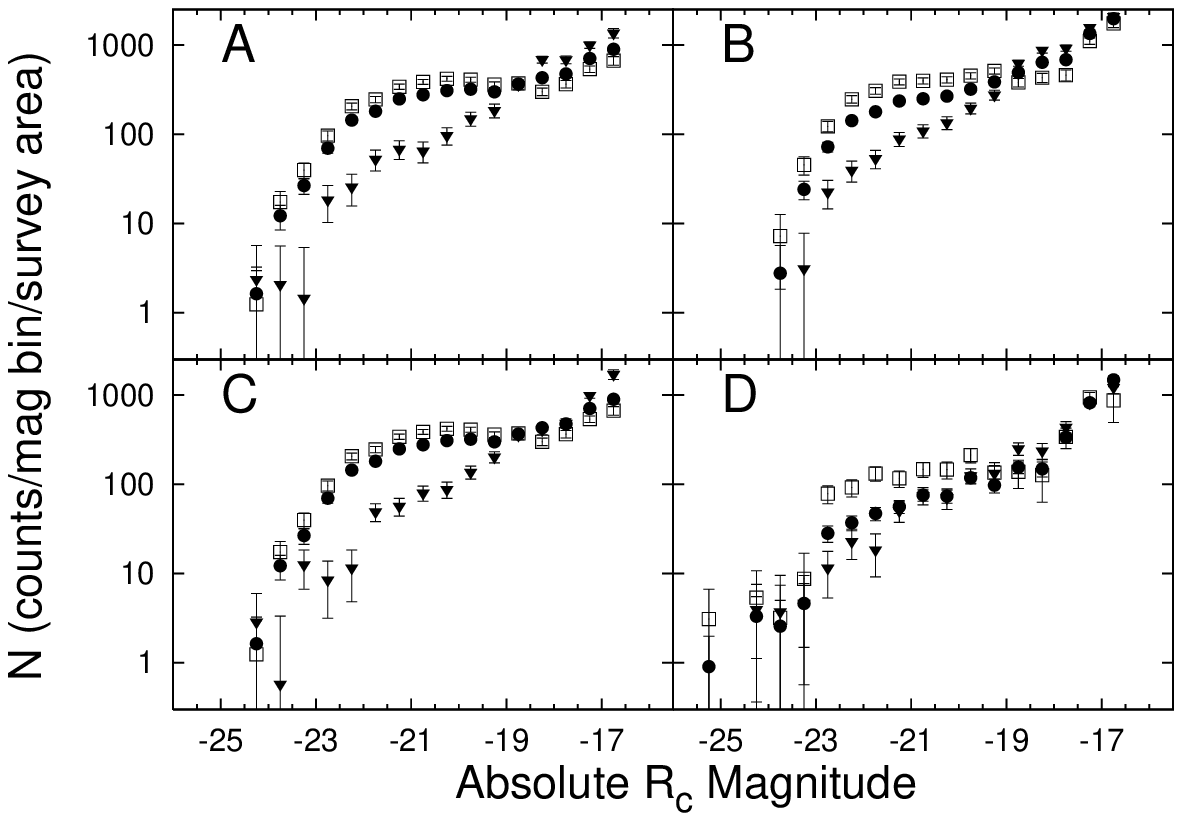}
\caption{Composite LF for the total, red sequence, and blue galaxy 
populations for four radial bins depicted in previous figures (e.g., see 
Figure~\ref{Comp-Blue-LF}). The red sequence and blue LFs (represented by 
open squares and solid triangles, respectively) have been scaled to have the 
same net galaxy counts as the total LF (solid circles) in the 
$-24.0\leq M_{R_c}\leq -16.5$ magnitude range.}
\label{Scale-Compare-LF}
\end{figure}

\clearpage

\begin{figure}
\figurenum{12}
\epsscale{1.0}
\plotone{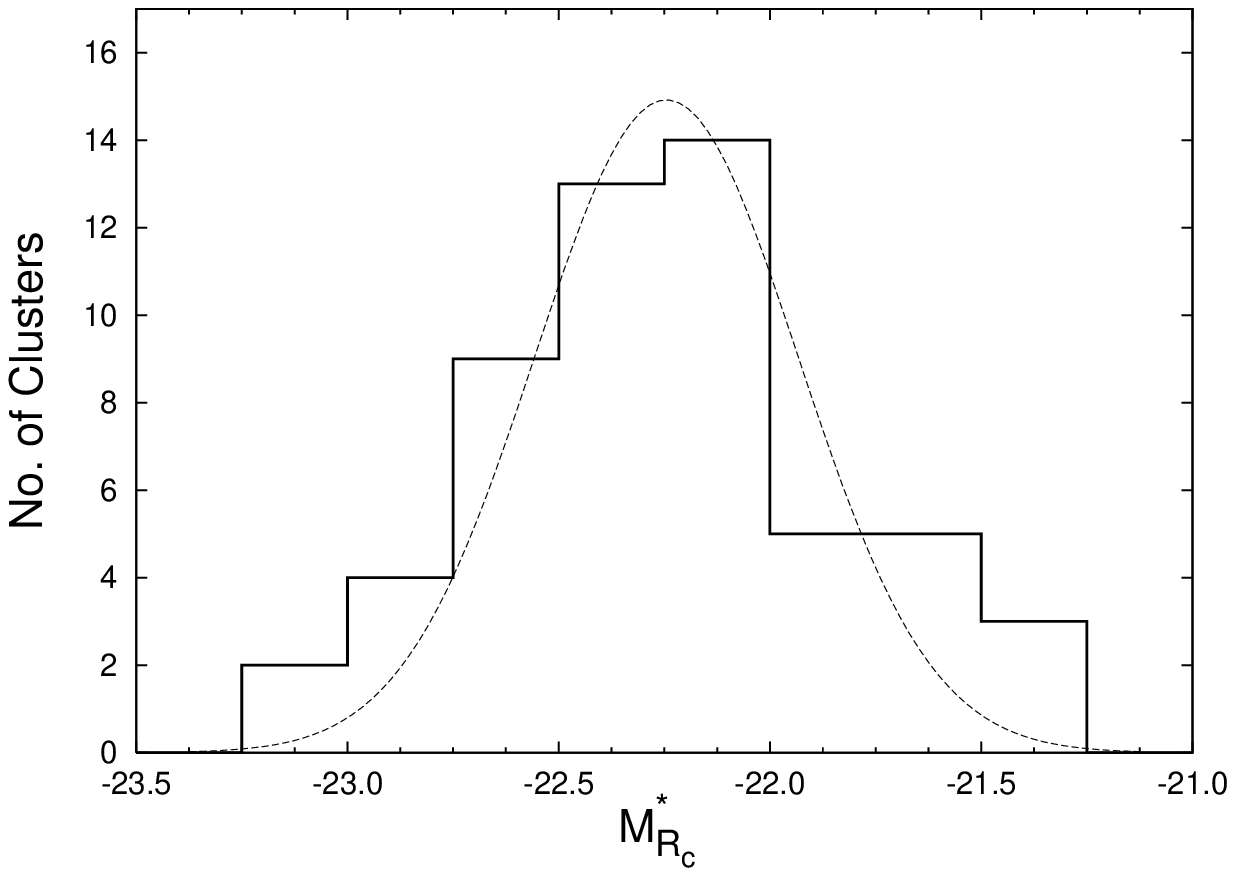}
\caption{Histogram distribution of $M_{R_c}^{\ast}$ measured for 55 clusters 
photometrically complete to $M_{R_c}=-20$ and covering a cluster-centric 
radius of $(r/r_{200})=0.4$. The distribution of $M_{R_c}^{\ast}$ is 
approximately Gaussian with $\langle M_{R_c}^{\ast}\rangle=-22.24\pm 0.06$ and 
a dispersion of $\sigma=0.31$ mag (dashed line).}
\label{Mstar-Gauss}
\end{figure}

\clearpage

\begin{figure}
\figurenum{13}
\epsscale{1.0}
\plotone{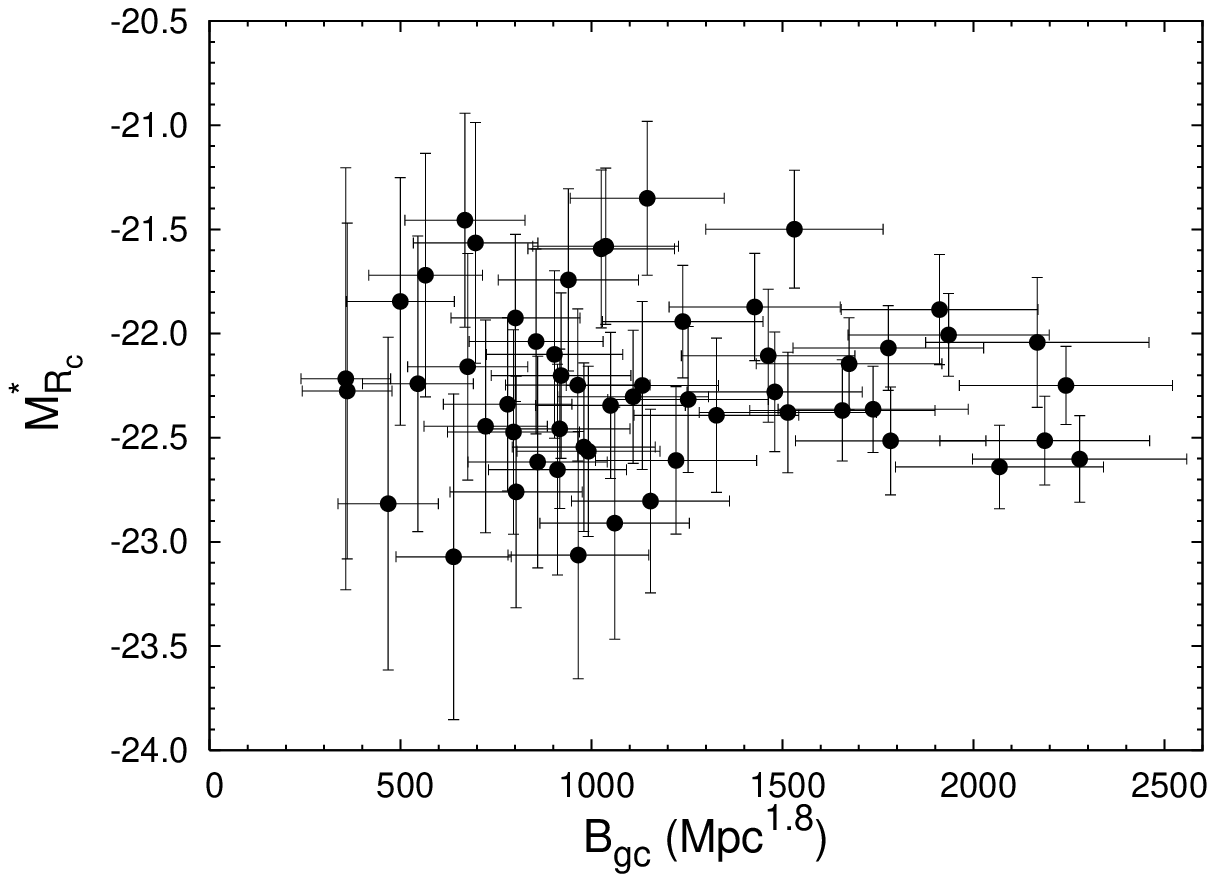}
\caption{Distribution of $M_{R_c}^{\ast}$ with cluster richness $B_{gc}$ for 
the 55 clusters depicted in Figure~\ref{Mstar-Gauss}. No significant 
correlation between $M_{R_c}^{\ast}$ and $B_{gc}$ was found for our sample.}
\label{Mstar-Bgc}
\end{figure}

\clearpage

\begin{figure}
\figurenum{14}
\epsscale{1.0}
\plotone{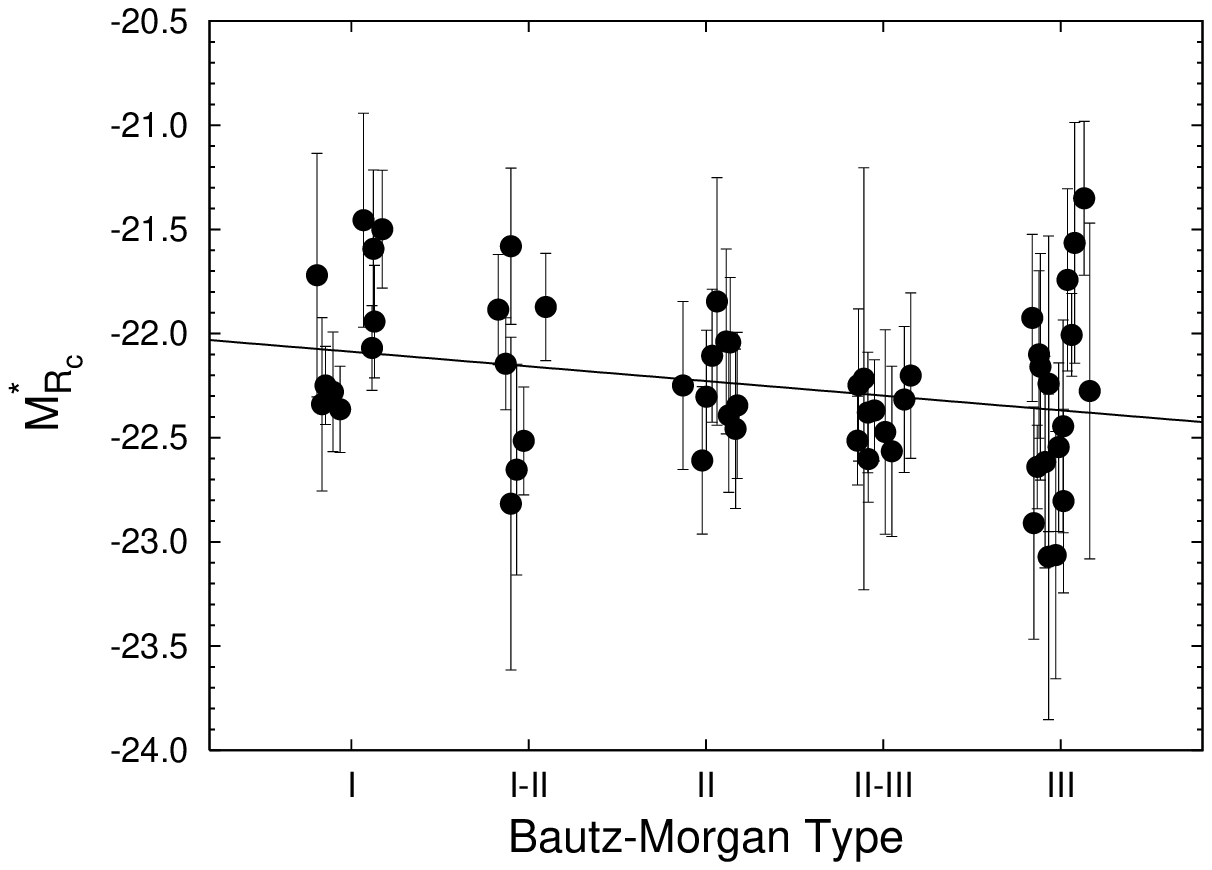}
\caption{$M_{R_c}^{\ast}$ as a function of BM-type for 54 of the 
55 clusters depicted in Figure~\ref{Mstar-Bgc}. The solid line represents 
a least-squares fit to the data and indicates that $M_{R_c}^{\ast}$ brightens 
for later BM-type. A Spearman rank-order correlation coefficient test yields 
that the two measurements are correlated at the 96\% significance level.}
\label{Mstar-Bgc-BM}
\end{figure}

\clearpage

\begin{figure}
\figurenum{15}
\epsscale{1.0}
\plotone{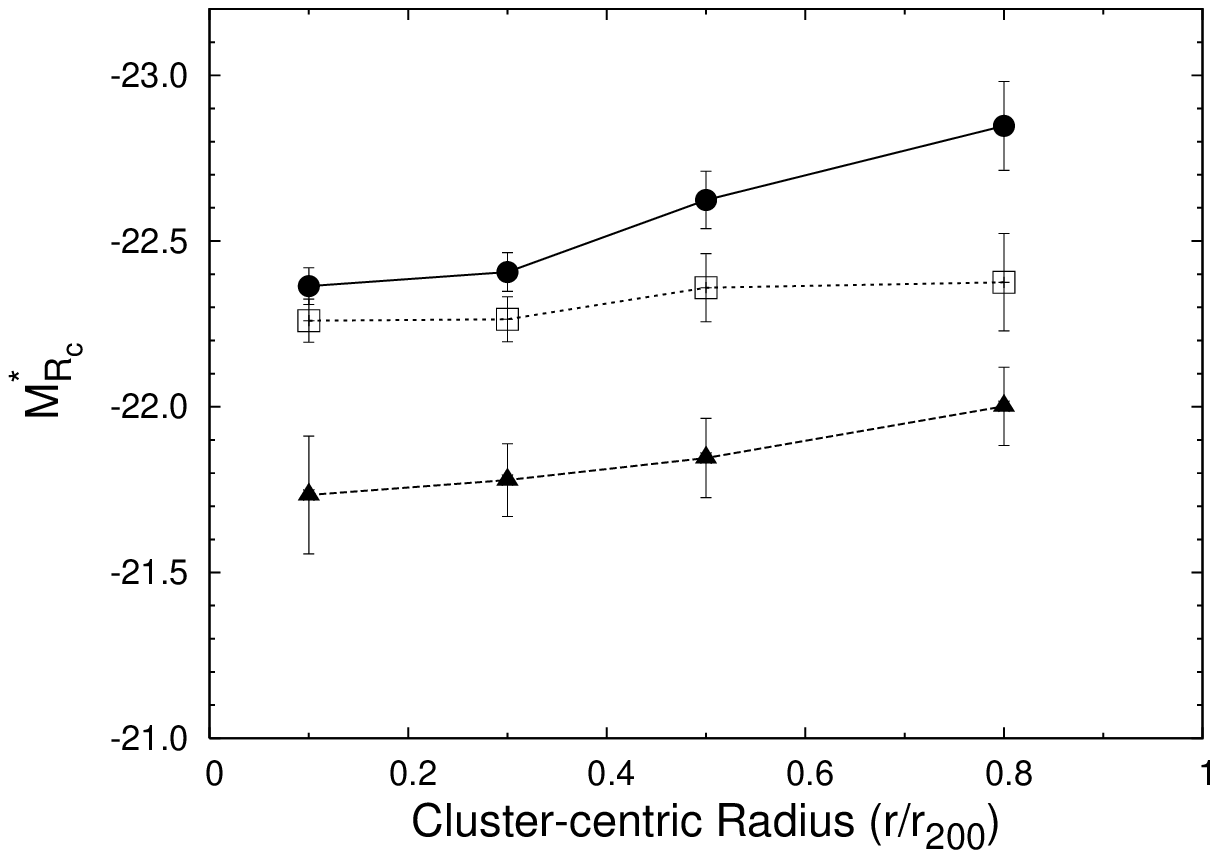}
\caption{Variation in $M_{R_c}^{\ast}$ with cluster-centric radius for the 
total (open squares), red sequence (solid circles), and blue galaxy 
populations (solid triangles). $M_{R_c}^{\ast}$ is measured for a composite 
sample of 57 clusters complete to $M_{R_c}=-20$. The red sequence and blue 
galaxy samples exhibit a trend in which $M_{R_c}^{\ast}$ becomes brighter 
with increasing cluster-centric radius.}
\label{Mstar-Radius}
\end{figure}

\clearpage

\begin{figure}
\figurenum{16}
\epsscale{1.0}
\plotone{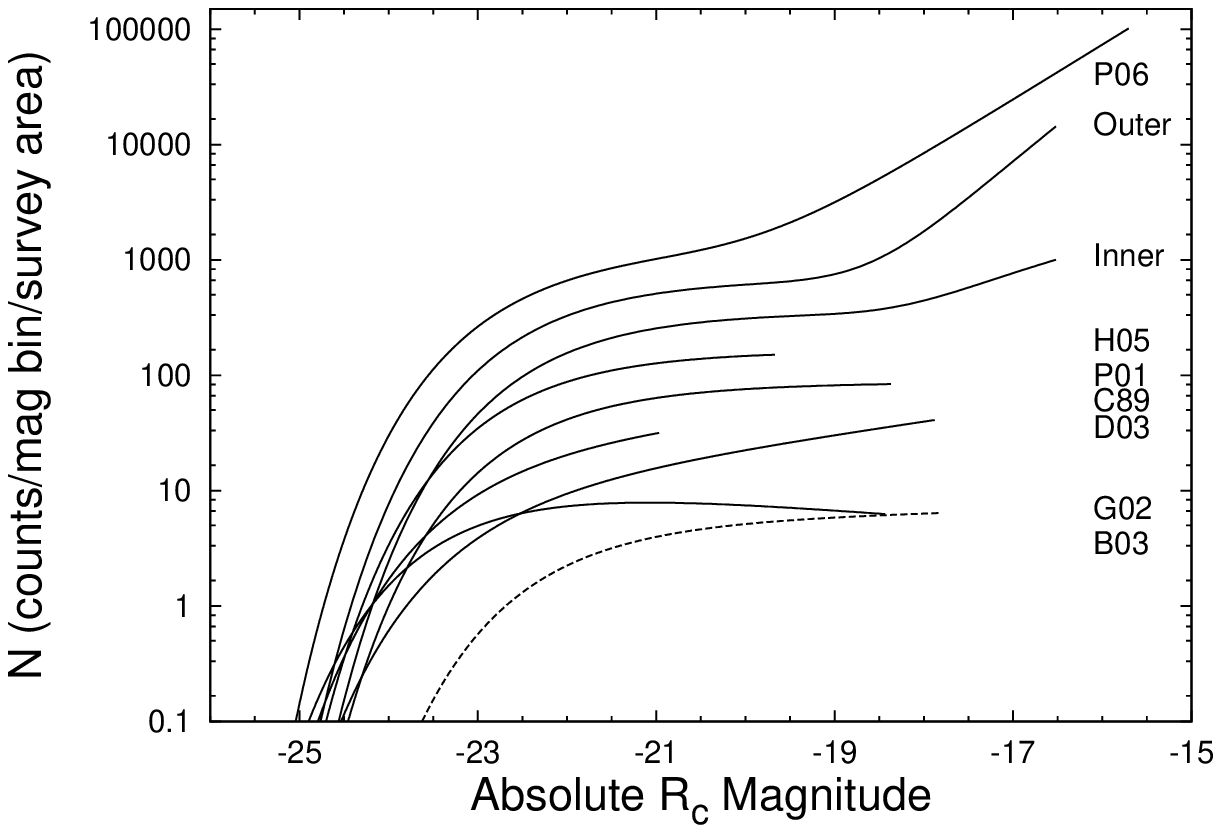}
\caption{Comparison of cluster LFs with published 
sources: P06 -- Popesso et al. 2006, Outer --- this paper, 
composite total LF for $0.6\leq (r/r_{200})\leq 1.0$, Inner -- this paper, 
composite total LF for $(r/r_{200})\leq 0.2$, H05 -- Hansen et al. 
2005, P01 -- Piranomonte et al. 2001, C89 -- Colless 1989, D03 -- De Propris 
et al. 2003, G02 -- Goto et al. 2002, and B03 -- Blanton et al. 2003 
(SDSS field LF; dashed line). The LFs have been scaled by 0.3 dex relative 
to each other for comparison purposes.}
\label{CompareLF}
\end{figure}

\clearpage

\begin{figure}
\figurenum{17}
\epsscale{1.0}
\plotone{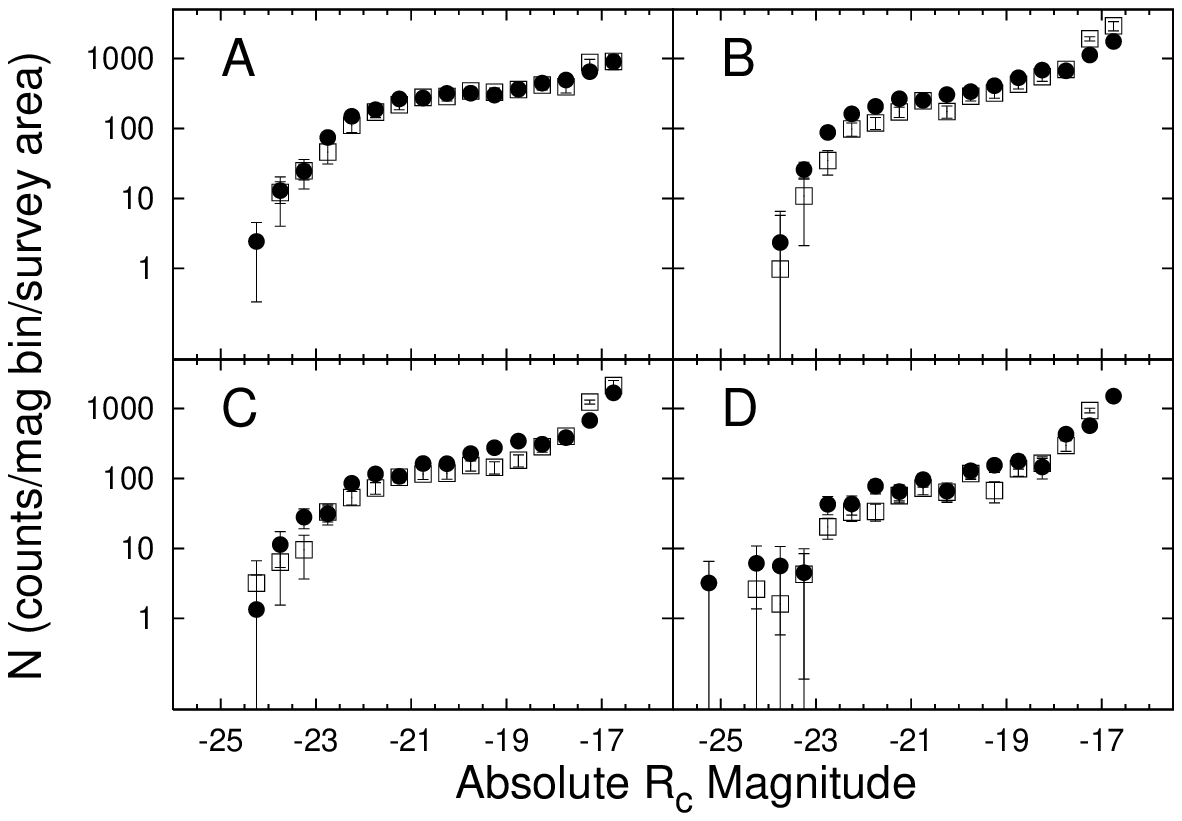}
\caption{Composite total $R_c$-band LF for two groups of 
clusters containing $\leq 10$ redshift-confirmed galaxies (open squares) and 
$\geq 25$ redshift-confirmed galaxies (solid circles). The LFs have 
been scaled to match the total composite LF in the $-24\leq M_{R_c}\leq -17$ 
magnitude range. The four radial bins are equivalent to those used in 
Figure~\ref{Comp-All-LF}.}
\label{Redshift10-25}
\end{figure}

\clearpage

\begin{figure}
\figurenum{18}
\epsscale{1.0}
\plotone{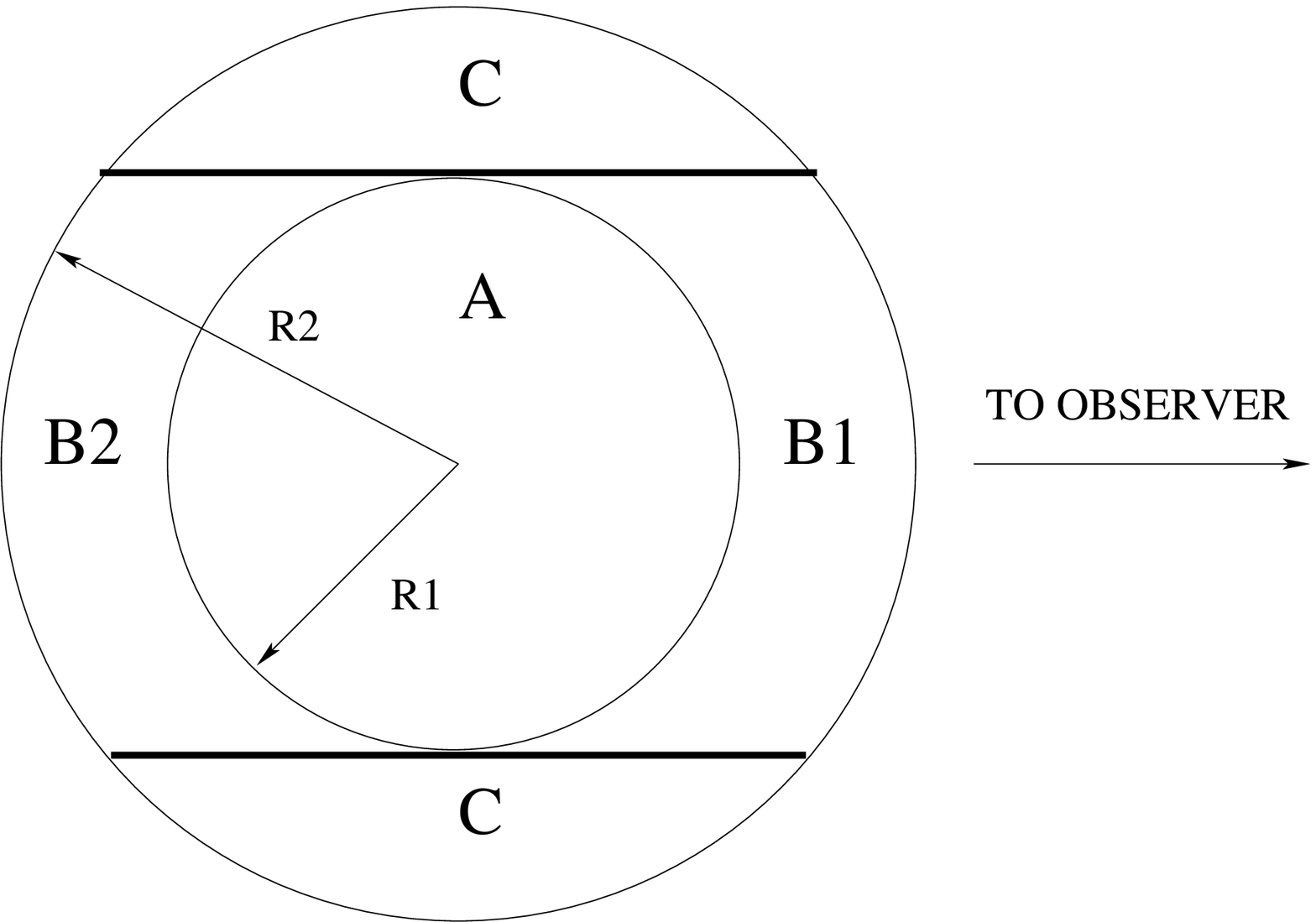}
\caption{Schematic diagram illustrating the geometry used to convert the 
projected LF to the deprojected LF. The projected central 
LF from region A will be contaminated by projected galaxies from regions B1 
and B2 that lie in the cluster outskirts. The LF in region C can be utilized 
to deproject the central LF and thus minimize the influence of the 
contaminating galaxies.}
\label{Deproject-Draw}
\end{figure}

\clearpage

\begin{figure}
\figurenum{19}
\epsscale{0.56}
\plotone{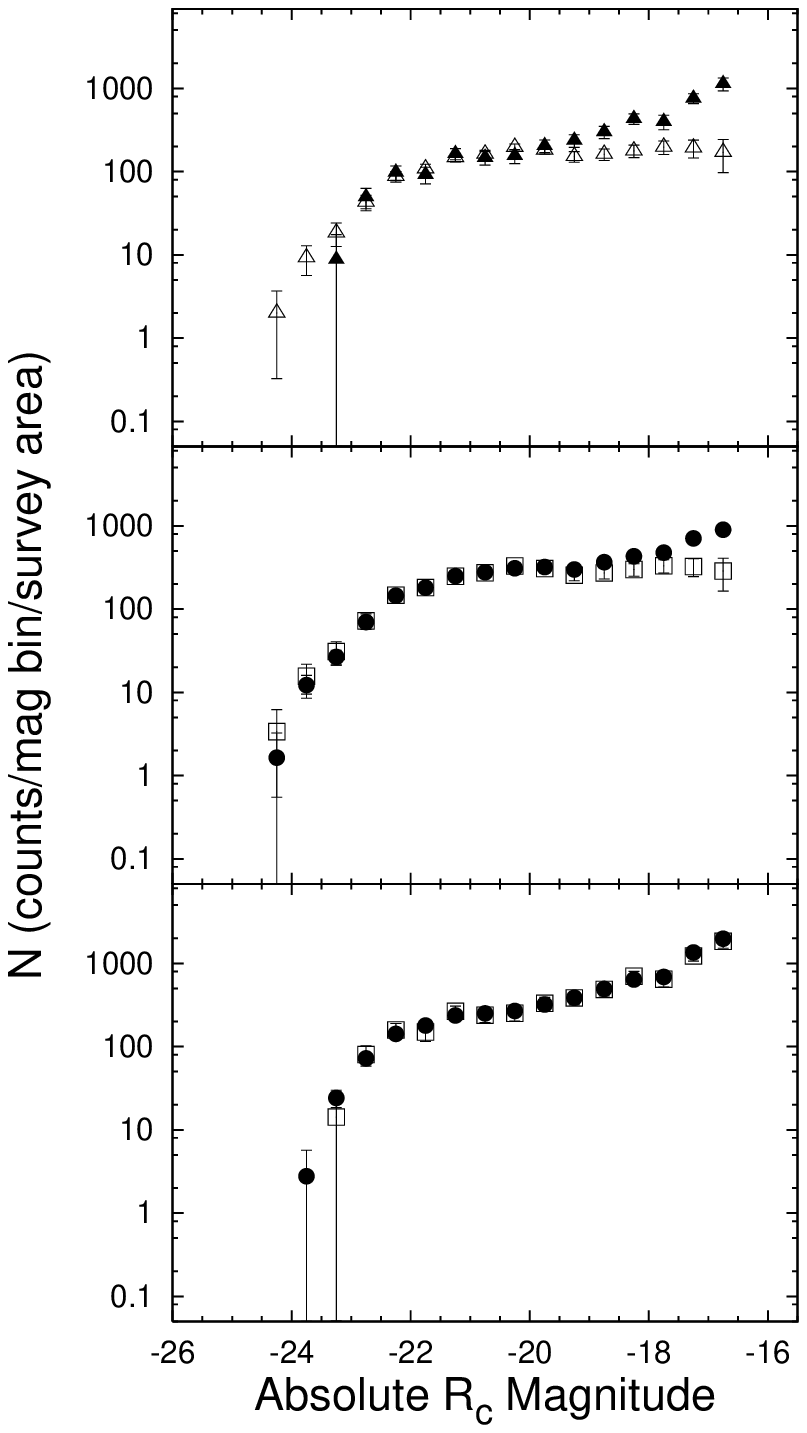}
\caption{{\bf Top Panel:} Deprojected total composite LFs are compared for the 
inner-most radial bin $(r/r_{200})\leq 0.2$ (open triangles) with the 
$0.2\leq (r/r_{200})\leq 0.4$ radial bin (solid triangles). The outer 
deprojected LF has been scaled to match the inner LF in the 
$-22\leq M_{R_c}\leq -21$ magnitude range. {\bf Middle Panel:} The deprojected 
total composite LF (open squares) is compared to the projected LF (solid 
circles) for the $(r/r_{200})\leq 0.2$ annulus. The deprojected LF has been 
scaled to match the projected LF in the $-22\leq M_{R_c}\leq -21$ magnitude 
interval. {\bf Bottom Panel:} The deprojected total composite LF 
(open squares) is compared to the projected LF (solid circles) for the 
$0.2\leq (r/r_{200})\leq 0.4$ annulus. The deprojected LF has been 
scaled to match the projected LF in the $-22\leq M_{R_c}\leq -21$ magnitude 
range.}
\label{All-Deproj}
\end{figure}

\clearpage

\begin{figure}
\figurenum{20}
\epsscale{0.60}
\plotone{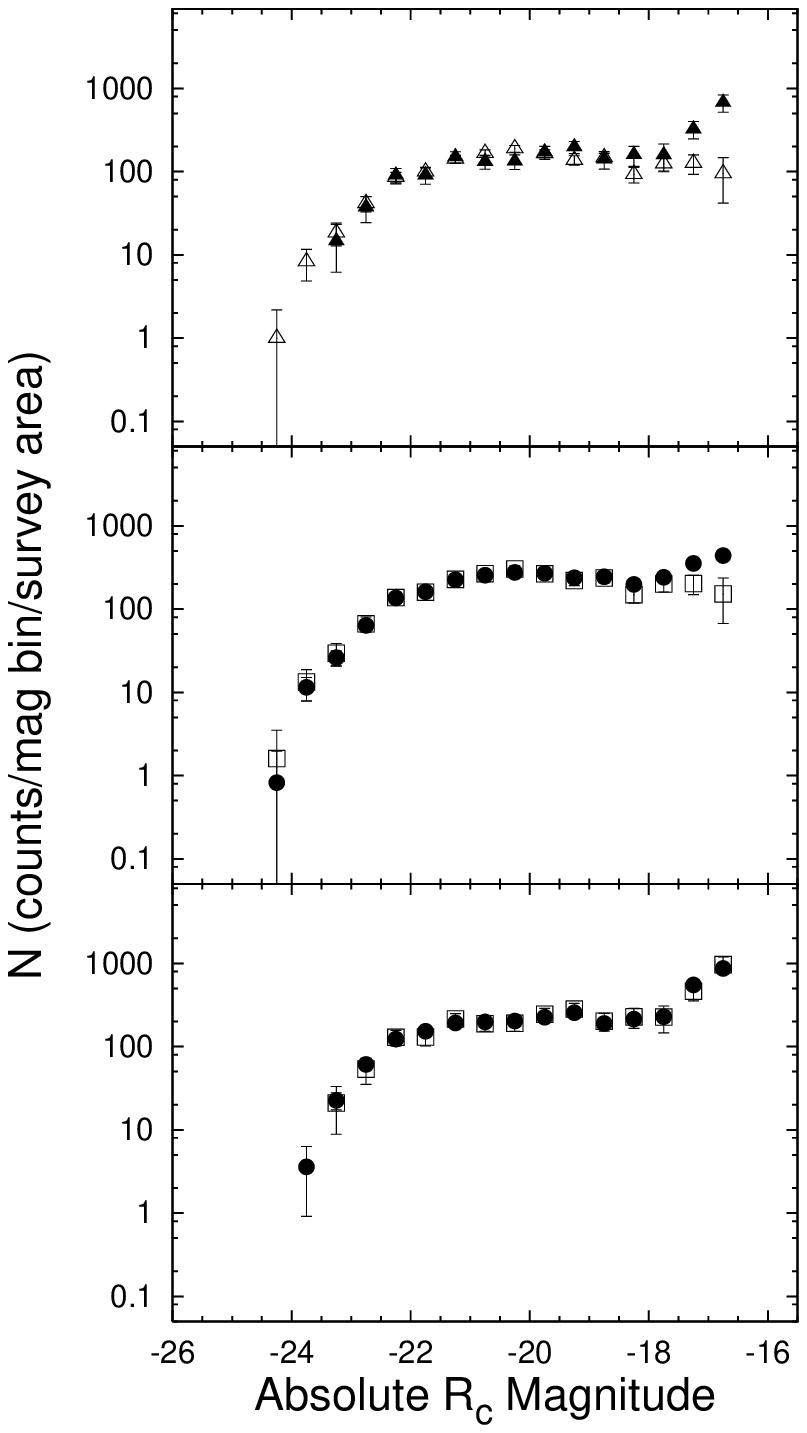}
\caption{Same as Figure~\ref{All-Deproj}, but for the red sequence composite 
LF.}
\label{Red-Deproj}
\end{figure}

\clearpage

\begin{figure}
\figurenum{21}
\epsscale{0.60}
\plotone{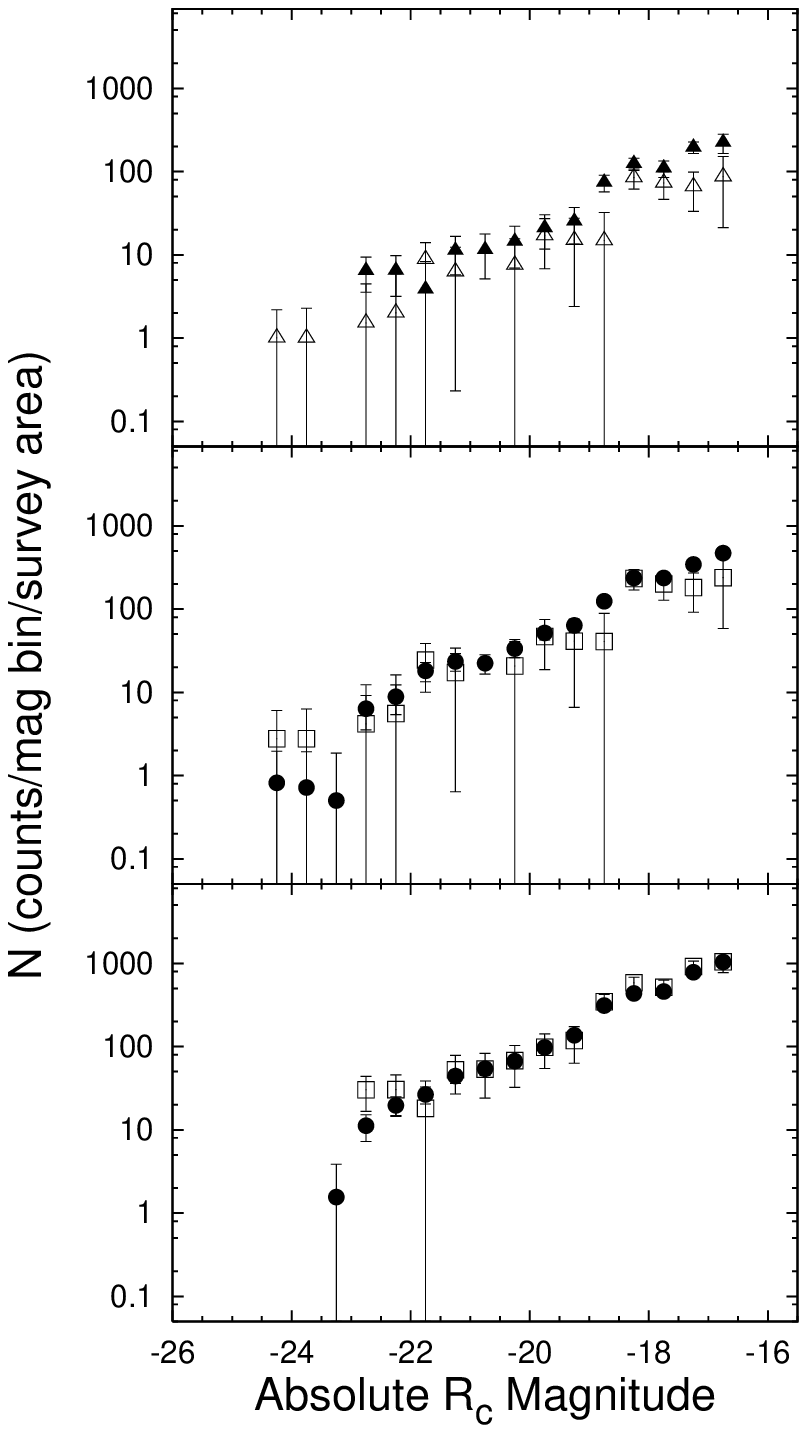}
\caption{Same as Figure~\ref{All-Deproj}, but for the blue composite LF.}
\label{Blue-Deproj}
\end{figure}

\clearpage

\begin{figure}
\figurenum{22}
\epsscale{1.0}
\plotone{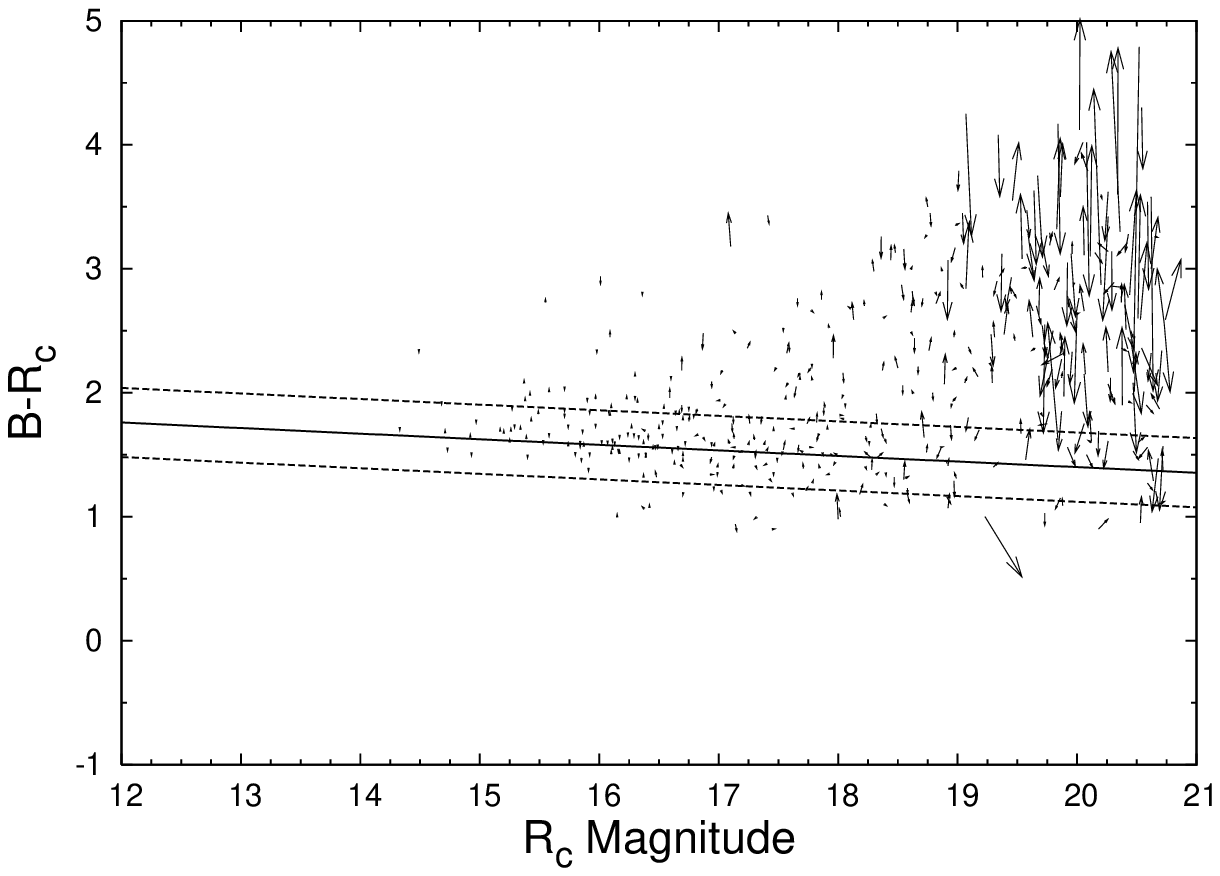}
\caption{Vector-style color-magnitude diagram for A260. The vectors define 
the relative change in $B-R_c$ color and $R_c$-mag for the position of the 
observed galaxies to the location of the simulated galaxies. The cluster 
CMR is depicted by the horizontal solid line and the $\pm 3\sigma$ limit by 
the dashed line. For clarity, only a fraction of the cluster galaxies 
brighter than the magnitude completeness limit for this cluster are 
displayed.}
\label{ColorBias}
\end{figure}

\clearpage

\begin{figure}
\figurenum{23}
\epsscale{1.0}
\plotone{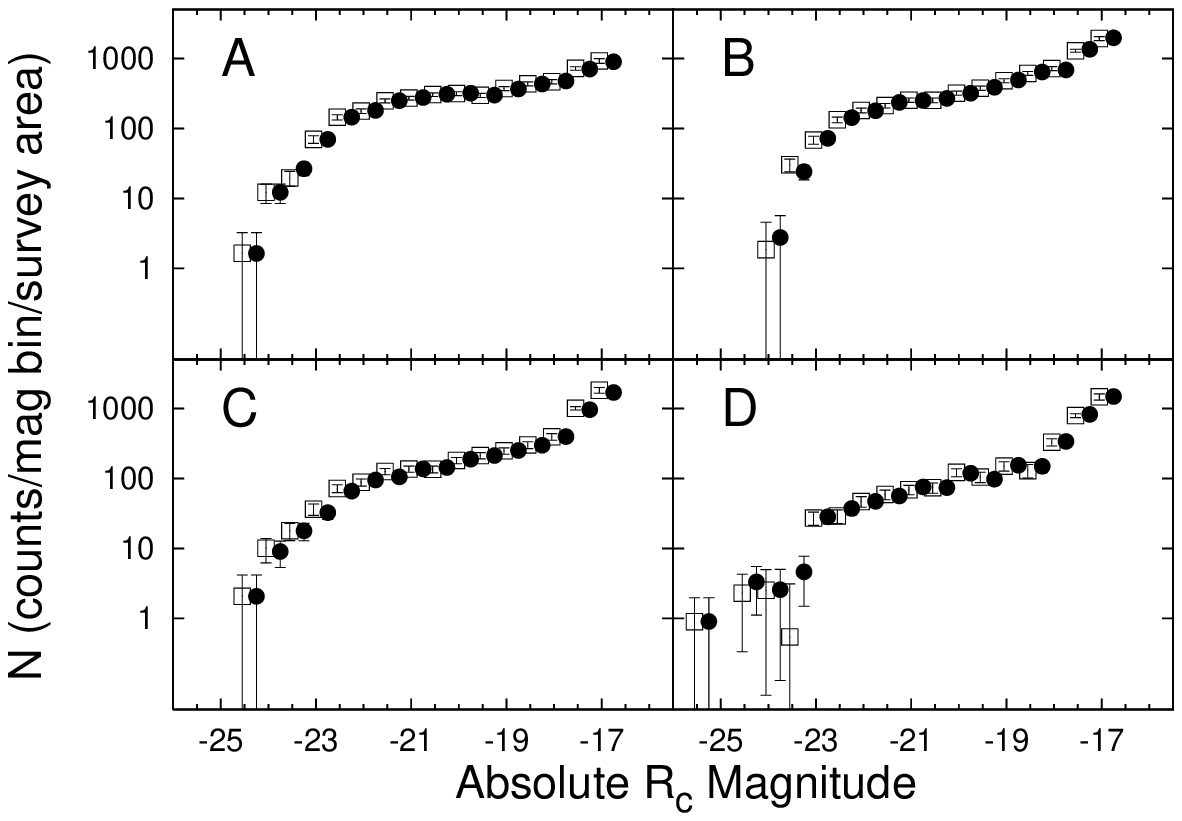}
\caption{Composite total observed LF (solid circles) is compared to the 
simulated LF (open squares) for the four radial bins used previously: 
A) $(r/r_{200})\leq 0.2$, B) $0.2\leq (r/r_{200})\leq 0.4$, 
C) $0.4\leq (r/r_{200})\leq 0.6$, and D) $0.6\leq (r/r_{200})\leq 1.0$. 
The simulated LF is constructed by randomly changing $r_{200}$ by 
$\pm 15\%$. The simulated LFs have been offset by $0.3$ mag in order to 
assist the comparison.} 
\label{Allr200}
\end{figure}

\clearpage

\begin{figure}
\figurenum{24}
\epsscale{1.0}
\plotone{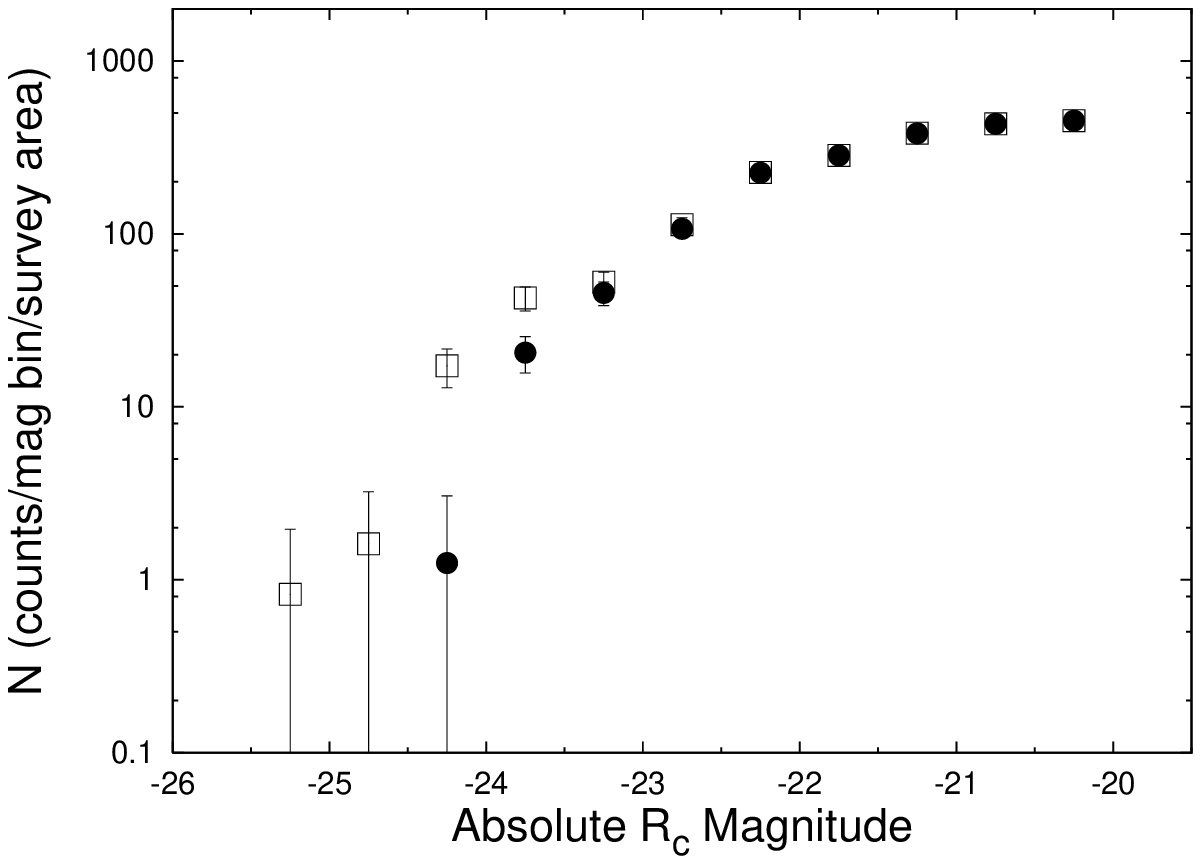}
\caption{Comparison of the total composite LF for 57 clusters that are 
photometrically complete to $M_{R_c}=-20$ and cover the $(r/r_{200})\leq 0.2$ 
annulus. The solid points represent the LF comprised by excluding the 
BCGs from the net galaxy counts. The inclusion of the 
BCGs in the LF is depicted by the open squares.}
\label{CompAll-BCG}
\end{figure}


\clearpage
\begin{thebibliography}{}
 
\bibitem[Abell(1962)]{Abell62} Abell, G. O. 1962, in IAU Symp. 15, Problems 
of Extra-Galactic Research, ed. G. C. McVittie (New York: Macmillan 
Press), 213

\bibitem[Abraham et~al.(1996)]{Abraham96} Abraham, R. G., et al. 1996, 
\apj, 471, 694

\bibitem[Akritas \& Bershady(1996)]{Akritas96} Akritas, M. G., \& Bershady, 
M. 1996, \apj, 470, 706

\bibitem[Akritas \& Siebert(1996)]{Akritas96b} Akritas, M. G., \& Siebert, 
J. 1996, \mnras, 278, 919

\bibitem[Andreon et~al.(2004)]{Andreon04} Andreon, S., Willis, J., 
Quintana, H., Valtchanov, I., Pierre, M., \& Pacaud, F. 2004, \mnras, 
353, 353

\bibitem[Aguerri et~al.(2007)]{Aguerri07} Aguerri, J. A. L., 
S\'anchez-Janssen, R., \& Mu\~noz-Tu\~n\'on, C. 2007, preprint 
(astro-ph/0704.1579)

\bibitem[Baldry et~al.(2004)]{Baldry04} Baldry, I. K., Glazebrook, K., 
Brinkmann, J. Ivezi\'{c}, \v{Z}., Lupton, R. H., Nichol, R. C., \& 
Szalay, A. S. 2004, \apj, 600, 681

\bibitem[Barkhouse(2003)]{Barkhouse03} Barkhouse, W. A. 2003, Ph.D. thesis, 
Univ. Toronto

\bibitem[Bautz \& Morgan(1970)]{BM70} Bautz, L. P., \& Morgan, W. W. 
1971, \apj, 162, L149

\bibitem[Beijersbergen et~al.(2002a)]{Beijersbergen02a} Beijersbergen, M., 
Hoekstra, H., van Dokkum, P. G., \& van der Hulst, T. 2002a, \mnras, 329, 385

\bibitem[Beijersbergen et~al.(2002b)]{Beijersbergen02b} Beijersbergen, M., 
Schaap, W. E., \& van der Hulst, J. M. 2002, \aap, 390, 817

\bibitem[Bernstein \& Bhavsar(2001)]{Bernstein01} Bernstein, J. P., \& 
Bhavsar, S. P. 2001, \mnras, 322, 625

\bibitem[Bhavsar(1989)]{Bhavsar89} Bhavsar, S. P. 1989, \apj, 338, 718

\bibitem[Binggeli et~al.(1988)]{Binggeli88} Binggeli, B., Sandage, A., \& 
Tammann, G. A. 1988, \araa, 26, 509

\bibitem[Biviano et~al.(1995)]{Biviano95} Biviano, A., Durret, F., 
Gerbal, D., Le F\'{e}vre, O., Lobo, C., Mazure, A., \& Slezak, E. 1995, 
\aap, 297, 610

\bibitem[Blanton et~al.(2003)]{Blanton03} Blanton, M. R., et al. 2003, 
\apj, 592, 819

\bibitem[Bower et~al.(1998)]{Bower98} Bower, R. G., Kodama, T., Terlevich, A. 
1998, \mnras, 299, 1193

\bibitem[Brown(1997)]{Brown97} Brown, J. P. 1997, Ph.D. thesis, Univ. Toronto

\bibitem[Burstein \& Heiles(1982)]{Burstein82} Burstein, D., \&  Heiles, C. 
1982, \aj, 87, 1165

\bibitem[Burstein \& Heiles(1984)]{Burstein84} Burstein, D., \& Heiles, C. 
1984, \apjs, 54, 33

\bibitem[Carlberg et~al.(1997)]{Carlberg97} Carlberg, R. G., Yee, H. K. C., 
\& Ellingson, E. 1997, \apj, 478, 462

\bibitem[Christlein \& Zabludoff(2003)]{Christlein03} Christlein, D., \& 
Zabludoff, A. I. 2003, \apj, 591, 764

\bibitem[Cole \& Lacey(1996)]{Cole96} Cole, S., \& Lacey, C. 1996, 
\mnras, 281, 716

\bibitem[Coleman et~al.(1980)]{Coleman80} Coleman, G. D., Wu, C. C., 
\& Weedman, D. W. 1980, \apjs, 43, 393

\bibitem[Colless(1989)]{Colless89} Colless, M. 1989, \mnras, 237, 799

\bibitem[De~Propris et~al.(2003)]{DePropris03} De Propris, R., et al. 
2003, \mnras, 342, 725

\bibitem[De~Propris et~al.(2004)]{dePropris99} De Propris, R., Stanford, 
S. A., Eisenhardt, P. R., Dickinson, M., \& Elston, R. 2004, \aj, 118, 719

\bibitem[Dressler(1978)]{Dressler78} Dressler, A. 1978, \apj, 223, 765

\bibitem[Dressler(1980)]{Dressler80} Dressler, A. 1980, \apj, 236, 351

\bibitem[Dressler et~al.(1999)]{Dressler99} Dressler, A., Smail, I., 
Poggianti, B. M., Butcher, H., Couch, W. J., Ellis, R. S., \& Oemler, 
A. 1999, \apjs, 122, 52

\bibitem[Driver et~al.(1994)]{Driver94} Driver, S. P., Phillipps, S., 
Davies, J. I., Morgan, I., \& Disney, M. J. 1994, \mnras, 268, 393

\bibitem[Driver et~al.(1998)]{Driver98} Driver, S. P., Couch, W. J., \& 
Phillipps, S. 1998, \mnras, 301, 369

\bibitem[Dubinski(1998)]{Dubinski98} Dubinski, J. 1998, \apj, 502, 141

\bibitem[Ebeling et~al.(1996)]{Ebeling96} Ebeling, H., Voges, W., 
B\"{o}hringer, H., Edges, A. C., Huchra, J. P., \& Briel, U. G. 1996, 
\mnras, 281, 799

\bibitem[Ebeling et~al.(2000)]{Ebeling00} Ebeling, H., Edges, A. C., 
Allen, S. W., Crawford, C. S., Fabian, A. C., \& Huchra, J. P. 2000, 
\mnras, 318, 333

\bibitem[Eddington(1940)]{edd40} Eddington, A. S. 1940, \mnras, 100, 354

\bibitem[Ellingson et~al.(2001)]{Ellingson01} Ellingson, E., Lin, H., 
Yee, H. K. C., \& Carlberg, R. G. 2001, \apj, 547, 609

\bibitem[Ellingson(2003)]{Ellingson03} Ellingson, E. 2003, \apss, 285, 9

\bibitem[Ferguson \& Binggeli(1994)]{Ferguson94} Ferguson, H. C., \& 
Binggeli, B. 1994, \aapr, 6, 67

\bibitem[Frei \& Gunn(1994)]{Frei94} Frei, Z., \& Gunn, J. E. 1994, \aj, 
108, 1476

\bibitem[Fukugita et~al.(1995)]{Fukugita95} Fukugita, M., Shimasaku, K., 
\& Ichikawa, T. 1995, \pasp, 107, 945

\bibitem[Gaidos(1997)]{Gaidos97} Gaidos, E. J. 1997, \aj, 113, 117

\bibitem[Gallagher \& Wyse(1994)]{Gallagher94} Gallagher, J. S., \& 
Wyse, R. F. G. 1994, \pasp, 106, 1225

\bibitem[Garilli et~al.(1999)]{Garilli99} Garilli, B., Maccagni, D., \& 
Andreon, S. 1999, \aap, 342, 408

\bibitem[Geller \& Peebles(1976)]{Geller76} Geller, M. J., \& Peebles, 
P. J. E. 1976, \apj, 206, 939

\bibitem[Gladders et~al.(1999)]{Gladders99} Gladders, M. J., 
L\'opez-Cruz, O., Yee, H. K. C., \& Kodama, T. 1999, \apj, 501, 571

\bibitem[Godwin \& Peach(1977)]{Godwin77} Godwin, J. G., \& Peach, 
J. V. 1977, \mnras, 181, 323

\bibitem[Gonz\'alez et~al.(2006)]{Gonzalez06} Gonz\'alez, R. E., Lares, M., 
Lambas, D. G., \& Valotto, C. 2006, \aap, 445, 51

\bibitem[Goto(2005)]{Goto05a} Goto, T. 2005, \mnras, 359, 1415

\bibitem[Goto et~al.(2002)]{Goto02} Goto, T., et al. 2002, \pasj, 54, 515

\bibitem[Goto et~al.(2005)]{Goto05} Goto, T., et al. 2005, \apj, 621, 188

\bibitem[Hansen et~al.(2005)]{Hansen05} Hansen, S. M., McKay, T. A., 
Wechsler, R. H., Annis, J., Sheldon, E. S., \& Kimball, A. 2005, 
\apj, 633, 122

\bibitem[Hausman \& Ostriker(1978)]{Hausman78} Hausman, M. A., \& Ostriker, 
J. P. 1978, \apj, 224, 320

\bibitem[Hilker et~al.(2003)]{Hilker03} Hilker, M., Mieske, S., \& 
Infante, L. 2003, \aap, 397, L9

\bibitem[Hilton et~al.(2005)]{Hilton05} Hilton, M., et al. 2005, \mnras, 
363, 661

\bibitem[Holden et~al.(2004)]{Holden04} Holden, B. P., Stanford, S. A., 
Eisenhardt, P., \& Dickinson, M. 2004, \aj, 127, 2484

\bibitem[Hubble(1936)]{Hubble36} Hubble, E. P. 1936, \apj, 84, 158

\bibitem[Humason et~al.(1956)]{Humason56} Humason, M. L., 
Mayall, N. U., \& Sandage, A. R. 1956, \aj, 61, 97

\bibitem[Jones \& Forman(1999)]{Jones99} Jones, C., \& Forman, W. 1999, 
\apj, 511, 65

\bibitem[Landolt(1992)]{Landolt92} Landolt, A. U. 1992, \aj, 104, 372

\bibitem[Loh \& Strauss(2006)]{Loh06} Loh, Y. -S., \& Strauss, M. A. 
2006, \mnras, 366, 373

\bibitem[L\'opez-Cruz(1997)]{Lopez97} L\'opez-Cruz O. 1997, Ph.D. thesis 
Univ. Toronto

\bibitem[L\'opez-Cruz(2001)]{Lopez01} L\'opez-Cruz, O. 2001, Rev. Mex. 
AA Conf. Ser., 11, 183

\bibitem[L\'opez-Cruz et~al.(1997)]{Lopez97b} L\'opez-Cruz, O., Yee, H. K. C., 
Brown, J. P., Jones. C., \& Forman, W. 1997, \apj, 475, L97

\bibitem[L\'opez-Cruz et~al.(2004)]{Lopez04} L\'opez-Cruz, O., Barkhouse, 
W. A., \& Yee, H. K. C. 2004, \apj, 614, 679 

\bibitem[Lugger(1986)]{Lugger86} Lugger, P. M. 1986, \apj, 303, 535

\bibitem[Madgwick et~al.(2002)]{Madgwick02} Madgwick,D. S. et al. 2002, 
\mnras, 333, 133

\bibitem[Mercurio et~al.(2003)]{Mercurio03} Mercurio, A., Massarotti, 
M., Merluzzi, P., Girardi, M., La Barbera, F., \& Busarello, G. 2003, 
\aap, 408, 57

\bibitem[Merritt(1984)]{merritt84} Merritt, D. 1984, \apj, 276, 26 

\bibitem[Miller et~al.(2005)]{Miller05} Miller, C. J., et al. 2005, \aj, 
130, 968

\bibitem[Molinari et~al.(1998)]{Molinari98} Molinari, E., Chincarini, G., 
Moretti, A., \& De Grandi, S. 1998, \aap, 338, 874

\bibitem[Moore et~al.(1998)]{Moore98} Moore, B, Lake, G., \& Katz, N. 1998,
\apj, 495, 139

\bibitem[N\"{a}slund et~al.(2000)]{Naslund00} N\"{a}slund, 
M., Fransson, C., \& Huldtgren, M. 2000, \aap, 356, 435

\bibitem[Oemler(1974)]{Oemler74} Oemler, A. 1974, \apj, 194, 1

\bibitem[Oke \& Sandage(1968)]{Oke68} Oke, J. B., \& Sandage, A. 1968, 
\apj, 154, 21

\bibitem[Ostriker \& Hausman(1977)]{Ostriker77} Ostriker, J. P., \& 
Hausman, M. A. 1977, \apj, 217, 125

\bibitem[Parolin et~al.(2003)]{Parolin03} Parolin, I., 
Molinari, E., \& Chincarini, G. 2003, \aap, 407, 823 

\bibitem[Piranomonte et~al.(2001)]{Piranomonte01} Piranomonte, S., Longo, G., 
Andreon, S., Puddu, E., Paolillo, M., Scaramella, R., Gal, R., \& 
Djorgovski, S. G. 2001, in ASP Conf. Ser. 225, Virtual Observatories of the 
Future, ed. R. J. Brunner, S. G. Djorgovski, \& A. S. Szalay 
(San Francisco: ASP), 73

\bibitem[Popesso et~al.(2005)]{Popesso05} Popesso, P., B\"{o}hringer, 
H., Romaniello, M., \& Voges, W. 2005, \aap, 433, 415

\bibitem[Popesso et~al.(2006)]{Popesso06} Popesso, P., Biviano, A., 
B\"{o}hringer, H., \& Romaniello, M. 2006, \aap, 445, 29

\bibitem[Popesso et~al.(2007)]{Popesso07} Popesso, P., Biviano, A., 
B\"{o}hringer, H., \& Romaniello, M. 2007, \aap, 461, 397

\bibitem[Press \& Schecter(1974)]{Press74} Press, W. H., \& Schechter, P. 
1974, \apj, 187, 425

\bibitem[Press et~al.(1992)]{Press92} Press, W. H., Teukolsky, S. A., 
Vetterling, W. T., \& Flannery, B. P. 1992, Numerical Recipes, The 
Art of Scientific Computing, (2d ed.; Cambridge: Cambridge University Press)

\bibitem[Rines et~al.(2003)]{Rines03} Rines, K., Geller, M. J., Kurtz, M. J., 
\& Diaferio, A. 2003, \apj, 126, 2152

\bibitem[Roediger et~al.(2006)]{Roediger06} Roediger, E., Br\"{u}ggen, M., 
\& Hoeft, M. 2006, \mnras, 371, 609

\bibitem[Sandage(1976)]{Sandage76} Sandage, A. 1976, \apj, 205, 6

\bibitem[Schechter(1976)]{Schechter76} Schechter, P. 1976, \apj, 203, 297

\bibitem[Spergel et~al.(2003)]{Spergel03} Spergel, D. N., et al., 2003, 
\apjs, 148, 175

\bibitem[Stanford et~al.(1998)]{Stanford98} Stanford, S. A., Eisenhardt, 
P. R., \& Dickinson, M. 1998, \apj, 492, 461

\bibitem[Struble \& Rood(1999)]{struble99} Struble, M. F., \& Rood, H. J. 
1999, \apj, 125, 35

\bibitem[Thompson \& Gregory(1993)]{Thompson93} Thompson, L. A., \&  Gregory, 
S. A. 1993, \aj, 106, 2197

\bibitem[Trentham et~al.(2001)]{Trentham01} Trentham, N., Tully, R. B., \& 
Verheijen, M. A. W. 2001, \mnras, 325, 385

\bibitem[Valotto et~al.(2001)]{val01} Valotto, C. A., Moore, B., \& 
Lambas, D. G. 2001, \apj, 546, 157

\bibitem[Yagi et~al.(2002)]{Yagi02} Yagi, M., Kashikawa, N., Sekiguchi, M., 
Doi, M., Yasuda, N., Shimasaku, K., \& Okamura, S. 2002, \aj, 123, 87

\bibitem[Yang et~al.(2004)]{Yang04} Yang, Y., Zhou, X., Yuan, Q., 
Jiang, Z., Ma, J., Wu, H., \& Chen, J. 2004, \apj, 600, 141

\bibitem[Yee(1991)]{Yee91} Yee, H. K. C. 1991, \pasp, 103, 396

\bibitem[Yee et~al.(1996)]{Yee96} Yee, H. K. C., Ellingson, E., \& 
Carlberg, R. G. 1996, \apjs, 102, 269

\bibitem[Yee \& L\'opez-Cruz(1999)]{Yee99} Yee, H. K. C., \& L\'opez-Cruz, 
O. 1999, \aj, 117, 1985

\bibitem[Yee \& Ellingson(2003)]{Yee03} Yee, H. K. C. \&  Ellingson, E. 
2003, \apj, 585, 215

\end{thebibliography}
\end{document}